\documentclass[twocolumn,superscriptaddress,notitlepage,longbibliography,floatfix,prx,aps]{revtex4-2}
\usepackage{color}
\usepackage{textcomp}
\usepackage{amsmath}
\usepackage{amssymb}
\usepackage{mathtools}
\usepackage{graphicx}
\usepackage{esint}
\usepackage{framed} 
\usepackage[dvipsnames]{xcolor}
\usepackage{hyperref}
\usepackage{enumerate}
\usepackage{enumitem}
\usepackage{moreenum}
\hypersetup{
  linkcolor = black,
  citecolor  = blue,
  urlcolor = black,
  colorlinks = true,
}

\usepackage{amsfonts}
\usepackage{bm}
\usepackage{bbold}
\usepackage{natbib}
\usepackage{svg}
\usepackage{float}
\usepackage[export]{adjustbox}
\hypersetup{    colorlinks,    linkcolor={blue!50!black},    citecolor={blue!50!black},    urlcolor={blue!80!black}}

\newcommand*{\bra}[1]{\ensuremath{\langle #1 \vert}}
\newcommand*{\ket}[1]{\ensuremath{\vert #1 \rangle}}
\newcommand{\braket}[2]{\langle #1 | #2 \rangle}

\renewcommand{\vec}[1]{{\boldsymbol{#1}}}

\DeclareMathOperator{\arccot}{arccot}
\DeclareMathOperator{\artanh}{artanh}

\usepackage{comment}
\usepackage{lipsum}

\usepackage[normalem]{ulem}

\begin{document}

\def \IBK{Institute for Theoretical Physics, University of Innsbruck, 6020 Innsbruck, Austria}
\def \IQOQI{Institute for Quantum Optics and Quantum Information of the Austrian Academy of Sciences, 6020 Innsbruck, Austria}
\def \SSM{Scuola Superiore Meridionale, Largo San Marcellino 10, 80138 Napoli, Italy}
\def \HongKong{Quantum Science Center of Guangdong-Hong Kong-Macao Greater Bay Area, 518045 Shenzhen, China}
\def \Harvard{Department of Physics, Harvard University, Cambridge, Massachusetts 02138, USA}
\def \Padua{Dipartimento di Fisica e Astronomia, Universit\`a degli Studi di Padova, I-35131 Padova, Italy}
\def \PaduaINFN{Istituto Nazionale di Fisica Nucleare (INFN), Sezione di Padova, I-35131 Padova, Italy}

\title{Quantum Adiabatic Optimization with Rydberg Arrays:  Localization Phenomena and Encoding Strategies}

\author{Lisa Bombieri}\thanks{These authors contributed equally to this work.}\affiliation{\IBK}\affiliation{\IQOQI}
\author{Zhongda Zeng}\thanks{These authors contributed equally to this work.}\affiliation{\IBK}\affiliation{\IQOQI}
\author{Roberto Tricarico}\thanks{These authors contributed equally to this work.}\affiliation{\IBK}\affiliation{\IQOQI}\affiliation{\SSM}
\author{Rui Lin}\affiliation{\IBK}\affiliation{\IQOQI}
\affiliation{\HongKong}
\author{Simone Notarnicola}\affiliation{\Harvard}\affiliation{\Padua}\affiliation{\PaduaINFN}
\author{Madelyn Cain}\affiliation{\Harvard}
\author{Mikhail~D.~Lukin}\affiliation{\Harvard}
\author{Hannes Pichler} \email{hannes.pichler@uibk.ac.at}\affiliation{\IBK}\affiliation{\IQOQI}
\preprint{}

\begin{abstract}
Quantum adiabatic optimization seeks to solve combinatorial problems using quantum dynamics, requiring the Hamiltonian of the system to align with the problem of interest. However, these Hamiltonians are often incompatible with the native constraints of quantum hardware, necessitating encoding strategies to map the original problem into a hardware-conformant form.
While the classical overhead associated with such mappings is easily quantifiable and typically polynomial in problem size, it is much harder to quantify their overhead on the quantum algorithm, e.g., the transformation of the adiabatic timescale. In this work, we address this challenge on the concrete example of the encoding scheme proposed in [Nguyen et al., PRX Quantum \textbf{4}, 010316 (2023)], which is designed to map optimization problems on arbitrarily connected graphs into Rydberg atom arrays. We consider the fundamental building blocks underlying this encoding scheme and determine the scaling of the minimum gap with system size along adiabatic protocols. Even when the original problem is trivially solvable, we find that the encoded problem can exhibit an exponentially closing minimum gap. We show that this originates from a quantum coherent effect, which gives rise to an unfavorable localization of the ground-state wavefunction. On the QuEra Aquila neutral atom machine, we observe such localization and its effect on the success probability of finding the correct solution to the encoded optimization problem. Finally, we propose quantum-aware modifications of the encoding scheme that avoid this quantum bottleneck and lead to an exponential improvement in the adiabatic performance. This highlights the crucial importance of accounting for quantum effects when designing strategies to encode classical problems onto quantum platforms.  
\end{abstract}

\maketitle

 \section{Introduction }\label{sec:introduction}
The past few decades have seen a growing interest in harnessing the dynamics of many-body quantum systems to explore potential quantum advantage in solving combinatorial optimization problems~\cite{Johnson_2011,Pagano_2020,Harrigan_2021,Ebadi_2022}.
For problems that can be reformulated in terms of a classical spin Hamiltonian whose ground state represents the solution, quantum adiabatic optimization stands out as one of the most promising quantum algorithms for pursuing such an advantage~\cite{Albash_2018,Lucas_2014,Kadowaki_1998,Farhi_2000,Farhi_2001,Arnab_2008,Hauke_2020}. 
Here, the Hamiltonian of the system is adiabatically evolved from one with a known and easy-to-prepare ground state to the target one.
In this way, the system follows the instantaneous ground state of the time-dependent Hamiltonian throughout the evolution and finally reaches the solution to the optimization problem.

In practice, many problems of interest do not naturally conform with the native constraints of the quantum hardware. Therefore, additional encoding strategies to map them into a hardware-conformant form are required~\cite{Andrew_2014,Mohseni_2022}. 
While the classical overhead associated with such strategies is easily quantifiable and typically polynomial in problem size, understanding their impact on the quantum dynamics of the adiabatic algorithm remains an outstanding challenge.

\begin{figure*}
    \centering
    \includegraphics[width=1.0\textwidth]{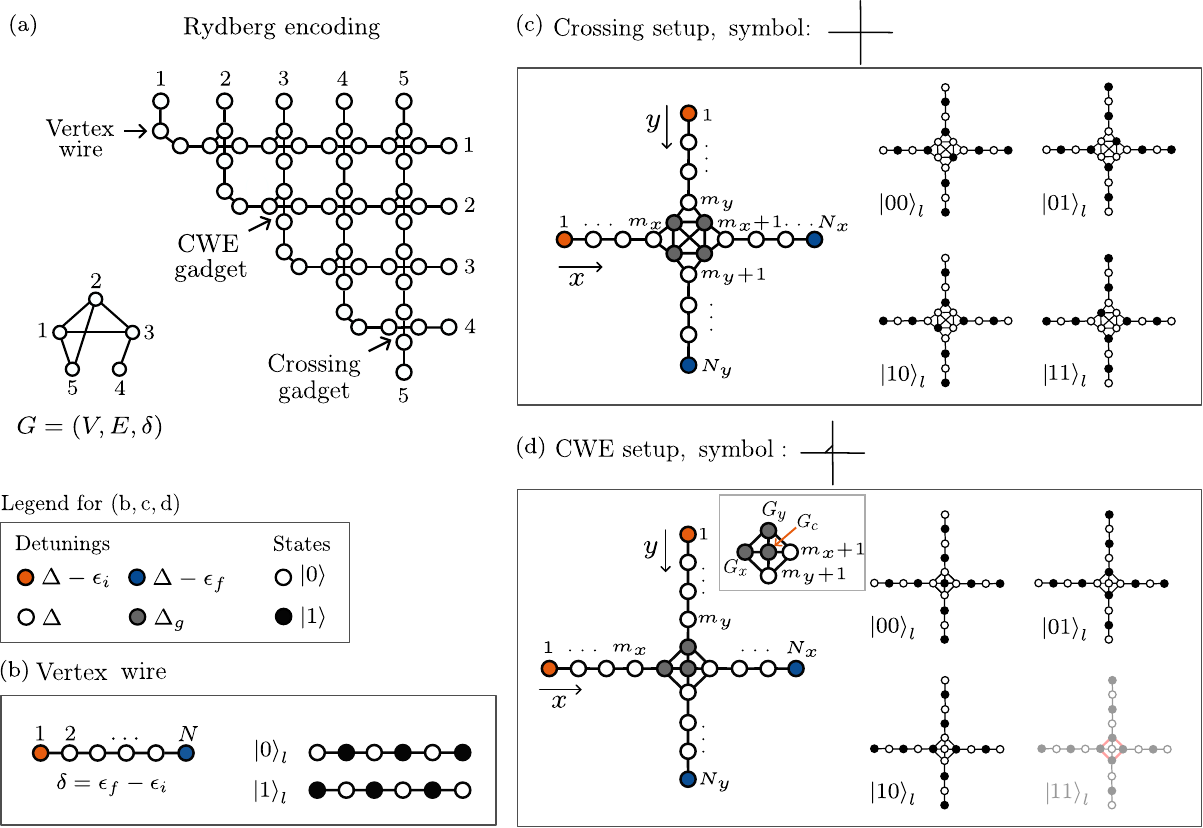}
    \caption{Encoding scheme of the maximum weighted independent set problem on arbitrarily connected graphs into Rydberg atom arrays~\cite{Nguyen_2023}. (a) Example for a non-unit-disk graph of five vertices $G=(V, E,\delta)$. (b) A vertex wire is a one-dimensional array of nearest-neighbor-blockaded
    atoms. (Left) The atoms are labeled from $1$ to $N$, with $N$ even. The detuning in the bulk is denoted by $\Delta$; the detuning on the initial and final atom by $\Delta-\epsilon_i$ and $\Delta-\epsilon_f$. The weight of a graph vertex $\delta$ is mapped to the boundary detuning difference: $\delta=\epsilon_f-\epsilon_i$. (Right) The logical states encoding the binary degrees of freedom of a graph vertex are the $\mathbb{Z}_2$-ordered states of the vertex wire. Here, the atoms in the ground (Rydberg) state are depicted in white (black). The (c) crossing and (d) CWE setups consist of two vertex wires intersecting via the crossing gadget (the four atoms in gray) and the CWE gadget (the three atoms in gray, labeled as in the inset). (Left) The atoms along the horizontal ($w=x$) and vertical ($w=y$) vertex wire are labeled from $1$ to $N_w$; the position of the last atom before the gadget is denoted by $m_w$. $N_w$ and $m_w$ are chosen to be even. The bulk detuning is denoted by $\Delta$; the detuning on the initial and the final atom of the $w$ vertex wire by $\Delta-\epsilon_{i,w}$ and $\Delta-\epsilon_{f,w}$; the detuning on the gadget atoms by $\Delta_g$. (Right) The logical states of the crossing setup are the four possible combinations of logical states of the vertex wires. In the CWE setup, the logical state $\ket{11}_l$ is forbidden due to the blockade mechanism. } \label{fig:setups_gadgets}
\end{figure*}

In this work we address this question, focusing on the example of quantum optimizers based on Rydberg atom arrays~\cite{Pichler_2018a,Ebadi_2022,Graham_2022, Kim_2022,Byun_2022, deOliveira_2024_new,Wurtz_2024a}.
Due to the blockade mechanism, Rydberg atom arrays naturally encode the paradigmatic maximum independent set problem on a specific class of graphs, so-called unit-disk graphs~\cite{Pichler_2018a}, and experiments on these graphs have already shown promising results~\cite{Ebadi_2022}. Several encoding strategies to go beyond the native hardware restriction of unit-disk graphs  have been proposed in the literature
~\cite{Pichler_2018b,Nguyen_2023,Qiu_2020, Lanthaler_2023, Stastny_2023,Goswami_2024,ZZ_2024}. In particular, Ref.~\cite{Nguyen_2023} proposed an encoding strategy that maps an optimization problem on an arbitrarily connected graph into an equivalent maximum weighted independent set (MWIS) problem on a unit-disk graph, which is directly implementable with Rydberg atom arrays (see Fig.~\ref{fig:setups_gadgets}).
Here, the binary degree of freedom associated with each graph vertex is encoded in the $\mathbb{Z}_2$-ordered states of a vertex wire of Rydberg atoms. The absence (presence) of an edge between a pair of vertices is reproduced by placing a crossing (crossing-with-edge) gadget at the intersection between the corresponding vertex wires. So far, this strategy has been mainly investigated on the classical level, and the quantum adiabatic performance, in particular for large systems, remains an outstanding open question. 

In this work, we investigate the quantum adiabatic performance of the main building blocks of this encoding scheme, consisting of two vertex wires intersecting either via the crossing or the crossing-with-edge (CWE) gadget, as depicted in Fig.~\ref{fig:setups_gadgets}(c,d).
When the two vertex wires intersect via the crossing-with-edge gadget, we find that an exponential closing of the minimum gap can occur for certain geometrical configurations, leading to exponentially long algorithm runtimes for large systems. We show that this adversarial gap closing originates from unfavorable localization of the ground-state wavefunction along the adiabatic protocol. Guided by these insights, we propose alternative strategies to overcome such a localization and ensure a polynomial closing of the minimum gap, thereby exponentially improving the adiabatic performance of the encoding scheme. 
In particular, we show how localization can be avoided by (\textit{i}) systematically varying the length of the intersecting vertex wires, or (\textit{ii}) modifying the structure of the crossing-with-edge gadget.

This work is organized as follows. In Sec.~\ref{sec: model}, we revisit the encoding of the MWIS problem on arbitrarily connected graphs into Rydberg atom arrays, as proposed in Ref.~\cite{Nguyen_2023}. After detailing its building blocks (namely, the crossing and the CWE setup), we introduce the two adiabatic protocols that we consider in this work, referred to as standard and logical protocols.
In Sec.~\ref{sec: numerics}, we employ the density-matrix renormalization group (DMRG) method to quantify the adiabatic performance of the two setups along the standard protocol. 
In Sec.~\ref{sec: effective th}, we develop an effective theory to characterize their performance along the logical protocol and explain the origin of the wavefunction localization.
In Sec.~\ref{sec:experiment}, using the QuEra Aquila neutral atom machine~\cite{Wurtz_2023}, we observe such localization phenomenon and its effect on encoded MWIS problems.
Finally, in Sec.~\ref{sec: Improvement}, we present the strategies to overcome the localization of the ground-state wavefunction and achieve an exponential improvement in the adiabatic performance.

\section{Rydberg encoding of maximum weighted independent set problems}\label{sec: model}
In this section, we recall the encoding of the maximum weighted independent set problem on unit-disk graphs into Rydberg atom arrays~\cite{Pichler_2018a,Ebadi_2022} and the gadget-based strategy of Ref.~\cite{Nguyen_2023} to go beyond the unit-disk requirement. We introduce the building blocks of this encoding strategy, and define notations and assumptions for the following discussion. Finally, we present the two adiabatic protocols under analysis.

\subsection{MWIS encoding for unit-disk graphs}
Consider a weighted graph $G = (V, E, \delta)$, where $V$ is the set of vertices (with $|V|$ indicating their total number), $\delta$ is the set of the corresponding weights, and $E$ is the set of edges. 
A set of vertices is said to be independent if no pair of vertices is connected by an edge.
The MWIS problem aims to find an independent set that maximizes the sum of the weights of its vertices.
This combinatorial problem can be reformulated as an optimization problem by assigning to each set of vertices $S\subseteq V$ a bitstring $\bm{\mu}_S=\left(\mu_1,\mu_2,\dots\right) \in \{0,1\}^{|V|}$, where ${\mu_v=1\:(\mu_v=0)}$ if the vertex ${v\in S\: (v\notin S)}$, and minimizing the following classical Hamiltonian over these bitstrings~\cite{Boros_2002,Pichler_2018a}:
\begin{equation}
    H_{\text{MWIS}} = - \sum_{v \in V} \delta_{v} \mu_{v} + \sum_{(v, u) \in E} U \mu_{v} \mu_{u},
    \label{eq:ch1_costfunct_MWIS}
\end{equation}
with $U \gg \delta_v>0$. In the remainder of this section, we consider the MWIS problem on unit-disk graphs, i.e., graphs where two vertices are connected by an edge if and only if their distance is smaller than a unit radius~\cite{Clark_1990}. This problem can be naturally encoded into Rydberg atom arrays, as detailed in the following.

Neutral atoms can be individually trapped by optical tweezers and arranged into arbitrary arrays~\cite{Kaufman_2021, Endres_2016, Barredo_2016, Saskin_2019, Singh_2022, Ebadi_2022, Gyger_2024, Manetsch_2024, Grün_2024_new}. Excitation to Rydberg states induces interactions between atoms, enabling applications not only in quantum optimization but also in quantum computation~\cite{Saffman_2010, Levine_2019, Madjarov_2020, Bluvstein_2022, Wu_2022, Evered_2023, Cesa_2023, Bluvstein_2024, Anand_2024_new, Radnaev_2024} and quantum simulation~\cite{Weimer_2010, Browaeys_2020, Labuhn_2016, Bernien_2017, Keesling_2019, Bluvstein_2021, Ebadi_2021, Scholl_2021, Semeghini_2021, Choi_2023, Eckner_2023, Fang_2024}. Specifically, by approximating the atoms as two-level systems, of ground state $\ket{0}$ and highly excited Rydberg state $\ket{1}$, the many-body dynamics of an array driven by coherent light can be described by the Hamiltonian
\begin{equation}
    H_{\text{Rydberg}} =  \Omega \sum_v \sigma^{x}_{v} - \sum_v \Delta_{v} n_{v} + \sum_{v< u } U_{v, u} n_{v} n_{u},
    \label{eq:hamiltonain_rydberg} 
\end{equation}
where $n_v=\ket{1_v}\bra{1_v}$ and $\sigma^{x}_v=\ket{1_v}\bra{0_v}+\ket{0_v}\bra{1_v}$. Here, $\Omega$ and $\Delta_v$ are the Rabi frequency and the detuning of the driving laser on atom $v$. ${U_{v,u}=C_6/d_{v,u}^6}$ is the van der Waals interaction strength between excited atoms placed at a distance $d_{v,u}$, with $C_6$ being the Van der Waals constant. The interaction strongly penalizes configurations where nearby atoms are simultaneously excited, giving rise to the so-called \textit{blockade mechanism}~\cite{Jaksch_2000,Gaetan_2009,Urban_2009}.
For simplicity, it is convenient to consider the blockade limit of $H_{\text{Rydberg}}$: $U_{v,u}=\infty$ for $d_{v,u}$ smaller than a characteristic distance $R_b=\left(C_6/\sqrt{(2\Omega)^2+\Delta^2}\right)^{1/6}$, known as Rydberg radius, and $U_{v,u}=0$ otherwise. 
In Appendix ~\ref{sec:Appendix_InteractionTails}, we show that the analyses developed in the blockade limit also apply when considering the tail interaction (i.e., $U_{v,u}=C_6/d_{v,u}^6$ for $d_{u,v}>R_b$).

Relying on the blockade mechanism, the mapping of a MWIS problem [Eq.~\eqref{eq:ch1_costfunct_MWIS}] on unit-disk graphs into Rydberg atom arrays is accomplished without any overhead. Here, each vertex of the original graph is represented by an atom of the array, with its binary degree of freedom corresponding to the quantum state of the atom, i.e., $\mu\in\{0,1\}\leftrightarrow\{\ket{0_v}, \ket{1_v}\}$. The Rydberg radius $R_b$ is identified with the unit-disk radius of the graph, such that an edge $(v,u)\in E$ between two vertices corresponds to the two atoms placed at a distance $d_{v,u}<R_b$. In this way, the minimum energy state(s) of the classical Hamiltonian in Eq.~\eqref{eq:ch1_costfunct_MWIS} corresponds to the ground state(s) of Eq.~\eqref{eq:hamiltonain_rydberg} for $\Delta_v=\delta_v$ and $\Omega=0$. 
\subsection{MWIS encoding for arbitrarily connected graphs}
A quadratic overhead in the atom number makes it possible to go beyond the unit-disk requirement and encode MWIS problems on arbitrarily connected graphs into two-dimensional atom arrays. We show the strategy of Ref.~\cite{Nguyen_2023} in Fig.~\ref{fig:setups_gadgets}.
In this scheme, each graph vertex is mapped into a vertex wire: a one-dimensional array of nearest-neighbor-blockaded atoms, whose ground states, in the $\Omega/\Delta\rightarrow0$ limit, are the $\mathbb{Z}_2$-ordered states~\cite{Bernien_2017}. The absence or the presence of an edge between two vertices is encoded by placing the crossing or the CWE gadget at the intersection of the corresponding vertex wires, respectively [see Fig.~\ref{fig:setups_gadgets}(a)]. 

\subsubsection{Single vertex wire}
We consider a vertex wire of $N$ atoms, with $N$ even [see Fig.~\ref{fig:setups_gadgets}(b)]. This is described by the Hamiltonian in Eq.~\eqref{eq:hamiltonain_rydberg}, with homogeneous detuning $\Delta>0$ in the bulk and smaller detuning on the two boundary atoms: $\Delta_1=\Delta-\epsilon_i>0$ and $\Delta_N=\Delta-\epsilon_f>0$. In the blockade limit, the binary states ${\mu \in \{0,1\}}$ of the graph vertex correspond to the lowest two energy states of the Hamiltonian for $\Omega=0$, namely $\ket{0}_l = \ket{0101\dots01}$ and $\ket{1}_l = \ket{1010\dots10}$. We refer to these two $\mathbb{Z}_2$-ordered states as \textit{logical states}. The corresponding energies are $-\Delta N/2+\epsilon_f$ and $-\Delta N/2+\epsilon_i$, respectively. The weight $\delta$ of the graph vertex (also referred to as \textit{wire weight}) is mapped to the difference of these energies ${\delta=\epsilon_f-\epsilon_i}$. 

\subsubsection{Two intersecting vertex wires}
The absence (presence) of an edge between two vertices is encoded by placing a crossing(-with-edge) gadget at the intersection between the corresponding vertex wires. In reference to Fig.~\ref{fig:setups_gadgets}(c,d), we set the following terminology and notation. We use the label $w$ to indicate the vertex wires: the horizontal one is denoted as $w=x$ and is oriented from left to right; the vertical one is denoted as $w=y$ and is oriented from top to bottom. The gadget divides each vertex wire into two \textit{legs}. The total number of atoms $N_w$ of the vertex wire $w$ is the sum of the atoms of its two legs. We indicate the length of the first leg of the vertex wire $w$ (the leg before the gadget, according to its orientation) with $m_w$. We choose both $N_w$ and $m_w$ to be even. We consider homogeneous detuning $\Delta>0$ on the whole system, except for $\Delta-\epsilon_{i,w}>0$ and $\Delta-\epsilon_{f,w}>0$ (with $w=x,y$) on the boundary atoms of the two vertex wires, and $\Delta_g\ge\Delta$ on the gadget atoms [see Fig.~\ref{fig:setups_gadgets}(c,d)].  

The four logical states of the intersecting vertex wires (i.e., the product of the logical states of the individual vertex wires) are denoted with $\ket{\mu_x\mu_y}_l$, where ${\mu_w\in\{0,1\}}$ refers to the logical state of the $w$ vertex wire. For $\Omega=0$ and equal boundary detuning (${\delta_x=\delta_y=0}$), these states constitute the ground states of the crossing setup [see Fig.~\ref{fig:setups_gadgets}(c)(right)], while the $\ket{11}_l$ state is prohibited in the CWE setup by the blockade mechanism [see Fig.~\ref{fig:setups_gadgets}(d)(right)]. Therefore, while the crossing gadget does not introduce any constraint among the logical degrees of freedom of the vertex wires, the CWE gadget effectively imposes $\mu_x \mu_y=0$. These gadgets thus represent the absence and the presence of an edge between two vertices of the original graph, respectively.

\subsection{Adiabatic protocols}
In quantum adiabatic optimization, we consider a time-dependent Hamiltonian $H(\tau)$, with $\tau\in [0, T]$, interpolating between an initial Hamiltonian, with known ground state, and a final one, with the ground state(s) encoding the solution(s) to the optimization problem. For the system to follow the instantaneous ground state of $H(\tau)$ along the evolution, the required algorithm runtime $T$ scales inversely with the minimum energy gap $\delta E_m$, according to the adiabatic theorem~\cite{Jansen_2007,Schaller_2006,Boixo_2010}. Here, $\delta E_m$ is defined as the minimum of the energy difference between the ground state and the first excited state of the time-dependent Hamiltonian. Therefore, if the minimum gap closes exponentially while increasing the system size, the required runtime will generally grow exponentially, leading to a failure of the adiabatic optimization for large systems~\cite{Albash_2018}.

\begin{figure}
    \centering      \includegraphics[width=1.0\columnwidth]{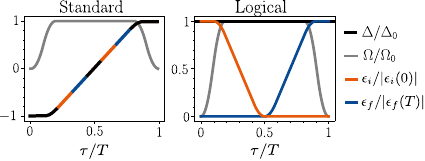}
    \caption{Sweep of Rabi frequency $\Omega(\tau)$, bulk detuning $\Delta(\tau)$, and boundary detuning $\epsilon_{i}(\tau)$ and $\epsilon_{f}(\tau)$ along standard and logical protocols. The sweeps are normalized to the modulus of their maximum value. At the final time, $\epsilon_{f}(T)-\epsilon_{i}(T) = \delta$, with $\delta$ being the wire weight, to ensure that the ground state encodes the solution to the MWIS problem. In the standard protocol, $\Delta$, $\epsilon_i$ and $\epsilon_f$ have the same normalized sweep profile.
    }
    \label{fig:protocol1and2}
\end{figure}

In this work, the target Hamiltonian is reached via two different adiabatic protocols, shown in Fig.~\ref{fig:protocol1and2}, which we refer to as \textit{standard} and \textit{logical} protocols. 
\subsubsection{Standard protocol}
The standard protocol, shown in Fig.~\ref{fig:protocol1and2}, is commonly implemented in experiments~\cite{Ebadi_2022}. All atoms are initialized in the $\ket{0}$ state, which is the ground state of the system for $\Delta(0)=-\Delta_0<0$ and $\Omega(0)=0$. Then, $\Omega(\tau)$ is ramped up to a finite value $\Omega_0$ and eventually down to zero, while $\Delta(\tau)$ is increased up to $\Delta_0$. The boundary detuning of each vertex wire are varied proportionally to $\Delta(\tau)$, keeping their difference fixed to $(\epsilon_f(\tau)-\epsilon_i(\tau))/\Delta(\tau)=\delta/\Delta_0$, with $\delta$ being the wire weight.  

\subsubsection{Logical protocol}
As an alternative to the standard protocol, we consider the logical protocol shown in Fig.~\ref{fig:protocol1and2}, which can be implemented with the same experimental resources as the standard one, and whose performance can be analyzed using perturbation theory. In this protocol, all vertex wires are initialized in the logical state $\ket{0}_l$, which is the ground state of the system for $\Delta(0)=\Delta_0>0$, $\Omega(0)=0$, and ${\epsilon_i(0) > \epsilon_f(0)}$. Then, $\Omega(\tau)$ is ramped up to a finite value $\Omega_0$ and eventually down to zero, while $\Delta(\tau)$ is kept constant along the protocol. On each vertex wire, $\epsilon_i(\tau)$ is ramped down while $\epsilon_f(\tau)$ is ramped up to satisfy the weight condition at final time: $\epsilon_f(T)-\epsilon_i(T)=\delta$, with $\delta$ being the wire weight.

\section{Adiabatic performance with standard protocol} \label{sec: numerics}
In this section, we investigate the adiabatic performance of two vertex wires intersecting via a gadget when considering the standard protocol of Fig.~\ref{fig:protocol1and2}. 
To perform this analysis, we develop an intersecting tensor-network method (see details in Appendix~\ref{sec:Appendix_DMRG}) and conduct a variational study using DMRG~\cite{White_1992,White_1993,Hubig_2015}. 

\subsection{Crossing setup} \label{sec:Crossing}

We consider the ground state of two vertex wires intersecting via the crossing gadget [see Fig.~\ref{fig:setups_gadgets}(c)]. Depending on the value of $\Delta_g/\Delta$, the gadget may or may not host a Rydberg excitation. This effectively imposes different boundary constraints on the legs, leading to various ground-state regimes. The different regimes can be classified from the ground-state properties, whose
changes can be tracked by computing the \textit{fidelity susceptibility}
\cite{Zanardi_2006,Gu_2010}:
\begin{equation}\label{eq:crossing_chi}
    \chi = -\frac{2}{\delta \lambda^2}\ln\left|\braket{\mathrm{GS}_{\lambda}}{\mathrm{GS}_{\lambda + \delta {\lambda}}}\right|,
\end{equation}
where $\lambda=\Delta/\Omega$, and $|\mathrm{GS}_{\lambda}\rangle$ denotes the ground state of the corresponding Rydberg Hamiltonian. 
\begin{figure}
    \centering \includegraphics[width=1\columnwidth]{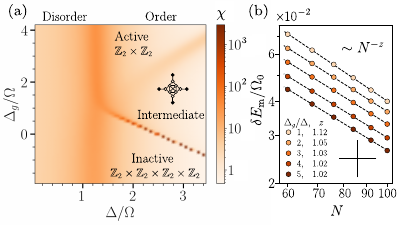}
    \caption{Analysis of the crossing setup. (a) Fidelity susceptibility $\chi$ in Eq.~\eqref{eq:crossing_chi} (computed with ${\delta \lambda = 0.05}$), as a function of $\Delta_g/\Omega$ and $\Delta/\Omega$. We consider two vertex wires of $N=100$ atoms intersecting at the middle (${m_x=m_y=N/2}$). (b) Scaling of the minimum gap $\delta E_m$ (normalized to the maximum Rabi frequency $\Omega_0$) along the standard protocol in Fig.~\ref{fig:protocol1and2} as a function of $N$, for different values of ${\Delta_g/\Delta}$.
    }
    \label{fig:crossing}
\end{figure}

In Fig.~\ref{fig:crossing}(a), we show $\chi$ as a function of $\Delta/\Omega$ and $\Delta_g/\Omega$ for the crossing setup in a symmetric configuration, where the four legs attached to the gadget have the same number of atoms (${m_x=m_y=N/2}$). We set ${\epsilon_{i,w}=\epsilon_{f,w}=0}$, with $w=x,y$, and consider a large system of $N=100$ atoms. 

As a dominant feature, we observe a high susceptibility around $\Delta/\Omega\sim 1.3$, where the system undergoes a phase transition between the \textit{disordered} phase and the \textit{ordered} one. This transition corresponds to the four legs undergoing (individually, or in pairs) the $\mathbb{Z}_2$ spontaneous symmetry breaking observed in one-dimensional Rydberg chains with nearest-neighbor blockade~\cite{Bernien_2017,Sachdev_2002,Fendley_2004}. 
More precisely, for $\Delta_g<0$, the gadget is ``empty'', i.e., all the gadget atoms are in the ground state. This has the effect of decoupling the four legs, which behave as independent Rydberg chains with open boundary conditions. In the transition from disorder to order, they thus individually break the $\mathbb{Z}_2$ symmetry, and the system enters an ``inactive'' ordered regime. Here, the ground state is given by one of the $2^4$ combinations of the $\mathbb{Z}_2$ states of the legs (and empty gadget). We describe the overall transition as a spontaneous breaking of a $\mathbb{Z}_2 \times \mathbb{Z}_2 \times \mathbb{Z}_2 \times \mathbb{Z}_2$ symmetry. In the opposite limit of $\Delta_g \gtrsim \Delta$, the gadget always hosts a Rydberg excitation. This has the effect of connecting opposite legs, which behave as single vertex wires. In this case, in the transition from disorder to order, the system spontaneously breaks a $\mathbb{Z}_2 \times \mathbb{Z}_2$ symmetry and enters an ``active'' ordered regime. Here, the ground state is given by one of the $2^2$ combinations of the $\mathbb{Z}_2$ states of the vertex wires. 

Within the ordered phase, an intermediate region separates the inactive regime from the active one. Here, the gadget is disentangled from the rest of the system and hosts a Rydberg excitation delocalized among its atoms. 
This has the effect of pinning the neighboring atoms to the ground state and thus picking a specific state for the legs from among the $2^4$ possible $\mathbb{Z}_2$ combinations. In Appendix~\ref{sec:Appendix_crossing}, we provide further insights into the different ordered regimes via second-order perturbation theory.

We now consider the scaling of the minimum gap, located around the transition from the disordered phase to the active regime at $\Delta/\Omega\sim 1.3$, along the standard protocol. We target the $\ket{11}_l$ state, encoding the solution to the trivial MWIS problem for a graph of two disconnected vertices with equal weights, by setting ${\epsilon_i(\tau)=0.8\Delta(\tau)}$ and ${\epsilon_f(\tau)=0.2\Delta(\tau)}$ on both vertex wires.   
In Fig.~\ref{fig:crossing}(b), we show that $\delta E_m$ scales as $1/L$ for $\Delta_g/\Delta>1$ (Ising transition, see Appendix~\ref{sec:Appendix_crossing}), and it still closes polynomially for $\Delta_g=\Delta$. We thus conclude that the crossing gadget does not constitute an obstruction to the adiabatic protocol. This result is shown for the symmetric configuration, but we confirmed that it is independent of where the two vertex wires intersect. 

\subsection{Crossing-with-edge setup}
\begin{figure}
    \centering
    \includegraphics[width=1.0\columnwidth]{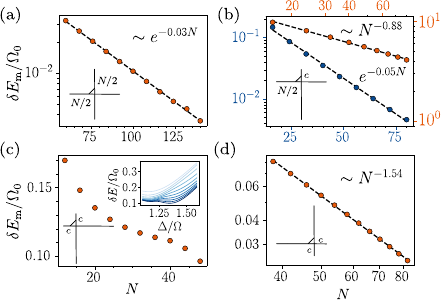}
    \caption{Scaling of the minimum gap $\delta E_m$ along the standard protocol for the crossing-with-edge setup. $\delta E_m$ (normalized to the maximum Rabi frequency $\Omega_0$) as a function of the number of atoms per vertex wire $N$, for the configurations:
    (a) $m_x=m_y=N/2$, (b) $m_x=N/2$, $m_y=4$, (c) $m_x=m_y=4$, and (d) $m_x=m_y=N-4$. For symmetric geometries (a,c,d), the ground state at final time is $\ket{01}_l$. In the asymmetric geometry (b), $\ket{10}_l$ (blue) and $\ket{01}_l$ (orange) are considered separately.
    In the top-right inset of (c), the gap $\delta E$ is shown as a function of $\Delta/\Omega$, for $N$ ranging from $8$ (light blue) to $48$ (dark blue).}     \label{fig:protocol1_scaling_different_geometries}
\end{figure}
Two vertex wires intersecting via the CWE gadget [see Fig.~\ref{fig:setups_gadgets}(d)] encode the simple graph consisting of two vertices connected by an edge. Even though the two vertices play a symmetric role in the original graph, this is not necessarily the case for the two vertex wires if $m_x\neq m_y$. We thus consider, for a given geometry (i.e., for a given choice of $m_x$ and $m_y$), two distinct MWIS problems, where $\delta_x>\delta_y$ and $\delta_x<\delta_y$. The solution to these problems is encoded in the $\ket{10}_l$ and $\ket{01}_l$ states of the two vertex wires, respectively.  Motivated by the possible positions of the gadget in a large network, we analyze four different geometrical configurations with $N_x=N_y=N$. We consider the gadget placed: (a) at the midpoint of the two vertex wires (symmetric case, ${m_x=m_y=N/2}$); (b) close to the beginning of the $y$ vertex wire ($m_x=N/2$, $m_y=c=4$); (c) close to the beginning of both vertex wires ($m_x=m_y=c=4$); (d) close to the end of both vertex wires (${m_x=m_y=N-c}$, $c=4$).

We characterize the adiabatic performance of the system, for $\Delta_g=\Delta$, by computing the scaling of the minimum gap along the standard protocol. To target the $\ket{10}_l$ solution, we set the boundary detuning at the end of the protocol to $(\epsilon_{i,x}(T),\epsilon_{f,x}(T))=(0.2,0.8)\Delta$ and $(\epsilon_{i,y}(T),\epsilon_{f,y}(T))=(0.5,0.5)\Delta$, corresponding to $\delta_x/\Delta_0=0.6$ and $\delta_y/\Delta_0=0$. Analogously, to target the $\ket{01}_l$ solution, we switch the profiles of $x$ and $y$ of the boundary detuning. 

In Fig.~\ref{fig:protocol1_scaling_different_geometries}, we illustrate the scaling of the minimum gap for the different geometries of interest. Even though solving the MWIS problem for two vertices connected by an edge is trivial, within the gadget-encoding strategy, the adiabatic performance strongly depends on the considered geometry and target solution. Indeed, for the same graph problem, we observe both polynomially and exponentially closing minimum gaps, as shown in Fig.~\ref{fig:protocol1_scaling_different_geometries}. For instance, in the (b) configuration, $\delta E_m$ closes polynomially if the target solution is $|01\rangle_{l}$ and exponentially if the target solution is $|10\rangle_{l}$: it is more challenging to reach the $|1\rangle_{l}$ state along the vertex wire with the longest leg before the gadget.

If this effect is extrapolated to larger networks with more than two intersecting wires, where the target solution is generally unknown and careful gadget positioning is thus not possible, one can expect that exponential gap closing will occur generically.
In Secs.~\ref{sec: effective th} and~\ref{sec: Improvement}, we, therefore, investigate the physical origin of this exponential gap and present strategies to circumvent it.

\section{Perturbation theory and adiabatic performance with logical protocol}\label{sec: effective th}
\begin{figure*}[ht]
    \centering
    \includegraphics[width=1.0\textwidth]{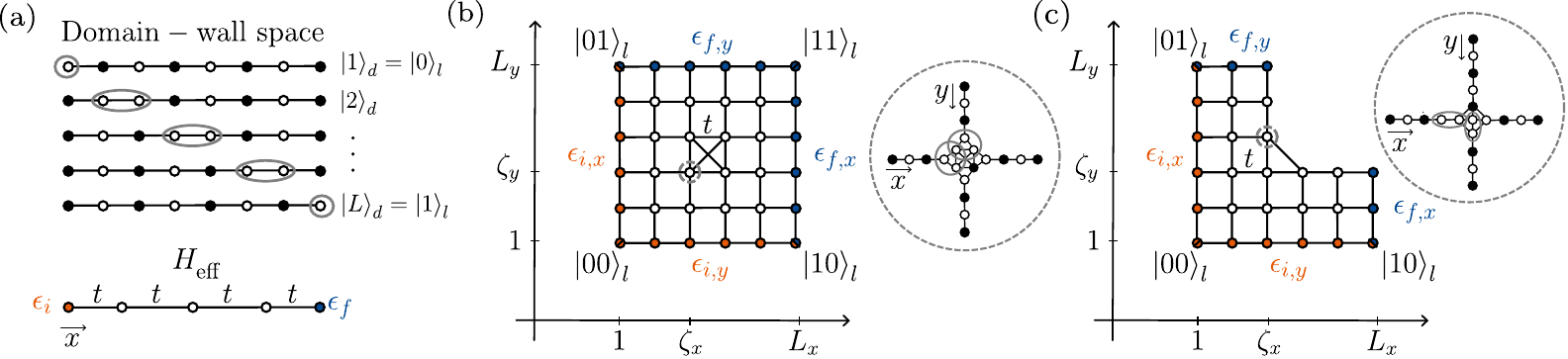}
    \caption{Schematic representation of the effective Hamiltonian. (a) (Top) For a vertex wire of $N$ atoms, the domain-wall states are characterized by two nearby atoms in the ground state (circled in gray), constituting the interface between two opposite $\mathbb{Z}_2$-ordered regions. They are labeled as $\ket{j}_d$, with $j=1,\dots, L=N/2+1$, according to the possible locations of the domain wall along the vertex wire. (Bottom) Illustration of the effective Hamiltonian in Eq.~\eqref{eq:chain_Heff_}. Each domain-wall state is represented as a site of a 1D lattice. Links represent hopping terms between the connected sites; diagonal terms are represented via colors: white for zero, orange for $\epsilon_i$, and blue for $\epsilon_f$. (b,c)  Illustration of the effective Hamiltonian in Eq.~\eqref{eq:Heff_2wires_general} for the (b) crossing and (c) crossing-with-edge setups. The two-domain-wall states, labeled by $\ket{\vec{j}}_d=\ket{j_x, j_y}_d$, form a 2D lattice. Links represent hopping terms; diagonal terms are represented via colors: white for zero, orange for $\epsilon_{i,w}$, and blue for $\epsilon_{f,w}$ (with $w=x,y$). The sites at the corners of the lattice correspond to the logical states, and the diagonal terms on them are the sum of two boundary detuning (e.g., $\epsilon_{i,x}+\epsilon_{i,y}$ for $\ket{00}_l$). The insets show the states $\ket{\zeta_x, \zeta_y}_d$ (for the crossing setup) and $\ket{\zeta_x, \zeta_y+1}_d$ (for the crossing-with-edge setup). }
    \label{fig:effective_th}
\end{figure*}
In this section, we develop a perturbative description to gain insights into the dynamics and adiabatic performance of two intersecting vertex wires, in the specific case of $\Delta_g=\Delta$. We develop an effective model within the lowest-energy subspace of the Rydberg Hamiltonian in  Eq.~\eqref{eq:hamiltonain_rydberg} for $\Omega,\epsilon\ll\Delta$, which consists of states with a single domain wall along each vertex wire. This model allows us to understand the relationship between the geometry of the CWE setup and the minimum gap scaling along the logical protocol.
\subsection{Vertex wire}
We first discuss the case of a single vertex wire [see Fig.~\ref{fig:setups_gadgets}(b)]. For $\Omega=\epsilon_i=\epsilon_f=0$, the lowest-energy subspace of the Rydberg Hamiltonian in Eq.~\eqref{eq:hamiltonain_rydberg} is spanned by the so-called \textit{domain-wall} states, shown in Fig.~\ref{fig:effective_th}(a). The domain-wall states are defined by the presence of two neighboring atoms in the ground state, i.e., a domain wall, constituting the interface between two opposite $\mathbb{Z}_2$-ordered regions. We label these states as $\ket{j}_d$, with ${j=1, \dots, L=N/2+1}$, according to the possible domain-wall locations along the vertex wire. We note that the two logical states correspond to $\ket{1}_d=\ket{0}_l$ and $\ket{L}_d=\ket{1}_l$.   

For finite $\Omega\ll\Delta$, the domain wall acquires kinetic energy and can change position. The dynamics reduce to a single-particle hopping on a one-dimensional (1D) line, described by the following effective Hamiltonian (see Appendix~\ref{sec:effective_th}):
\begin{equation}
    H_{\text{eff}} = - \epsilon_{i} \ket{1}_d\bra{1} - \epsilon_{f} \ket{L}_d \bra{L} - t \sum_{j=1}^{L-1} \left(\ket{j}_d\bra{j+1} + \text{h.c}\right),
    \label{eq:chain_Heff_}
\end{equation}
with hopping amplitude $t=\Omega^2/\Delta$. This equation is valid for $\epsilon_i,\epsilon_f\ll\Delta$, and it is pictorially illustrated in Fig.~\ref{fig:effective_th}(a). One can easily see that, within the logical protocol (which explores only the ordered phase of the system), the minimum gap scales as $1/L^2$ (see Appendix~\ref{sec:Appendix_chain}), contrary to the $1/L$ scaling in the standard protocol.

\subsection{Crossing setup}

We now describe the crossing setup with an analogous formalism. The ground states of Eq.~\eqref{eq:hamiltonain_rydberg} for ${\Omega=\epsilon_{i,w}=\epsilon_{f,w}=0}$ are the product states between single-domain-wall states. They can thus be labeled as $\ket{\bm{j}}_d=\ket{j_x,j_y}_d$, where ${j_w=1,\dots,L_w=N_w/2+2}$ corresponds to the position of the domain wall along the vertex wire $w$ (with $w=x,y$). Thus, the ground states span a two-domain-wall Hilbert space, which can be represented as a two-dimensional (2D) rectangular lattice, whose corners correspond to the four logical states [see Fig.~\ref{fig:effective_th}(b)]. 

\subsubsection{Effective Hamiltonian}
For finite ${\Omega\ll\Delta}$, the domain walls acquire kinetic energy. The dynamics reduce to a single-particle hopping on the 2D lattice, described by the following effective Hamiltonian (see Appendix~\ref{sec:effective_th}):
\begin{equation}
    H_{\rm eff} = -t \sum_{(\bm{j},\bm{j}')} \left(\ket{\bm{j}}_d\bra{\bm{j}'}+ \text{h.c}\right)-\sum_{\bm{j}} \epsilon(\bm{j})\ket{\bm{j}}_d\bra{\bm{j}},
    \label{eq:Heff_2wires_general}
\end{equation}
with hopping amplitude $t=\Omega^2/\Delta$. This equation is valid for $\epsilon_{i,w},\epsilon_{f,w}\ll\Delta$, and it is pictorially illustrated in Fig.~\ref{fig:effective_th}(b). Here, $(\bm{j},\bm{j}')$ denotes the sum over all pairs of states connected by a link, and $\epsilon(\bm{j})$ indicates a site-dependent diagonal term. Furthermore, we denote with $\zeta_w=m_w/2+1$ the last domain-wall position before the gadget along the vertex wire $w$.

We note that the dynamics along the vertical and horizontal directions are not completely decoupled. Indeed, the gadget introduces two diagonal hopping matrix elements between $\ket{\zeta_x,\zeta_y}_d$ and $\ket{\zeta_x+1,\zeta_y+1}_d$, and between $\ket{\zeta_x+1,\zeta_y}_d$ and $\ket{\zeta_x,\zeta_y+1}_d$. These additional hoppings effectively lead to an interaction between domain walls inside the gadget. We further note that $\epsilon(\bm{j})$ is nonzero only on the boundaries of the domain-wall lattice (orange and blue circles), where its value can be controlled via local detuning on the extremes of the vertex wires. In particular, the down, up, left, and right boundary lines are controlled via $\epsilon_{i,y}$, $\epsilon_{f,y}$, $\epsilon_{i,x}$, and $\epsilon_{f,x}$, respectively. At the four corners, the diagonal terms are the sum of those of the two adjacent boundary lines [e.g., {$\epsilon(0,0)=\epsilon_{i,x}+\epsilon_{i,y}$}].

\subsubsection{Polynomial runtime via diabatic evolution}
\begin{figure}[b!]
\centering
\includegraphics[width=1.0\columnwidth]{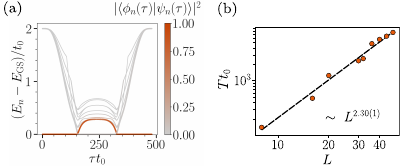}
    \caption{Analysis of the performance of the crossing setup, when considering the logical protocol.  (a) Energy difference between the instantaneous eigenstates ($E_n$) and the instantaneous ground state ($E_{\rm GS}$) of the effective Hamiltonian, normalized to $t_0=\Omega_0^2/\Delta_0$ (where $\Omega_0$ and $\Delta_0$ are the maximum Rabi frequency and detuning), as a function of time $\tau$, when targeting $\ket{11}_l$ as the final ground state. We consider two vertex wires of $N=28$ atoms intersecting at the middle ($m_x=m_y=N/2$). The spectrum is colored according to the overlap between the evolved state $\ket{\psi(\tau)}$ and the instantaneous eigenstates $\ket{\phi_n(\tau)}$ of the Hamiltonian: ${|\braket{\phi_n(\tau)}{\psi(\tau)}|^2}$. The final time is set to $Tt_0=480$, such that $|\braket{\psi(T)}{11}_l|^2\geq0.9$. (b) Scaling of the algorithm runtime needed to achieve $|\braket{\psi(T)}{11}_l|^2\geq0.9$, as a function of $L=N/2+2$. }
    \label{fig:crossing_H_evolution}
\end{figure}
Since the crossing gadget has been designed to effectively decouple the logical degrees of freedom of two intersecting vertex wires, we expect the crossing setup to behave as two independent vertex wires during the adiabatic computation. This behavior is confirmed in Fig.~\ref{fig:crossing_H_evolution}(b), where we observe an $L^2$ scaling of the algorithm runtime for the logical protocol (see Fig.~\ref{fig:protocol1and2}), when targeting $\ket{11}_l$ as the final ground state [with $\epsilon_{i,w}(0)=\epsilon_{f,w}(T)=2t_0$, where $w=x,y$ and $t_0=\Omega_0^2/\Delta_0$]. 

While this polynomial runtime is expected, it is nevertheless interesting to investigate the instantaneous energy spectrum of the system along the adiabatic protocol [see Fig.~\ref{fig:crossing_H_evolution}(a)]. Perhaps surprisingly, we observe two exponentially closing gaps. These gaps are a direct consequence of the exponentially small overlap between the ground state at half-protocol (i.e., at $\tau \simeq T/2$) and the initial and final ground states of the adiabatic protocol. Indeed, at half-protocol, the two-domain-wall ground-state wavefunction is localized around the gadget and decays exponentially along the legs of the vertex wires, forming a so-called \textit{bound state}. Despite these exponential gaps, the optimization problem is still solved in a polynomial runtime by exploiting a diabatic evolution. Indeed, at a polynomial timescale, the system does not couple to the unfavorable bound state; instead, it transitions from the instantaneous ground state to the first excited state at the first gap closing point and then returns to the ground state at the second one~\cite{Siddharth_2016, Crosson_2021,Albash_2018, Somma_2012, FryBouriaux_2021}.
\subsection{Crossing-with-edge setup}
The crossing-with-edge gadget [see Fig.~\ref{fig:setups_gadgets}(d)] effectively imposes the $\mu_x \mu_y=0$ constraint between the logical states of two intersecting vertex wires ($\mu_w=0,1$ with $w=x,y$). This constraint induces a mutual restriction on the position of the domain walls along the vertex wires, allowing at most one domain wall to go through the gadget. As a result, the two-domain-wall Hilbert space acquires the $L$-shaped geometry depicted in Fig.~\ref{fig:effective_th}(c), where, compared to the case of the crossing gadget, the lattice sites with both $j_x>\zeta_x$ and $j_y>\zeta_y$ are forbidden. 
This two-domain-wall space can be divided into three regions: one \textit{lead} extending along $x$, $\left[\zeta_x+1,L_x\right]\times \left[0,\zeta_y\right]$; one \textit{lead} extending along $y$, $\left[0,\zeta_x\right]\times \left[\zeta_y+1,L_y\right]$; and the \textit{inner} region, $\left[0,\zeta_x\right]\times \left[0,\zeta_y\right]$. Here, $L_w=N_w/2+1$ with $w=x,y$.

In the $\Omega,\epsilon\ll\Delta$ limit, the dynamics are described by the effective Hamiltonian in Eq.~\eqref{eq:Heff_2wires_general}, with links and diagonal terms represented in Fig.~\ref{fig:effective_th}(c). This Hamiltonian is analogous to the one of the crossing setup, except for the removal of the sites in the forbidden region.
Depending on whether or not the diagonal terms in Eq.~\eqref{eq:Heff_2wires_general} can be controlled by applying a detuning to the atoms at the extremes of the vertex wires, we subdivide the boundary of the Hilbert space into a \textit{logical} component [colored sites in Fig.~\ref{fig:effective_th}(c)] and a \textit{nonlogical} one. In particular, the logical component consists of states where one (or both) vertex wire is in a logical state. 

\subsubsection{Minimum gap scaling and domain-wall localization}
\begin{figure}
    \centering
    \includegraphics[width=1.0\columnwidth]{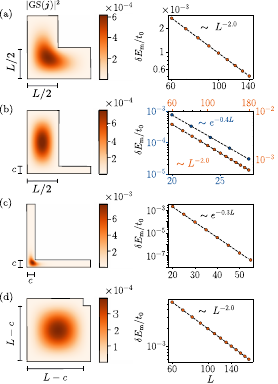}
    \caption{Analysis of the performance of the crossing-with-edge setup for the four configurations in Fig.~\ref{fig:protocol1_scaling_different_geometries}, when considering the logical protocol. (Left) Ground-state wavefunction $|\mathrm{GS}(\bm{j})|^2=|\braket{\bm{j}}{ \mathrm{GS}}|^2$ at half-protocol ($\tau=T/2$), when ${\epsilon_{i,w}=\epsilon_{f,w}=0}$ for ${w={x,y}}$. For illustration purposes, we use $c=12$.  (Right) Scaling of the minimum gap $\delta E_m$, normalized to $t_0=\Omega_0^2/\Delta_0$ (where $\Omega_0$ and $\Delta_0$ are the maximum Rabi frequency and detuning) as a function of $L=N/2+1$ (with $N$ being the number of atoms per vertex wire), setting $c=3$ (analogous to Fig.~\ref{fig:protocol1_scaling_different_geometries}). For symmetric geometries (a,c,d), the ground state at final time is $\ket{01}_l$. In the asymmetric geometry (b), $\ket{10}_l$ (blue) and $\ket{01}_l$ (orange) are considered separately.}
    \label{fig:protocol2_scalings}
\end{figure}
As observed in Sec.~\ref{sec: numerics}, in the CWE setup, the minimum gap scaling strongly depends on the geometry of the system and target state. Here, we aim to clarify this dependence by studying the two-domain-wall dynamics along the logical protocol, for the same geometrical configurations considered in the standard protocol (see Fig.~\ref{fig:protocol1_scaling_different_geometries}). In Fig.~\ref{fig:protocol2_scalings}, we show (right) the scaling of the minimum gap and (left) the ground-state wavefunction at half-protocol, where ${\epsilon_{i,w}=\epsilon_{f,w}=0}$. 
Specifically, we consider two vertex wires with the same number of atoms, i.e., $L_x=L_y=L=N/2+1$, and perform the logical protocol of Fig.~\ref{fig:protocol1and2} by choosing: ${\epsilon_{i,x}(0)=\epsilon_{i,y}(0)=2t_0}$, ${\epsilon_{f,x}(0)=\epsilon_{f,y}(0)=0}$, ${\epsilon_{i,x}(T)=\epsilon_{i,y}(T)=4t_0}$ and $\epsilon_{f,x}(T)=\frac{12}{5}\epsilon_{f,y}(T)=12t_0$ ($\epsilon_{f,y}(T)=\frac{12}{5}\epsilon_{f,x}(T)=12t_0$)
to target $\ket{10}_l$ ($\ket{01}_l$) as the final ground state. 

As for the standard protocol, we observe both polynomial and exponential scaling, depending on the geometry and the target state. In particular, $\delta E_m$ scales polynomially in configurations (a), (d), exponentially in (c), and in (b), depending on the target state, it exhibits both behaviors. 
We can predict whether the scaling is exponential or polynomial based on the overlap between the ground-state wavefunction before the minimum gap point and the final target state~\cite{Amin_2009,Jorg_2008,Suzuki_2013,Cain_2023,Schiffer_2024}. In particular, 
when the wavefunction at half-protocol decays exponentially in the lead where the target state is located --- as in (c) and in (b) for the target state $\ket{1 0}_l$ --- we find an exponential scaling of the minimum gap. Here, in contrast to the crossing setup, there is only a single exponentially closing gap dominating the dynamics. Therefore, diabatic transitions to higher excited states cannot be exploited to speed up the algorithm runtime.

The probability distribution of the ground-state wavefunction at half-protocol is a consequence of the minimization of the domain-wall kinetic energy. This can be clearly seen in (b), where the ground state localizes in the $\mathcal{R}_1=[0, L/2]\times[0, L]$ region. Indeed, expanding into the narrow region $\mathcal{R}_2=[L/2, L]\times[0,c]$, of width $c \ll L$, would require a higher kinetic-energy cost. However, it is less clear how these arguments apply to the (c) configuration, where the wavefunction localizes in the inner region $[0, c]\times[0, c]$. Understanding this behavior requires a more detailed analysis, which can be conducted within the following continuum limit. 

In the continuum limit of the system, the eigenvalue problem $H_{\rm eff}\ket{\psi}=E_k\ket{\psi}$ [$H_{\rm eff}$ in Eq.~\eqref{eq:Heff_2wires_general}] is mapped to the following Helmholtz equation (see Appendix~\ref{sec:Appendix_chain}):
\begin{equation}
    \nabla^2 \psi (\bm{r}) + k^2 \psi(\bm{r}) =0,
    \label{eq:Helmoltz_2d}
\end{equation}
with $\bm{r}=(x,y)$ being defined in a continuous $L$-shaped 
domain, and $k^2\simeq E_k/t+4$ for $k\ll 1$. The boundary conditions of Eq.~\eqref{eq:Helmoltz_2d} are determined by $\epsilon(\vec{j})$. At ${\tau=T/2}$, $\epsilon(\vec{j})=0$ maps to homogeneous Dirichlet boundary conditions (DBCs). In the following, we use this model to infer the ground-state wavefunction in the (b) and (c) configurations.  

In the (b) configuration, the ground-state energy can be estimated by neglecting $\mathcal{R}_2$ and solving the Helmholtz equation in $\mathcal{R}_1$ only, with homogeneous DBC on the entire boundary: ${k^2_{\rm GS}\sim (\pi/L)^2 + (2\pi/L)^2}$. Then, the exponential decay of the wavefunction profile in the $\mathcal{R}_2$ region can be inferred by considering Eq.~\eqref{eq:Helmoltz_2d} in this region, with $k^2=k^2_{\rm GS}$ and free boundary condition at the interface with $\mathcal{R}_1$. By separating the variables, we obtain $\psi(x,y)\sim\sinh(\kappa_x(x-L))\sin(k_yy)$, where ${k_y=\pi/c\gg k_{\rm GS}}$ and $\kappa_x^2=k_y^2-k_{\rm GS}^2$. 

In the (c) configuration, populating the leads would require a high kinetic cost, scaling as ${(\pi/c)^2}$. 
The domain-wall wavefunction thus minimizes the kinetic energy by localizing in the inner region. This effectively traps the domain walls before the gadget, forming a so-called \textit{bound state}.
The emergence of such a bound state in an $L$-shaped Hilbert space is a purely quantum phenomenon, which was first noted for quark confinement and hadronic interactions~\cite{Lenz_1986}, and extensively studied in the field of quantum waveguides~\cite{Avishai_1991,Londergan_1999,Exner_2015}. This bound state is separated from the continuum spectrum by a finite energy gap $\delta E$, which does not close in the infinite-$L$ limit. In Appendix~\ref{sec:Appendix_CWE}, we analytically describe such phenomenon within the so-called single-mode approximation and via conformal mapping methods. We predict the energy gap and compute the geometric range of $\zeta_x/\zeta_y$ for which it remains finite in the $L\to\infty$ limit.

\begin{figure}
    \centering
    \includegraphics[width=1.0\columnwidth]{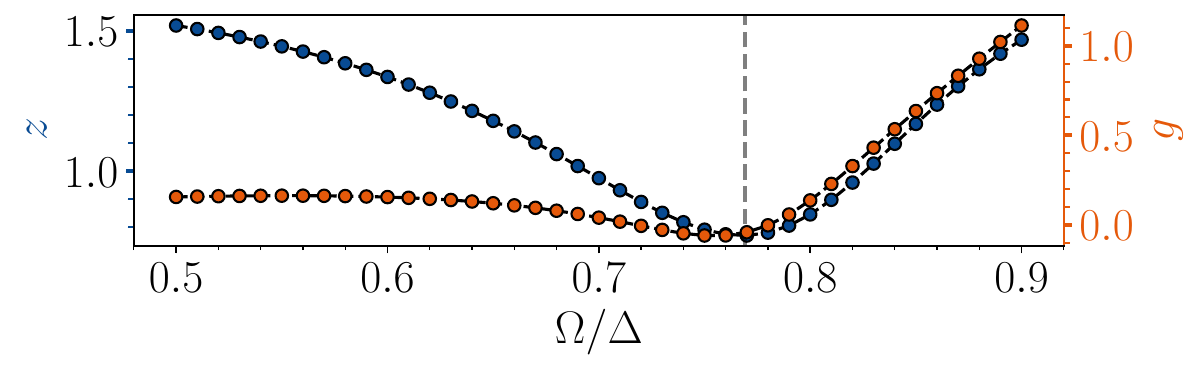}
    \caption{Bound-state analysis in the nonperturbative regime. Parameters of the gap scaling in Eq.~\eqref{eq:gap_dirichlet_DMRG} for ${\zeta=3}$ as a function of $\Omega/\Delta$. The dashed gray line indicates the transition point known for one-dimensional Rydberg chains with nearest-neighbor blockade, $\Omega/\Delta \sim 1/1.3$~\cite{Bernien_2017}.}
    \label{fig:gap_dirichlet_DMRG}%cwe_gap_DBC
\end{figure}

While the appearance of such a bound state can be analytically proven in the perturbative $\Omega/\Delta\ll 1$ limit, we study its existence for larger ratios of $\Omega/\Delta$ using DMRG. We find the following scaling of the energy gap $\delta E$ as a function of $\Omega/\Delta$ (in the symmetric case $\zeta_x=\zeta_y=\zeta$):
\begin{equation}
    \delta E/\Delta\:\overset{L \to \infty}{\sim}  \:\frac{g\left(\Omega/\Delta\right)}{\zeta^2}+\frac{G\left(\Omega/\Delta\right)}{L^{z\left(\Omega/\Delta\right)}} ,
    \label{eq:gap_dirichlet_DMRG}
\end{equation}
where $g\left(\Omega/\Delta\right)$ and $z\left(\Omega/\Delta\right)$ are shown in Fig.~\ref{fig:gap_dirichlet_DMRG}. This expression generalizes the scaling of the minimum gap in the perturbative limit, where $z=2$ (see Appendix~\ref{sec:Appendix_CWE} for details). For $L\to\infty$, the system presents a bound state if $g\neq0$. For $\Omega/\Delta \ll 1$, $g$ approaches the energy gap predicted by the effective theory (see Appendix~\ref{sec:Appendix_CWE}). As $\Omega/\Delta$ increases, the bound state survives throughout the ordered phase, vanishing only upon entering the disordered phase at $\Omega/\Delta \sim 1/1.3$~\cite{Bernien_2017}.  

\section{Observation of localization and optimization performance on neutral atom  hardware}\label{sec:experiment}
\begin{figure}[ht!]
    \centering
    \includegraphics[width=1.0\columnwidth]{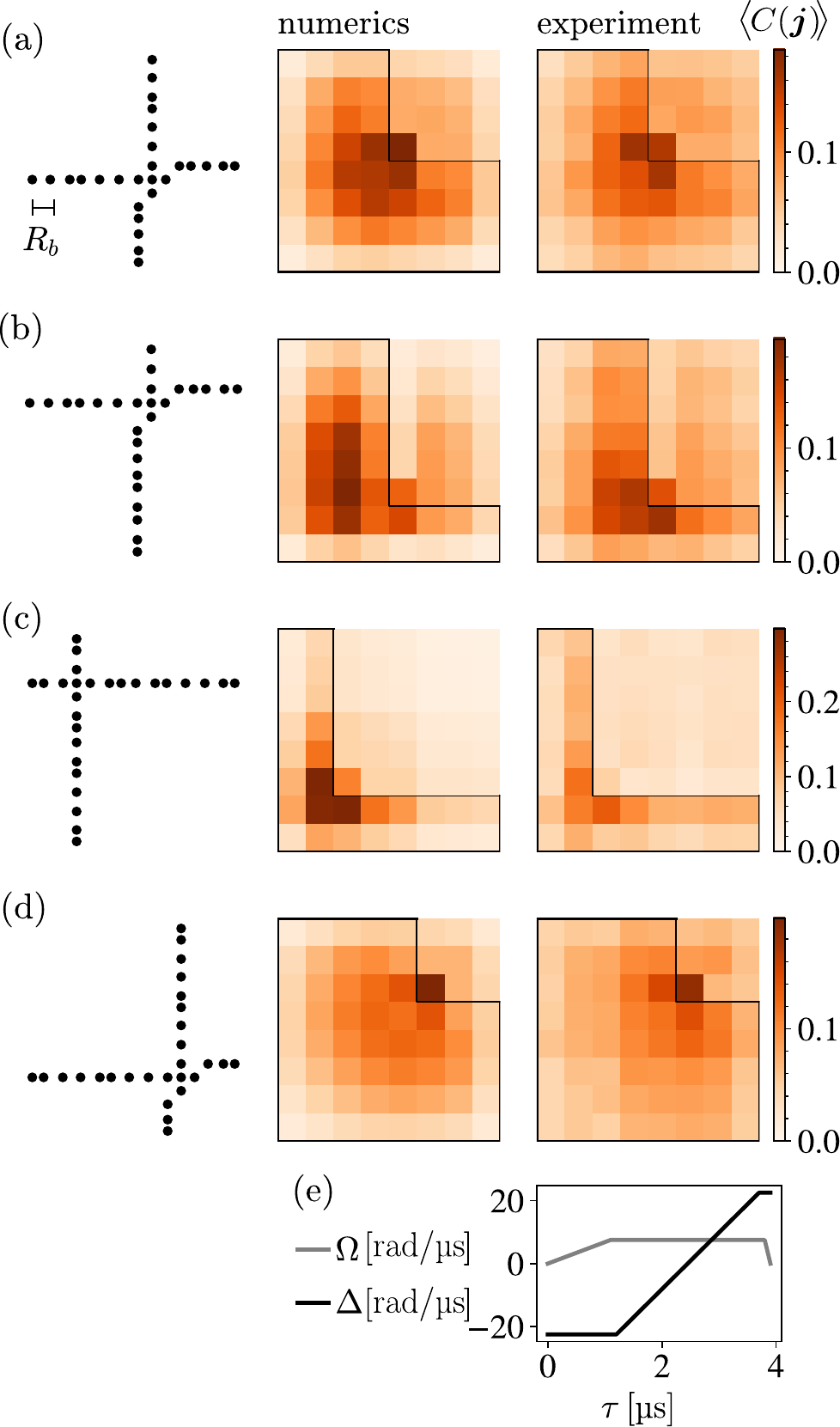}
    \caption{Observation of the localization phenomenon on the QuEra Aquila neutral atom hardware. We consider the four configurations of Figs.~\ref{fig:protocol1_scaling_different_geometries} and \ref{fig:protocol2_scalings}, for $\Omega=\Delta/3=7.5$ rad/\textmu s, $C_6 = 5.42 \times 10^6$ ~\textmu m$^6$~$\times$~rad/\textmu s, and $R_b=\left(C_6/\sqrt{(2\Omega)^2 + \Delta^2}\right)^{1/6}=7.65$~\textmu m. The total number of atoms is limited by the area of the platform of  ${75\:\mbox{\textmu m} \,\times\, 75\: \mbox{\textmu m}}$~\cite{Wurtz_2023}; we use $N=12$ atoms per wire.  (Left column) Optimized position of the atoms to counteract the effect of the tail interaction. (Central column) Expectation value of the domain-wall correlation operator in Eq.~\eqref{eq:dw_correlation} for the ground state computed via exact diagonalization including tail interaction among nonblockaded atoms. (Right column) Reconstruction of $\langle C(\bm{j})\rangle$ for 4500 measurements in the $\left\{\ket{0_v},\ket{1_v}\right\}$ basis. The measurement errors are reported in Fig.~\ref{fig:exp_errors} of Appendix~\ref{sec:Appendix_InteractionTails}. (e) Sweep of $\Omega(\tau)$ and $\Delta(\tau)$ for the ground-state preparation.}
    \label{fig:expermient}
\end{figure}
In the previous sections, we showed how the geometry of the CWE setup determines the performance of the adiabatic optimization, and how this relates to the localization of the ground-state wavefunction along the protocol. In this section, we further investigate these phenomena on the QuEra Aquila neutral atom hardware~\cite{Wurtz_2023}. First, we directly observe the ground-state localization and its dependence on the geometry of the setup. Then, we measure how it affects the probability of finding the solution to the encoded MWIS problem when executing the standard protocol. 

To efficiently probe these phenomena on the hardware, we include the tail interaction  [i.e., $U_{v, u}$ in Eq.~\eqref{eq:hamiltonain_rydberg} beyond the blockade radius] in modeling the system, and elaborate appropriate strategies to mitigate its effect on the quantum dynamics. In particular, the tail interaction leads to site-dependent diagonal terms $\epsilon(\bm{j})$ in the two-domain-wall effective Hamiltonian in Eq.~\eqref{eq:Heff_2wires_general}. In Appendix~\ref{sec:Appendix_InteractionTails}, we discuss how we can minimize the variance of these terms (and thus recover the blockade-limit physics) by adjusting either the position of the atoms along the vertex wires or the local detuning on the atoms. In this section, we employ the position-based mitigation strategy to counteract the tail interaction. Therefore, on the physical hardware, we implement the four configurations of the CWE setup with optimized positions shown in the left column of Fig.~\ref{fig:expermient}. 

\subsection{Localized ground states}
The domain-wall localization is a purely quantum phenomenon, as it stems from coherent hopping dynamics. Here, 
we directly observe such a phenomenon on the neutral atom hardware. As observable to measure localization, we consider the two-domain-wall correlation operator $C(\bm{j})$, which generalizes the two-domain-wall wavefunction and reduces to that in the perturbative regime. Specifically, it is defined as $C(\bm{j})=C(j_x)C(j_y)$, where: 
\begin{equation}\label{eq:dw_correlation}
    C(j_x) = \begin{cases}
        p_{2j_x-2} p_{2j_x-1}  & j_x \neq \zeta_x-1, \zeta_x  \\
        p_{m_x+1} p_{m_y+1} p_{G_x} p_{G_c} & j_x=\zeta_x+1 \\
        p_{m_x} p_{G_x} & j_x=\zeta_x,
    \end{cases}
\end{equation}
and $C(j_y)$ is obtained by exchanging $x$ and $y$. In this expression, $p_v=1-n_v$, while $G_x$, $G_y$, and $G_c$ label the gadget atoms [as shown in the inset of Fig.~\ref{fig:setups_gadgets}(d)].

In the central column of Fig.~\ref{fig:expermient}, we show the expectation value of $C(\bm{j})$, evaluated for the ground state of the system computed via exact diagonalization for  $\Delta=3\Omega$. Due to the contribution of states with more than two domain walls, $\langle C(\bm{j})\rangle$ slightly deviates from the perturbative ground states of Fig.~\ref{fig:protocol2_scalings}, while preserving their fundamental localization features (in agreement with the discussion at the end of Sec.~\ref{sec: effective th}B). In particular, the nonzero value of $\langle C(\bm{j})\rangle$ outside of the $L$-shaped domain is due to states with more than two domain walls, and it tends to zero as $\Omega/\Delta$ decreases.

On the QuEra Aquila neutral atom machine, we adiabatically prepare these four ground states by varying $\Omega(t)$ and $\Delta(t)$ as shown in Fig.~\ref{fig:expermient}(e). In particular, we begin by following the standard protocol of Fig.~\ref{fig:protocol1and2}, then suddenly switch $\Omega(t)$ off to freeze the system in the instantaneous ground state at finite $\Omega$. 
To ensure adiabaticity and prepare the final ground state with high fidelity within the maximum time of $4$~\textmu s, we set $\Omega$ to the maximum available value of 7.5 rad/\textmu s (and $\Delta=3\Omega$).
In the right column of Fig.~\ref{fig:expermient}, we reconstruct $\langle C(\bm{j})\rangle$ by performing 4500 measurements in the $\left\{\ket{0_v},\ket{1_v}\right\}$ basis for each configuration. The errors are shown in Fig.~\ref{fig:exp_errors} of Appendix ~\ref{sec:Appendix_InteractionTails}. The measurements recover the numerical predictions and display the localization phenomenon occurring in configurations (b) and (c), which is responsible for the exponential slowdown of the algorithm runtime. 
\subsection{Success probability}
\begin{figure}
    \centering
    \includegraphics[width=\columnwidth]{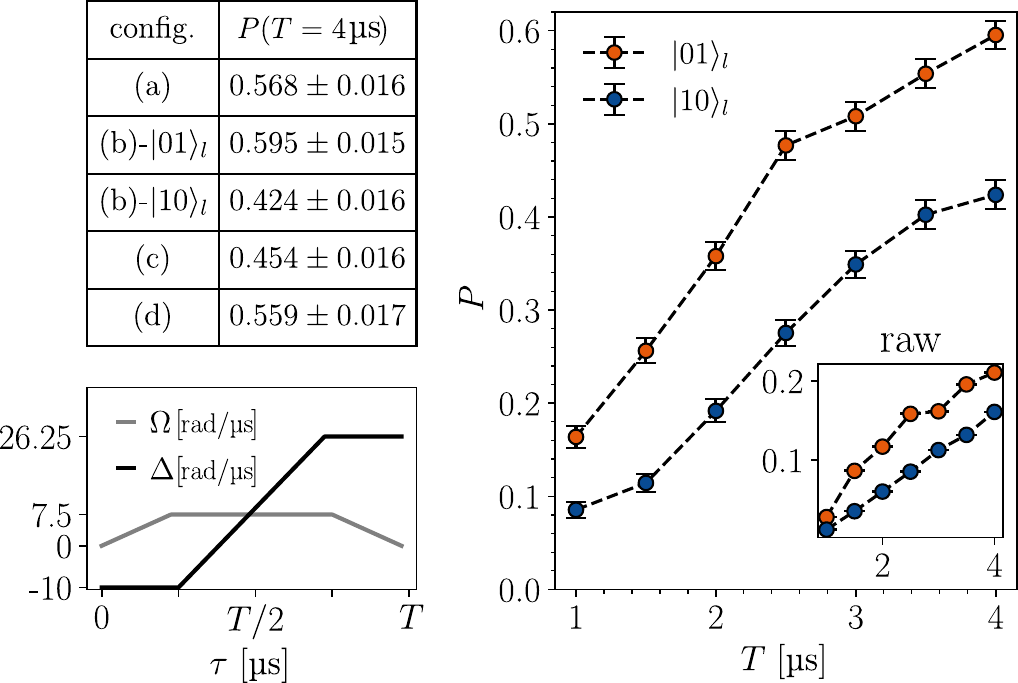}
    \caption{Observation of the success probability on the QuEra Aquila neutral atom hardware. (Left) Success probability $P$ of the four configurations in the left column of Fig.~\ref{fig:expermient} when executing the standard protocol (shown in the bottom) for $T=4$~\textmu s. We target the final state by removing the final atom of the vertex wire in the $\ket{1}_l$ state. (Right) $P$ as a function of the algorithm runtime $T$ for the (b) configuration of Fig.~\ref{fig:expermient}. For each configuration and target state, we perform 1500 measurements of the final state. See Appendix~\ref{App:exp_MIS} for details on the data analysis.}
    \label{fig:MWIS_experiment}
\end{figure}

Here, we directly observe how the geometry of the CWE setup (and the associated localization physics) determines the success probability on the neutral atom hardware. We define the success probability of adiabatic optimization as the probability of measuring the solution to the encoded MWIS problem at the end of the protocol. 
In particular, we consider the four configurations of the CWE setup in the left column of Fig.~\ref{fig:expermient} and execute the standard protocol in the bottom left of Fig.~\ref{fig:MWIS_experiment}. 
This setup encodes the simple graph of two vertices connected by an edge, and the solution to the MWIS problem is known based on the weight of the vertices. We can target this solution by properly removing one atom from the system, without resorting to the local detuning on the boundary atoms. Specifically, when $\delta_x>\delta_y$ ($\delta_x<\delta_y$), we target the $\ket{10}_l$ ($\ket{01}_l$) state by removing the last atom from the $x$ ($y$) vertex wire. 

In the top-left of Fig.~\ref{fig:MWIS_experiment}, we show the probability $P$ of detecting the strings encoding the solution to the MWIS problem, after proper postprocessing of the data (see Appendix~\ref{App:exp_MIS} for details). For the maximum available time of $T=4$~\textmu s, $P$ varies between $0.4$ and $0.6$, depending on configuration and target state. Importantly, the lowest values are observed in the cases where an unfavorable localization of the ground-state wavefunction with respect to the target solution is expected to occur. In particular, in the (b) configuration of Fig.~\ref{fig:expermient}, the success probability of finding the correct solution to the MWIS problem is equal to $0.59$ for the ``easy" target state (i.e., $\ket{01}_l$) and to $0.42$ for the ``hard" one (i.e., $\ket{10}_l$). 

In Fig.~\ref{fig:MWIS_experiment}(right), we deepen the analysis of this configuration by varying the algorithm runtime $T$ and measuring the corresponding probability $P$. As expected, at any $T$, we observe a higher success probability for the easy case than for the hard one. Notably, this difference can also be directly seen in the raw data (i.e., the data before postprocessing), which are shown in the inset of the figure. 

\section{Exponential improvement in the adiabatic performance} \label{sec: Improvement}
In this section, we present two different strategies to avoid the exponential closing of the minimum gap observed in the analysis of the CWE setup. This results in an exponential improvement in the adiabatic performance and suggests a `quantum-aware' modification of the encoding strategy of Ref.~\cite{Nguyen_2023}.  

The first approach leverages the effect of the geometry of the two-domain-wall Hilbert space on the minimum gap. In particular, we show that polynomial scaling along the logical protocol can be guaranteed through an unbiased extension of the first legs of the vertex wires.    
The second approach exploits the domain-wall boundary condition to delocalize the ground-state wavefunction along the logical protocol.
Specifically, we show that polynomial scaling can be guaranteed by modifying the structure of the CWE gadget to achieve such delocalization. 
In Appendix~\ref{sec:CWE_Rabi}, we present a third approach to achieve such a polynomial scaling by exploiting the control of the Rabi frequency on the atoms at the extreme of the vertex wires. 

\subsection{Modification of the domain-wall space geometry via vertex wire extension}
The first strategy relies on the dependence of the minimum gap scaling on the geometry of the two-domain-wall Hilbert space (see Fig.~\ref{fig:protocol2_scalings}).
Specifically, we notice that the gap closes exponentially with $L=N/2+1$ when the width of one (both) lead(s) is kept constant during the scaling (i.e., $\zeta_x$ and/or $\zeta_y$). In the physical system, this corresponds to keeping the first leg(s) of the two vertex wires fixed when increasing $L$. 
This observation suggests a modification of the setup, where the first legs of both vertex wires are simultaneously and unbiasedly extended by adding $2d$ atoms, chosen to be proportional to the original number of atoms per vertex wire $N$. This increases the total number of domain-wall states per vertex wire to $L+d$. 

\begin{figure}
	\centering
	\includegraphics[width=1.0\columnwidth]{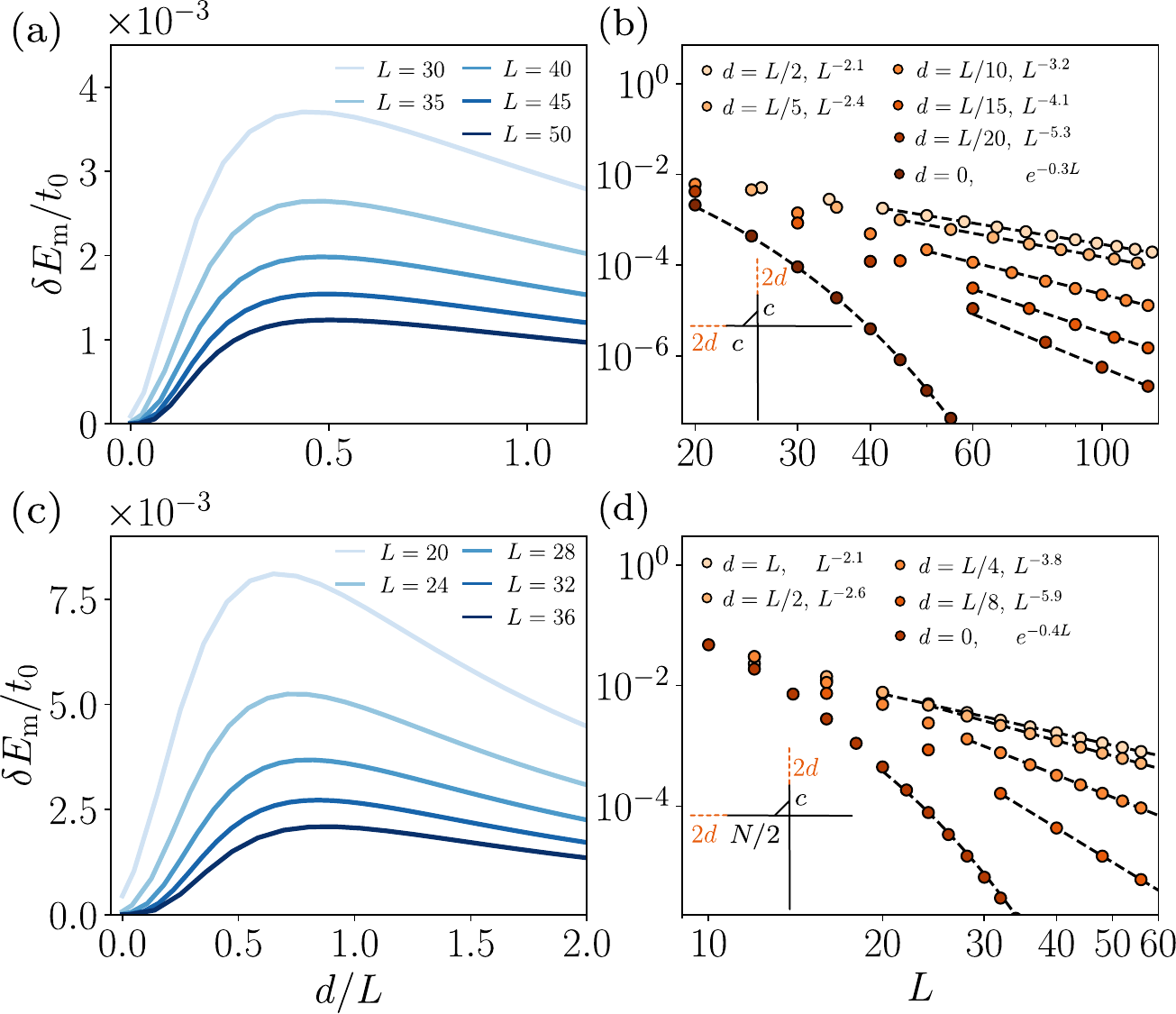} 
	\caption{
    Effectiveness of the vertex wire-extension strategy. This strategy is applied (a,b) to the configuration of Fig.~\ref{fig:protocol1_scaling_different_geometries}(c) and (c,d) to the configuration of Fig.~\ref{fig:protocol1_scaling_different_geometries}(b). The vertex wires (with $N$ atoms) are extended by adding $2d$ atoms to the legs before the gadget so that $L=N/2+1$ is extended to $L+d$ [see bottom-left insets in (b,d)]. 
    (a,c) Minimum gap $\delta E_m$, normalized to $t_0=\Omega_0^2/\Delta_0$ (where $\Omega_0$ and $\Delta_0$ are the maximum Rabi frequency and detuning), along the logical protocol as a function of $d/L$ for a fixed $L$.
    (b,d) Scaling of the minimum gap for different extension ratios $d/L$ as a function of $L$.}
	\label{fig:gadget_scaling}
\end{figure}

The effectiveness of this strategy is verified in Fig.~\ref{fig:gadget_scaling}. In Fig.~\ref{fig:gadget_scaling}(a), we consider the (c) configuration of Figs.~\ref{fig:protocol1_scaling_different_geometries} and \ref{fig:protocol2_scalings} and compute $\delta E_m$ for different original lengths $L$ and extensions $d$. At fixed $L$, the minimum gap grows substantially as $d$ increases compared to the configuration with no extension ($d=0$), reaching a maximal value for $d\approx 0.3L$. In Fig.~\ref{fig:gadget_scaling}(b), we show how $\delta E_m$ scales with the original length $L$, when fixing the extension ratio $d/L$. We observe that, as soon as $d/L\neq0$, the scaling goes from exponential to polynomial. In particular, $\delta E_m\sim L^{-z}$, with $z$ starting from $5.6$ at $d/L=1/20$ and converging to $2$ for $d/L\gtrsim 1/2$. The same strategy applies to the (b) configuration, as shown in Fig.~\ref{fig:gadget_scaling}(c,d).

This strategy could be generalized to a network of gadgets encoding an arbitrarily connected graph of $|V|$ vertices [see Fig.~\ref{fig:setups_gadgets}(a) and Ref.~\cite{Nguyen_2023}] by adding $2d$ atoms at the beginning of each wire. Assuming that a similar analysis holds in this general case, the optimal choice of $d/L\simeq0.5$, suggested by Fig.~\ref{fig:gadget_scaling}(c), would increase the atom-number overhead of the encoding scheme from the original scaling of $4|V|^2$ to $4|V|^2+2d|V|\simeq6|V|^2$.

\subsection{Domain-wall delocalization via ancillary atoms}
As observed in Sec.~\ref{sec: effective th}C, exponential closing of the minimum gap are detected when the ground-state wavefunction unfavorably localizes with respect to the solution to the optimization problem during the adiabatic evolution. 
The second strategy leverages the ground-state delocalization to improve the adiabatic performance~\cite{Albash_2018,Cain_2023,Schiffer_2024}. In detail, we aim to fully delocalize the ground state at $\tau=T/2$ in the logical protocol by controlling the boundary conditions of the two-domain-wall Hilbert space. 

The relationship between boundary conditions and ground-state delocalization can be understood in the continuum limit. In particular, when homogeneous Neumann boundary conditions (NBCs) are assigned on the whole boundary of the $L$-shaped domain, the ground state of the Helmholtz equation in Eq.~\eqref{eq:Helmoltz_2d} is the constant function, independently of the configuration. 

For the discrete lattice of Fig.~\ref{fig:effective_th}(c), homogeneous NBC translates into the following relationship between diagonal and hopping terms in Eq.~\eqref{eq:Heff_2wires_general}:
\begin{equation}
    \epsilon(\bm{j}) = t\left[4 - \# \text{links}(\bm{j})\right],
    \label{eq:condition_flat_GS}
\end{equation}
where $\# \text{links}(\bm{j})$ indicates the number of links connected to site $\bm{j}$. This equation is obtained by imposing the equal-weight superposition state to be the ground state of the Hamiltonian in Eq.~\eqref{eq:Heff_2wires_general}. We refer to it as \textit{hopping-detuning} balance condition.

To satisfy this condition, the diagonal terms $\epsilon(\bm{j})$ need to compensate for the lack of links on the boundary sites [where $\# \text{links}(\bm{j})<4$]. With the current structure of the gadget, this can be achieved on the logical component of the boundary by controlling the local detuning on the atoms at the extreme of the vertex wires (setting ${\epsilon_{i,w}=\epsilon_{f,w}=t}$, with $w= x,y$). However, it cannot be achieved on the nonlogical component, where $\epsilon(\bm{j})$ are forced to be identically zero.

To enforce the hopping-detuning balance condition on the nonlogical component of the boundary, we propose to modify the structure of the CWE gadget by introducing properly designed ancillary atoms, as depicted in Fig.~\ref{fig:CWE_ancilla_combined_text}(a) (see details in Appendix~\ref{sec:Appendix_CWE_ancillae}). In this way, we extend the boundary of the two-domain-wall Hilbert space by creating additional domain-wall states, whose diagonal terms can be controlled via local detuning on the ancillae. The effectiveness of this strategy is demonstrated in Fig.~\ref{fig:CWE_ancilla_combined_text}(b), where we employ this modification of the gadget to impose Eq.~\eqref{eq:condition_flat_GS} at half-protocol. We observe a polynomial scaling of the minimum gap for all configurations and target states. 

\begin{figure}
\centering
\includegraphics[width=1\columnwidth]{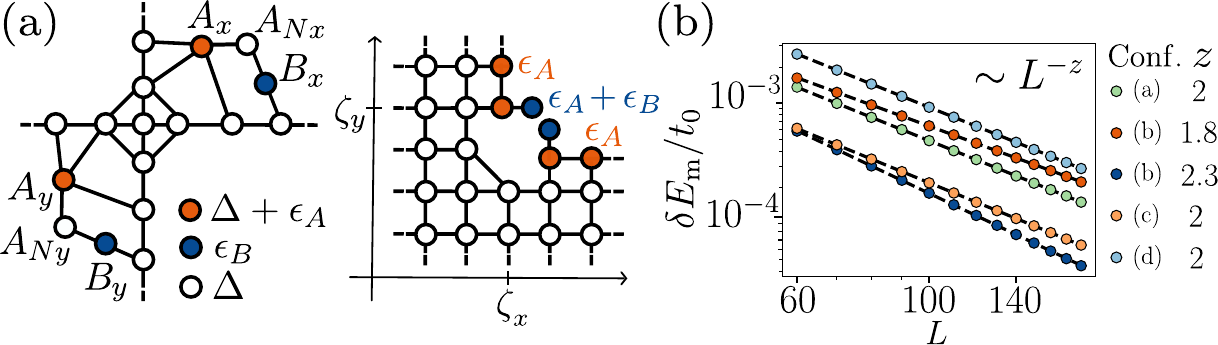}
    \caption{Modification of the crossing-with-edge gadget via the introduction of ancillary atoms. (a) Six ancillae are added to the gadget structure, with detuning $\Delta+\epsilon_A$ on $A_{w}$, $\Delta$ on $A_{Nw}$, and $\epsilon_B$ on $B_{w}$, where $w=x,y$. This graph can be drawn explicitly as a unit-disk graph. This modification effectively extends the nonlogical component of the two-domain-wall Hilbert space boundary by introducing additional virtual sites depicted in orange and blue. Here, the diagonal terms are given by the detuning on the ancillae: $\epsilon_A$ on the orange sites and $\epsilon_A+\epsilon_B$ on the blue sites. (b) Minimum gap scaling as a function of $L=N/2+1$ (with $N$ being the number of atoms per vertex wire) for the four configurations of Fig.~\ref{fig:protocol2_scalings} when modifying the gadget structure as in (a) and imposing Eq.~\eqref{eq:condition_flat_GS} at $\tau=T/2$ in the logical protocol. For the configuration of Fig.~\ref{fig:protocol2_scalings}(b), the data corresponding to the two final ground states $\ket{01}_l$ and $\ket{10}_l$ are shown in orange and blue, respectively.}
    \label{fig:CWE_ancilla_combined_text}
\end{figure}

This strategy could also be applied to a network of gadgets. In this case, since the minimal unit-disk realization of the modified CWE gadget in Fig.~\ref{fig:CWE_ancilla_combined_text}(a) requires 17 atoms, the atom-number overhead of the encoding scheme would increase from the original scaling of $4|V|^2$ to $8.5|V|^2$.

\section{Conclusion}\label{sec:conclusion}
In this work, we analyzed the quantum properties of the Rydberg encoding scheme for maximum weighted independent set problems on arbitrarily connected graphs introduced in Ref.~\cite{Nguyen_2023}. In particular, we focused the analysis on its fundamental building blocks --- two vertex wires intersecting via a gadget --- for various geometrical configurations. 

We quantified the adiabatic performance of these systems by looking at the minimum gap scaling with system size along two representative adiabatic protocols.
Our analysis shows that the interplay between geometry and domain-wall kinetic energy can lead to an unfavorable localization of the ground-state wavefunction with respect to the solution to the optimization problem, and to an exponential closing of the minimum gap. Finally, we proposed two strategies to address this localization and provide exponential improvement in the adiabatic performance. Our findings are summarized in Table.~\ref{tab:summary}.

\begin{table}
\centering
\begin{tabular}{|c|c|c|c|}
    \hline
    Setup & Standard & Logical & Logical/improv.  \\ \hline
    Crossing &  poly & poly & poly  \\ \hline 
    CWE---(a) & exp & poly &  poly \\
    CWE---(b) & poly/exp$^*$  & poly/exp$^*$ & poly \\
    CWE---(c) & unclear & exp & poly \\
    CWE---(d) & poly & poly & poly \\ 
    \hline
\end{tabular}
\caption{ Summary of the minimum gap scaling for the two setups along the standard protocol, the logical protocol, and the logical protocol with improvement strategies. For the CWE setup, we refer to the geometrical configurations of Figs.~\ref{fig:protocol1_scaling_different_geometries} and \ref{fig:protocol2_scalings}. The asterisk denotes that the scaling depends on the target state.}
\label{tab:summary}
\end{table}

While we discussed  the localization phenomenon in the simplest instances of the encoding scheme in Ref.~\cite{Nguyen_2023}, we expect this effect to also arise in more complex instances and in other encoding schemes.
 For example, in the case of a network of $|V|$ intersecting vertex wires, the $L$-shaped domain-wall space is generalized to a truncated $|V|$-dimensional hypercube. In such a domain, we anticipate the presence of analogous localization phenomena. While it is possible to extend our improvement strategies to this general case, quantifying their effectiveness for large networks remains an open problem. In addition, it would be interesting to consider alternative strategies designed to prevent quantum optimization algorithms from being trapped in suboptimal solutions, such as diabatic sweeps or counteradiabatic driving \cite{Crosson_2014,Sels_2017,Odelin_2019,Crosson_2021,Lukin_2024}.

In summary, our work shows, through a concrete example, that a careful design of encoding strategies for quantum adiabatic optimization requires going beyond a mere classical analysis of the target Hamiltonian. It also necessitates a detailed consideration of the interplay between the encoding strategy and the quantum mechanical effects that occur during the adiabatic algorithm.  Our work provides an example of such a `quantum-aware' approach, showing how the encoding strategy of Ref.~\cite{Nguyen_2023} can be adapted to avoid quantum bottlenecks.

\section*{Acknowledgment}
We thank G. Giudici for valuable discussions on developing the intersecting tensor network method. We also thank S.-T. Wang, M. Kornjaca, and J. Long for their helpful discussions. We acknowledge Amazon Web Services for providing credits to access the Aquila neutral atom machine. 
This work is supported by the European Union’s Horizon Europe research and innovation program under Grant Agreement No.~101113690 (PASQuanS2.1), the ERC Starting grant QARA (Grant No.~101041435), the EU-QUANTERA project TNiSQ (N-6001), the Austrian Science Fund (FWF) (Grant No.~DOI 10.55776/COE1), the DARPA ONISQ program (Grant No.~W911NF2010021), the US Department of Energy (DOE Quantum Systems Accelerator Center, Grant No.~DE-AC02-05CH11231), the Center for Ultracold Atoms (an NSF Physics Frontiers Center), and the National Science Foundation. SN acknowledges funding from the European Union’s Horizon Europe research and innovation program under the Marie Skłodowska-Curie Grant No.~101059826 (ETNA4Ryd) and support from the Quantum Computing and Simulation Center (QCSC) of Padova University.
 
%%%%%%%%%%%%%%%%%%%%%%%%%%%%%%%
%%%%%%%%%%%%%%%%%%%%%%%%%%%%%%

\appendix
\section{Intersecting tensor network }\label{sec:Appendix_DMRG}
Here, we illustrate the intersecting tensor network (ITN) method designed for simulating two vertex wires intersecting via a gadget [see Fig.~\ref{fig:setups_gadgets}(c,d)].

Let us start by considering the case of a single vertex wire, where it is possible to apply the MPS-DMRG method. This method consists of approximating the state of the system with a matrix product state (MPS) ansatz and variationally searching for the ground state via the DMRG algorithm~\cite{Schollwock_2011}.

\begin{figure}
    \centering
    \includegraphics[width=1.0\columnwidth]{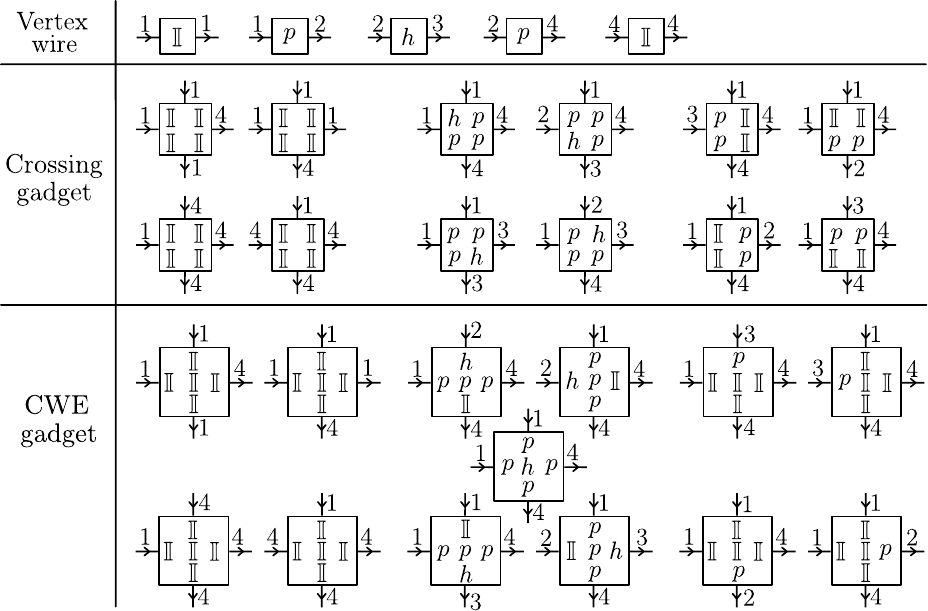}
    \caption{Graphical representation of MPOs for the vertex wire, the crossing, and the CWE gadget in Fig.~\ref{fig:setups_gadgets}. We follow the notation of Ref.~\cite{Hubig_2017}, where the numbers indicate the position of the operators within the element. For instance, $2\rightarrow \boxed{h} \rightarrow3$ denotes that operator $h$ is located at the second row and third column of $W$ in Eq.~\eqref{eq:mpo_chain}. For the gadgets, the placing of the operators inside the boxes corresponds to the position of the atoms in the physical structure. The physical indices are omitted for simplicity.}
    \label{fig:App_MPO}
\end{figure}

To construct the matrix product operator (MPO) for the Rydberg Hamiltonian within the blockade-constraint subspace, we employ an approach that can be easily extended from the vertex wire to the gadget setups.  
Let us start from the Rydberg Hamiltonian in the generic form of Eq.~\eqref{eq:hamiltonain_rydberg}. In the blockade limit, it can be reformulated into the following ``PXP'' form: $H_{\rm Rydberg} =  \sum_v \mathcal{P} h_v \mathcal{P}$, where $h_{v}=~\Omega\sigma^{x}_{v} - \Delta_{v} n_{v}$, and $\mathcal{P}$ is the projector into the blockade-constraint subspace, which comprises of all states with no neighboring atoms in $\ket{1}$. Practically, within the sum, we can substitute the global projector  $\mathcal{P}$ with the local ones $\mathcal{P}_v\equiv\prod_{u\in \mathcal{B}(v)}p_u$, where $p_u=|0\rangle_u\langle0|$ and $\mathcal{B}(v)$ denotes the set of atoms blockaded with the atom $v$. 
For a single vertex wire, we can express the Rydberg Hamiltonian exactly in the form of an MPO, where we associate to each atom $v$ the following tensor:
\begin{equation}\label{eq:mpo_chain}
    W^{[v]}=\begin{pmatrix} \mathbb{I}_{v} & p_{v} & 0 & 0\\
        0 & 0 & h_{v} & 0\\
        0 & 0 & 0 & p_{v}\\
        0 & 0 & 0 & \mathbb{I}_{v}
        \end{pmatrix},
\end{equation}
where $\mathbb{I}_v=\ket{0}_v\bra{0}+\ket{1}_v\bra{1}$ is the identity. $W^{[v]}$ has two virtual indices of dimension $4$, and two physical indices of dimension $d=2$. Pictorially, we represent the entries of $W^{[v]}$ as in Fig.~\ref{fig:App_MPO}.

\begin{figure}
    \centering
    \includegraphics[width=1.0\columnwidth]{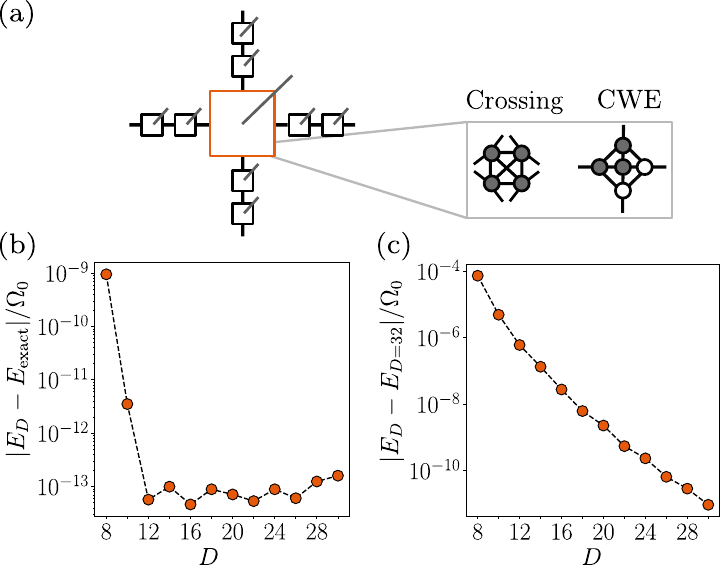}
    \caption{Intersecting-tensor-network method for the gadget setups. (a) Graphical representation of the state as a product of tensors. Tensors of two virtual indices (in black) and one physical index (in gray) correspond to the vertex-wire atoms. The tensor depicted in orange of four virtual indices (in black) and one physical index (in gray) corresponds to the atoms at the intersection of the vertex wires, shown in the inset. (b,c) Analysis of the convergence of the DMRG method for the crossing setup with vertex wires (of $N$ atoms) intersecting at the middle ($m_x=m_y=N/2$), $\Delta=1.3\Omega$, and $\Delta_g = 3\Delta$. (b) Difference between the ground-state energy obtained via exact diagonalization ($E_{\rm exact}$) and DMRG ($E_D$) as a function of the maximum bond dimension $D$, for $N=10$. (c) Difference between the ground-state energy obtained using DMRG with bound dimension $D<32$ and $D=32$, for $N=102$.}
    \label{fig:tensor_network}
\end{figure}

We now consider the case of two intersecting vertex wires. Here, the state of the system can be expressed as a generalization of the MPS for a vertex wire, as shown in Fig.~\ref{fig:tensor_network}(a). Here, the atoms on the vertex wires are represented by tensors with two virtual indices and one physical index. The atoms at the intersection of the two vertex wires --- the gadget atoms for the crossing setup and the gadget atoms plus two other atoms for the CWE setup --- are grouped into a \textit{gadget} tensor with four virtual indices and one physical index. In representing the state, we truncate the dimension of the virtual indices to a maximum \textit{bond dimension} of $D$. The dimension of the physical indices $d$ is equal to $2$ for the vertex wire, $5$ for the crossing gadget tensor, and $8$ for the CWE gadget tensor. Importantly, for this structure, it is feasible to employ the single-site update DMRG algorithm to variationally search for the ground states and low-excited states~\cite{White_1992,White_1993,Hubig_2015,Stoudenmire_2012}.

From this representation of the state, we can then construct the corresponding MPO form of the Rydberg Hamiltonian for the gadget setups. As for the single vertex wire, the tensor corresponding to a vertex wire atom $v$ is given by $W^{[v]}$ in Eq.~\eqref{eq:mpo_chain}, while a larger tensor $W^{[\mathrm{gad}]}$ corresponds to the gadget atoms. This tensor has four virtual indices of dimension $4$ and two physical indices of dimension $d=5$ for the crossing setup and $d=8$ for the CWE setup. We graphically represent its nonzero entries in Fig.~\ref{fig:App_MPO}.   

In Fig.~\ref{fig:tensor_network}, we test the convergence of the algorithm for increasing $D$ in the case of the crossing setup. In (b), we show the difference between the ground-state energy computed using DMRG (as $D$ varies) and that computed via exact diagonalization, for a small system of $N=10$ atoms per vertex wire. In (c), we compare it with the ground-state energy computed using DMRG with the largest accessible bound dimension $D=32$ for a large system of $N=102$ atoms per vertex wire. The largest accessible bound dimension is mainly limited by the dimension of the gadget tensor ($d D^4$). From this analysis, we see that a bond dimension of $D=16$ allows us to achieve sufficiently small errors. In this work, we set $D=16$.  Similar considerations hold for the CWE setup.

\section{Effective Hamiltonian for vertex wires}\label{sec:effective_th}
Here, we detail the derivation of the effective Hamiltonian for a single vertex wire [Eq.~\ref{eq:chain_Heff_}] and for two vertex wires intersecting via a gadget [Eq.\ref{eq:Heff_2wires_general}]. For these setups, it is convenient to rewrite $H_{\rm Rydberg}$ in Eq.~\eqref{eq:hamiltonain_rydberg} in the blockade limit as
\begin{equation}
H_{\text{Rydberg}}=\mathcal{P}\left(H_0+H_{\Omega}+H_{\epsilon}\right)\mathcal{P} 
\label{eq:hamiltonian_dec}, 
\end{equation} 
where: $\mathcal{P}$ is the projection operator into the blockaded subspace introduced in Appendix~\ref{sec:Appendix_DMRG};
\begin{equation}
    H_0 =  - \Delta \sum_{v\notin G} n_{v}  - \Delta_g \sum_{v\in G} n_{v}, \:\:\ H_{\Omega} =  \Omega\sum_v \sigma_v^x,
    \label{eq:hamiltonian_0}
\end{equation}
with $v \in(\notin) G$ indicating if an atom belongs (or not) to the gadget [gray atoms in Fig.~\ref{fig:setups_gadgets}(c,d)]; and $H_{\epsilon}$ accounts for local detuning inhomogeneities at the boundaries.

We treat ${H_1 = H_{\Omega}+ H_{\epsilon}}$ as small correction of $H_0$. In second-order perturbation theory, the dynamics within the lowest-energy subspace of $H_0$ (of energy $E_0$) are described by the effective Hamiltonian
\begin{equation}
    H_{\text{eff}} = PH_0P + PH_1P
    + P   H_1 Q \frac{1}{E_0 - Q H_0 Q} Q H_1 P,
    \label{eq:Heff}
\end{equation}
where $P$ the projector onto the lowest energy subspace, and $Q=\mathbb{I}-P$ (with $\mathbb{I}$ being the identity). In the following, we will discuss $H_{\text{eff}}$ for the setups under study.  
\begin{figure}
    \centering
    \includegraphics[width=1.0\columnwidth]{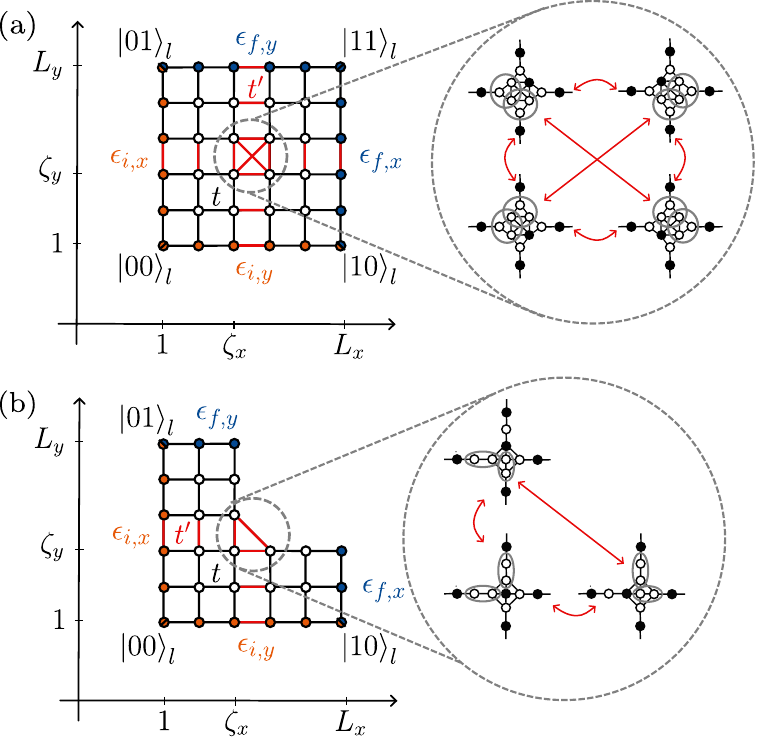}
    \caption{Graphical representation [according to the conventions in Eq.~\eqref{eq:Heff_2wires_general}] of the effective Hamiltonian in Eq.~\eqref{eq:Heff_2wires} for (a) the crossing and (b) the crossing-with-edge setups with detuning $\Delta_g$ on the gadget atoms. Here, $t=\Omega^2/\Delta$ and $t'=\Omega^2/\Delta_g$. The insets show the diagonal hopping matrix elements between $\ket{\zeta_x+1,\zeta_y}_d$ and $\ket{\zeta_x,\zeta_y+1}_d$, and, only for the crossing setup, between $\ket{\zeta_x,\zeta_y}_d$ and $\ket{\zeta_x+1,\zeta_y+1}_d$.}
    \label{fig:Heff_appendix}
\end{figure}

\subsection{Vertex wire}
For the vertex wire in Fig.~\ref{fig:setups_gadgets}(b), the lowest energy subspace of $H_0$ comprises the ${L=N/2+1}$ single-domain-wall states $\ket{j}_d$ shown in Fig.~\ref{fig:effective_th}(a) (with energy ${E_0 = -N\Delta/2}$), and ${P=\sum_{j=1}^{L=N/2+1}\ket{j}_d\bra{j}}$. The effective Hamiltonian in Eq.~\eqref{eq:Heff} takes the form
\begin{eqnarray}\label{eq:H_eff_wire}
    {H_{\text{eff}} = P\left(H_0+H_\epsilon-\frac{1}{\Delta} H_\Omega^2\right)P}.
\end{eqnarray}

Here, $H_0$ gives rise to a constant energy term $E_0$, while $H_\epsilon= \epsilon_i n_1 + \epsilon_f n_N$ generates both a constant energy term $(\epsilon_i+\epsilon_f)$ and a nontrivial diagonal term ${-(\epsilon_i\ket{1}_d\bra{1}+\epsilon_f \ket{L}_d\bra{L})}$.
Then, $H_\Omega^2$ is responsible for two different second-order processes. In the first process, an atom in the Rydberg state is de-excited, while a neighboring atom transitions from the ground to the Rydberg state. This effectively results in a domain-wall hopping of amplitude $t=\Omega^2/\Delta$. In the second process, an atom in the Rydberg state is de-excited to the ground state and then re-excited. This lowers the energy by a constant term $-Nt/2$.
Thus, Eq.~\eqref{eq:Heff} returns Eq.~\eqref{eq:chain_Heff_} up to a constant diagonal term of $\epsilon_i+\epsilon_f-\left(\Delta+t\right)N/2$. This effective Hamiltonian holds in the limit of $\left|\epsilon_i\right|,\left|\epsilon_f\right|,\Omega\ll\Delta$.

\subsection{Two intersecting vertex wires}
For two vertex wires intersecting via a gadget [see {Fig.~\ref{fig:setups_gadgets}(c,d)}], the lowest-energy eigenstates of $H_0$ are characterized by one domain wall per vertex wire and are depicted in Fig.~\ref{fig:effective_th}(b,c). The effective Hamiltonian in Eq.~\eqref{eq:Heff} takes the form
\begin{eqnarray}\label{eq:Heff_2wires}
    {H_{\text{eff}} = P\left(H_0+H_\epsilon-\frac{1}{\Delta} H_\Omega Q_w H_\Omega-\frac{1}{\Delta_g} H_\Omega Q_g H_\Omega\right)P}, \nonumber 
\end{eqnarray}
where $Q_w$ ($Q_g$) is the projector onto the subspace constituted by the states derived from the domain-wall states by de-exciting an excited atom to the ground state along the vertex wire (in the gadget). 

Here, $H_0$ generates a constant energy term $E_0$, while $H_{\epsilon}=\sum_{w = x,y} \left(\epsilon_{i,w} n_{1,w} + \epsilon_{f,w} n_{N,w}\right)$ --- where the subscripts `$1,w$' (`$N,w$') label the first (last) atom on the vertex wire $w$ --- generates both a constant energy term $\sum_{w=x,y}(\epsilon_{i,w}+\epsilon_{f,w})$ and a nontrivial diagonal term $-(\sum_{j_y}\epsilon_{i,x} \ket{1,j_y}_d\bra{1,j_y}+\epsilon_{f,x} \ket{L_x,j_y}_d\bra{L_x,j_y} + \sum_{j_x}\epsilon_{i,y} \ket{j_x,1}_d\bra{j_x,1}+\epsilon_{f,y} \ket{j_x,L_y}_d\bra{j_x,L_y})$. Then, $H_\Omega Q_w H_\Omega$ and $H_\Omega Q_g H_\Omega$ generate the two second-order processes described below. 

In the first process, $H_\Omega Q_w H_\Omega$ ($H_\Omega Q_g H_\Omega$) gives rise to the hopping of a single domain wall along one vertex wire (across the gadget), with amplitude $t=\Omega^2/\Delta$ ($t'=\Omega^2/\Delta_g$). 
Notably, when both domain walls are located at the gadget, $H_\Omega Q_g H_\Omega$ generates a simultaneous hopping of the two domain walls. Indeed, for these states, one of the gadget atoms is excited, while all four vertex-wire atoms attached to the gadget are in the ground state. As a result, $H_\Omega Q_g H_\Omega$ can de-excite the excited gadget atom and excite any other gadget atom, giving rise to the following hopping matrix elements: $\ket{\zeta_x,\zeta_y}_d\bra{\zeta_x+1,\zeta_y+1}$ and  $\ket{\zeta_x+1,\zeta_y}_d\bra{\zeta_x,\zeta_y+1}$ for the crossing gadget, and only $\ket{\zeta_x+1,\zeta_y}_d\bra{\zeta_x,\zeta_y+1}$ for the CWE gadget. 

In the second process, $H_\Omega Q_w H_\Omega$ ($H_\Omega Q_g H_\Omega$) de-excites a vertex-wire (gadget) atom in the Rydberg state to the ground state and then re-excites it back. This process lowers the energy by a constant term $-N_xt/2-N_yt/2-t'$. 

Eventually, Eq.~\eqref{eq:Heff} returns the hopping Hamiltonian illustrated in Fig.~\ref{fig:Heff_appendix}. For $\Delta_g=\Delta$ (and thus $t=t'$), we retrieve Eq.~\eqref{eq:Heff_2wires_general} and its graphical representations in Fig.~\ref{fig:effective_th}(b,c).  This effective Hamiltonian holds in the limit of $\left|\epsilon_i\right|,\left|\epsilon_f\right|,\Omega\ll\Delta, \Delta_g$.

\section{vertex wire and continuum model}\label{sec:Appendix_chain}
Here, we discuss the eigenvalue problem for the effective Hamiltonian of a vertex wire in Eq.~\eqref{eq:chain_Heff_} and show that, for $\epsilon_{i}=\epsilon_{f}=0$ and $\epsilon_{i}=\epsilon_{f}=t$, it can be mapped to a 1D-Helmholtz equation with homogeneous Dirichlet boundary conditions (DBCs) and homogeneous Neumann boundary conditions (NBCs), respectively.
\subsection{Eigenvalue problem}
The eigenvalue problem $H_{\text{eff}} \ket{\psi_k} = E_k \ket{\psi_k}$ for Eq.~\eqref{eq:chain_Heff_} can be reformulated by treating the local diagonal terms $\epsilon_i$ and $\epsilon_f$ as boundary conditions on two additional virtual domain-wall sites, $\ket{0}_d$ and $\ket{L+1}_d$, within a purely hopping dynamics. Indeed, given $\ket{\psi_k}=\sum_{j=0}^{L+1}\psi_k(j)\ket{j}_d$, the original eigenvalue problem is recovered by imposing the following hopping equation for $j=1,\dots, L$:
\begin{subequations}
 \label{eq:chain_eq_psi_BC}
 \begin{equation}
     \label{eq:chain_eq_psi}
    -t \left[\psi_k(j-1) + \psi_k(j+1)\right] = E_k \psi_k(j),
     \end{equation}
and the following boundary conditions on the virtual sites $j=0,L+1$:
 \begin{align}
 \begin{split}
     \label{eq:chain_eq_BC}
    \psi_k(0) = r_i \psi_k(1), \:\:
    \psi_k(L+1) = r_f \psi_k(L),
 \end{split}
 \end{align}
 \end{subequations}
where $r_{i} = \epsilon_{i}/t$ and $r_{f} = \epsilon_{f}/t$. 
\subsubsection{Solutions to the eigenvalue problem}
We solve Eq.~\eqref{eq:chain_eq_psi_BC} via the ansatz $\psi_k(j) = e^{ikj} - z e^{-ikj}$. Eq.~\eqref{eq:chain_eq_psi} returns the dispersion relation ${E_k = - 2 t \cos{k}}$, while the allowed $k$  are determined from Eq.~\eqref{eq:chain_eq_BC}, which returns the following conditions:
\begin{align}
    &z \left(1 - r_i e^{-ik}\right) = 1- r_i e^{ik}, \nonumber\\
    &z e^{-ikL}\left( e^{-ik} - r_f \right) = e^{ikL} \left( e^{ik}- r_f \right).
\label{eq:BC_z_k}
\end{align}
The real $k$ solutions of this equation form the propagating band of the system and correspond to eigenstates delocalized along the whole system; the imaginary $k$ solutions have energy below the propagating band, and the associated eigenstates are localized at the boundaries. 

Let us first discuss the spectrum in the paradigmatic cases of ${r_i=r_f=0}$ and ${r_i=r_f=1}$. In these cases, Eq.~\eqref{eq:BC_z_k} admits only real $k$ solutions. For ${r_i=r_f=0}$, the solutions are
\begin{equation}
\label{eq:Appendix_chain_seno}
   \psi_{k_m}(j) = \sin(k_m j), \:\:\:\ \text{with}\:\:\:\  k_m  = \frac{m \pi}{L+1}
\end{equation}
and $m=1,\dots,L$. For ${r_i=r_f=0}$, the solutions are
\begin{equation}
\label{eq:Appendix_chain_coseno}
   \psi_{k_m}(j) = \cos(k_m (j-1/2)), \:\:\:\ \text{with}\:\:\:\  k_m  = \frac{m \pi}{L}
\end{equation}
and $m=0,\dots,L-1$. Boundary conditions strongly affect the ground state of the system. For $r_i=r_f=0$, the ground-state population is minimum at the extremes and maximum in the middle of the vertex wire. In contrast, for $r_i=r_f=1$,  the ground-state population is constant along the entire system (namely, the ground state is the equal-weight superposition state).  

For arbitrary boundary conditions, we derive an analytic expression of the smallest roots of Eq.~\eqref{eq:BC_z_k} in the \textit{large}-$L$ limit by rewriting it as 
\begin{equation}
     \sin{\left(kL+k\right)} - \left(r_i + r_f\right) \sin{\left(kL\right)} + r_i r_f \sin{\left(kL-k\right)} = 0. 
     \label{eq:condition_k_chain_1}
\end{equation}
When $L \gg 1$, since the sine functions behave the same at low values of $k$, the roots of the equation are in the vicinity of $kL \simeq m \pi$. We thus Taylor expand the left-hand side of the equation around $k=m\pi/L$, $m=0,1,2,\dots, L$, and derive the following expression of the associated $k$ to the order of $1/L^2$:
\begin{equation}
k_m  \approx \frac{m \pi}{L} + \frac{1}{L}\frac{\left(1-r_i r_f\right)\sin(m\pi/L)}{ \left(r_i+r_f\right)-\left(r_i r_f  +1\right)\cos(m\pi/L)}.
    \label{eq:km_cont_2}
\end{equation}
At the lowest $1/L$ order, the propagating band is therefore not affected by the boundary conditions, and the energy difference between the first two conduction modes scales as $1/L^2$. 

For $r_i$ or(and) $r_f$ larger than $1$, Eq.~\eqref{eq:BC_z_k} admits one(two) imaginary $k = i \kappa$ solutions. In the $\kappa L \to \infty$ limit, we find $\kappa_i = \log{r_i}$ when $r_i > 1$, and $\kappa_f = \log{r_f}$ when $r_f > 1$. The associated wavefunctions are
\begin{equation}
    \psi(j)_{i} = r_i^{-j}, \quad
    \psi(j)_{f} = r_f^{-j} - \frac{1 - r_i / r_f}{1 - r_i r_f} r_{f}^{j},
    \label{eq:chain_BS_right}
\end{equation}
respectively.
\subsection{Continuum model}
For the special cases of ${r_i=r_f=0}$ and ${r_i=r_f=1}$, it is instructive to consider an equivalent continuum model of Eq.~\eqref{eq:chain_eq_psi_BC}, which is valid for the low values of $k$ in the large-$L$ limit. For $L\gg1$, the dispersion relation associated with the first values of $k$ in Eqs.~(\ref{eq:Appendix_chain_seno}, \ref{eq:Appendix_chain_coseno}) can be approximated as $E_k=-2t\cos(k)\simeq-2t+tk^2$. Thus, Eq.~\eqref{eq:chain_eq_psi} simplifies to
\begin{equation}
    \label{eq:chain_eq_psi_BC_approx}
    \psi_k(j-1) + \psi_k(j+1) = \left(2-k^2\right) \psi_k(j). 
\end{equation}

The continuum model is then obtained by replacing the lattice operator $\hat{\mathcal{D}}: \psi(j) \to \psi(j-1) + \psi(j+1)$, defined on the discrete set $j=1,\dots, L$, with the continuous operator $\hat{\mathcal{C}}: \psi(x) \to \psi'' (x) + 2\psi(x)$, defined on the interval $[1, L]\simeq[0, L]$. In this way, Eq.~\eqref{eq:chain_eq_psi_BC_approx} is mapped to a 1D Helmholtz equation:
\begin{equation}
      \psi_k''(x) + k^2 \psi_k(x) = 0,\:\:\:\:x\in\left[0,L\right].
\end{equation}
The boundary conditions of the hopping model are then mapped to the boundary conditions of the Helmholtz equation. In particular, ${r_i=r_f=0}$ is mapped to homogeneous DBC  ${\psi_k(0)=\psi_k(L)=0}$, and ${r_i=r_f=1}$ is mapped to homogeneous NBC $\psi'_k(0)=\psi'_k(L)=0$.

In detail, this mapping relies on Taylor expansions of $\psi_k(x)$, for $k\ll 1$ and $L\gg1$. From
\begin{equation}
    f(x\pm1) \simeq f(x) \pm f'(x) + \frac{1}{2} f''(x),
\end{equation} 
it follows the equivalence between $\hat{\mathcal{D}}$ and $\hat{\mathcal{C}}$; from 

\begin{widetext}
\begin{equation}
\begin{split}
    f(x_0) &\simeq f(x_0+1/2)-\frac{1}{2}f'(x_0+1/2)+\frac{1}{8}f''(L+1/2), \\
    f(x_0+1) &\simeq f(x_0+1/2)+\frac{1}{2}f'(x_0+1/2)+\frac{1}{8}f''(L+1/2),
\end{split}
\end{equation} 
\end{widetext}
it follows the correspondence between the condition ${r_i=r_f=1}$ and the homogeneous NBC. The correspondence between the condition $r_i=r_f=0$ and the homogeneous DBC is straightforward.

\section{Further analyses on the crossing setup}\label{sec:Appendix_crossing}
Here, we analyze the phase diagram of the crossing setup, presented in Fig.~\ref{fig:crossing}, in more detail. First, we develop an effective 1D model, which allows us to describe the different regimes observed in the ordered phase. Second, we employ this model to explain the results obtained using DMRG. Third, we show that the transition between the ordered and disordered phase is of Ising-type.

\subsection{Effective 1D model: single vertex wire with two modified detuning}

Here, we develop an effective theory for a vertex wire of ${N+2}$ atoms with inhomogeneous detuning: $\Delta_g$ on the two atoms at the middle of the vertex wire and $\Delta$ elsewhere. The two midpoint atoms mimic a gadget and divide the vertex wire into two legs, each with an even number of atoms $m=N/2$. This model effectively describes the dynamics of a single domain wall along a vertex wire in the crossing setup (when the position of the domain wall along the other vertex wire is considered to be fixed). 

\subsubsection{Effective theory for \texorpdfstring{$\left|\Delta_g\right|\gg\Omega$}{deltagggomega}}

For $\Omega\ll\Delta,|\Delta_g|$ (with $\Delta>0$), we can treat $H_1=\Omega \sum_{v} \sigma^x_{v}$ as a perturbation to $H_0=- \Delta \sum_{v \notin G} n_v - \Delta_g \sum_{v \in G} n_v$ in Eq.~\eqref{eq:Heff}.

For $\Delta_g>0$, the lowest-energy subspace of $H_0$ comprises single-domain-wall states of the vertex wire, and Eq.~\eqref{eq:Heff} returns the following effective Hamiltonian (up to a constant diagonal term $-\left[\left(\Delta+t\right)N/2+\Delta_g+t'\right]\mathbb{I}$):
\begin{equation}
    H^{\Delta_g>0}_{\text{eff}} = - t' (\ket{\zeta}_d\bra{\zeta+1}+\text{h.c}) 
    - t \sum_{j\neq\zeta} (\ket{j}_d\bra{j+1} + \text{h.c.})
    ,
    \label{eq:Heffwire_deltagpositive}
\end{equation}
with  $j=1,\dots,L$ (where $L=N/2+2$), $\zeta=N/2$, $t=\Omega^2/\Delta$, and $t'=\Omega^2/\Delta_g$. This effective Hamiltonian is the same as the one of the crossing setup, represented in Fig.~\ref{fig:Heff_appendix}(a), when fixing the position of one of the two domain walls.
When $0<\Delta_g<\Delta$, the ground state is localized at $\zeta$ and $\zeta+1$ (around the gadget) and exponentially decays in the two legs. In particular, for an infinite vertex wire, the ground state has energy $-(t'^2+t^2)/t'$, and it is given by (up to a normalization factor):
\begin{equation}
    \ket{\psi}= \sum_{j=-\infty}^{+\infty} \left(t'/t\right)^{-\left|j-\zeta-1/2\right|}\ket{j}.
    \label{eq:boundstate_wire}
\end{equation}
Here, the most populated domain-wall states are $\ket{\zeta}$ and $\ket{\zeta+1}$, where $\left|\psi(\zeta)\right|^2=\left|\psi(\zeta+1)\right|^2=\Delta_g/\Delta$. 
When $\Delta_g\gg\Delta$, the occupation of the states $\ket{\zeta}$ and ${\ket{\zeta+1}}$ is penalized. In this case, the ground state is the equal superposition of two sine wavefunctions [as in Eq.~\eqref{eq:Appendix_chain_seno}] with supports on the two legs, and its energy is given by $E_{\Delta_g\gg\Delta}=-2t$. 

When $\Delta_g<0$, the gadget atoms are pinned to the ground state. This effectively decouples the two legs. Indeed, the lowest energy subspace of $H_0$ comprises states with two domain walls, each on a separate leg, undergoing independent hopping dynamics. We label these states as $\ket{j_1,j_2}$, where $j_1$ and $j_2$ run from $1$ to $\zeta$, indicating the position of the first and the second domain wall on their respective leg. The effective Hamiltonian reads
\begin{equation}
    H^{\Delta_g<0}_{\rm eff} = -t \sum_{j_1, j_2} \ket{j_1, j_2} \bra{j_1+1, j_2} + \ket{j_1, j_2} \bra{j_1, j_2+1} + \text{h.c},
    \label{eq:Heffwire_deltagnegative}
\end{equation}
up to a to a diagonal term $-\left[\left(\Delta+t\right)N/2\right]\mathbb{I}$. The energy of the ground state is now given by ${E_{\Delta_g<0}=-4t}$.

\subsubsection{Effective theory for  \texorpdfstring{$\left|\Delta_g\right| \ll\Delta$}{Deltaglldelta}}
For $\left|\Delta_g\right|, \Omega\ll \Delta$, we can treat $H_1 = -\Delta_g \sum_{v \in G} n_v + \Omega \sum_{v} \sigma^x_{v}$ as a perturbation to $H_0 = -\Delta \sum_{v \notin G} n_v$ in Eq.~\eqref{eq:Heff}. Here, the lowest energy states of the system consist of (\textit{i}) the states $\ket{j}_d$ considered in Eq.~\eqref{eq:Heffwire_deltagpositive}, and (\textit{ii}) the states $\ket{j_1, j_2}$ considered in Eq.~\eqref{eq:Heffwire_deltagnegative}. 
The effective Hamiltonian reads
\begin{equation}
    H_{\rm eff} = H_{i} +  H_{ii} + H_{i,ii},
    \label{eq:eff_crossing_deltag0}
\end{equation}
up to a diagonal term $-\left[\left(\Delta+t\right)N/2\right]\mathbb{I}$, where $\mathbb{I}=\mathbb{I}_i+\mathbb{I}_{ii}$, with  $\mathbb{I}_{i}=\sum_{j}\ket{j}_d\bra{j}$ and $\mathbb{I}_{ii}=\sum_{j_1,j_2}\ket{j_1,j_2}_d\bra{j_1,j_2}$. Here, $H_i = - \Delta_g \mathbb{I}_i + H^{\Delta_g>0}_{\rm eff}\left(t'=0\right)$ describes the dynamics among the (\textit{i})-states, $H_{ii}=H^{\Delta_g<0}_{\rm eff}$ describes the dynamics among the (\textit{ii})-states, and 
\begin{equation}
    H_{i,ii} = \Omega   \sum_{j \leq \zeta} \ket{j}\bra{j,1} + \Omega  \sum_{j > \zeta} \left(\ket{j}\bra{\zeta,\zeta+2-j} + \text{h.c.}\right)
\end{equation}
couples the two sets of states. 

As seen in the previous paragraph, for $\Delta_g<\Delta$, the ground state predominantly populates the (\textit{i})-states $|\zeta\rangle$ and $|\zeta+1\rangle$. These states are coupled to the (\textit{ii})-state $|\zeta,1\rangle$ only, which corresponds to the state where the two domain walls are adjacent to the empty gadget. Therefore, the ground-state energy can be predicted via a simplified three-level model [instead of the complete Hamiltonian in Eq.~\eqref{eq:eff_crossing_deltag0}], which solely includes these three dominant states: $H_\mathrm{3-levels} \approx \Omega (|\zeta\rangle\langle \zeta,1| + |\zeta+1\rangle\langle \zeta,1|+\text{h.c.}) - \Delta_g \left(|\zeta\rangle\langle \zeta|+|\zeta+1\rangle\langle \zeta+1|\right)$. The ground-state energy is given by $E_{\left|\Delta_g\right|\ll\Omega} =-\left(\Delta_g+\sqrt{\Delta_g^2+8\Omega^2}\right)/2$.

\subsection{Phase diagram}
\begin{figure}
    \centering  \includegraphics[width=1.0\columnwidth]{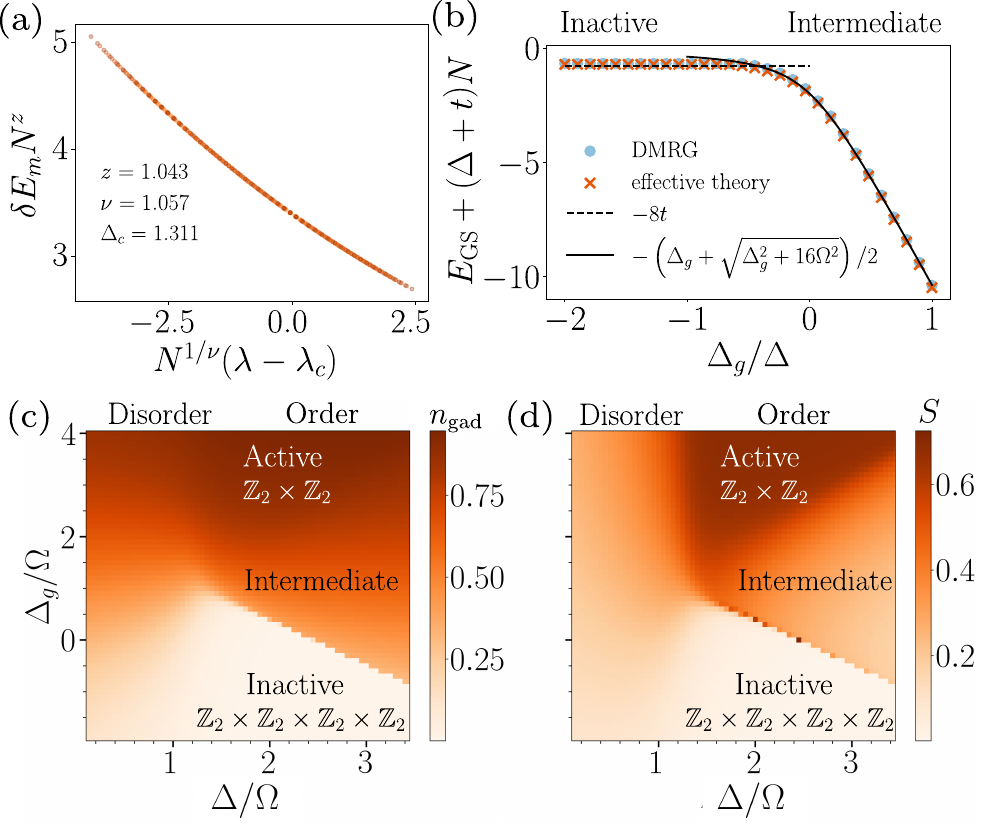}
    \caption{(a) Finite-size scaling analysis of the minimum energy gap of the crossing setup along the standard protocol. We consider two vertex wires of $N$ atoms, with $N$ ranging from $100$ (light orange) to $172$ (dark orange), intersecting at the middle ($m_x=m_y=N/2$), and $\Delta_g=3\Delta$. The data collapse to Eq.~\eqref{eq:crossing_ising_fss} for $z\approx0.91$, $\nu\approx1.07$, and $\lambda_c\approx1.32$.
    (b) Ground-state energy, up to a constant $(\Delta+t)N$, where $t=\Omega^2/\Delta$, as a function of $\Delta_g/\Delta$ for $N=50$ and $\Delta=10\Omega$ obtained using DMRG (blue dots) and the effective theory in Eq.~\eqref{eq:eff_crossing_deltag0} (orange crosses). The black lines are the analytical predictions $E_{\Delta_g<0}=-8t$ and  $E_{\Delta_g>0}=-\left(\Delta_g+\sqrt{\Delta_g^2+16\Omega^2}\right)/2$. The values in the plot are shown for $\Omega=1$. (c) Rydberg excitation density in the gadget $n_{\rm gad}$ as a function of $\Delta/\Omega$ and $\Delta_g/\Omega$. (d) Entanglement entropy $S$ between one leg and the rest of the system as a function of $\Delta/\Omega$ and $\Delta_g/\Omega$.}
    \label{fig:Crossing_FSS}
\end{figure}
Here, we investigate the regimes of the ordered phase for the crossing setup, shown in Fig.~\ref{fig:crossing}.
Using DMRG (see Appendix~\ref{sec:Appendix_DMRG}), we compute two quantities: the Rydberg excitation density in the gadget, denoted as ${n_{\rm gad}=\bra{{\rm GS}} \sum_{v \in G} n_{v}\ket{{\rm GS}}}$, and the entanglement entropy between one leg and the rest of the system, defined as~\cite{Bennett_1996}
\begin{equation}
    S = - \sum_i |s_i|^2 \log |s_i|^2.
\end{equation}
Here, $s_i$ is the singular value at the bond that connects one leg to the gadget tensor. We show $n_{\mathrm{gad}}$ in Fig.~\ref{fig:Crossing_FSS}(c) and $S$ in Fig.~\ref{fig:Crossing_FSS}(d) as a function of $\Delta/\Omega$ and $\Delta_g/\Omega$. In the following, we provide an interpretation of these quantities based on the effective 1D model developed in the previous subsection for a single-vertex wire with two modified detuning.

For $\Delta_g<0$, the gadget is empty ($n_{\rm gad}\approx0$). When going from the disordered to the ordered phase, each leg is decoupled from the rest of the system  (since $S\approx0$) and independently undergoes the $\mathbb{Z}_2$ spontaneous symmetry breaking of a one-dimensional Rydberg chain. Consequently, the system enters an ordered regime characterized by a $\mathbb{Z}_2\times\mathbb{Z}_2\times\mathbb{Z}_2\times\mathbb{Z}_2$ symmetry. 
Within this ordered regime, a perturbative description is then possible when $\Omega\ll\Delta,\left|\Delta_g\right|$. The effective Hamiltonian is obtained by generalizing the single-vertex wire Hamiltonian in Eq.~\eqref{eq:Heffwire_deltagnegative} to two vertex wires. In this regime, the ground-state energy is given by $E_{\Delta_g<0}=-8t$ (twice that of the single vertex wire).

At the opposite limit $\Delta_g\gtrsim\Delta$, the gadget hosts a Rydberg excitation ($n_{\rm gad}\approx1$). This has the effect of coupling opposite legs, which behave as a single vertex wire and undergo (in pair) the $\mathbb{Z}_2$ symmetry breaking. Consequently, at $\Delta/\Omega\sim1.3$, the system enters an ordered regime with a $\mathbb{Z}_2\times\mathbb{Z}_2$ symmetry. Here, the entanglement entropy between a single leg and the rest of the system (specifically, with the opposite leg) is maximum. In the $\Delta_g\gg\Delta$ limit, a perturbative treatment of the system generalizes the results of the single-wire analysis [see Eq.~\eqref{eq:Heffwire_deltagpositive}] and predicts a ground-state energy equal to $E_{\Delta_g\gg\Delta}=-4t$ (twice that of the single vertex wire).

The active and inactive regimes are separated by an intermediate region ($0<\Delta_g<\Delta$). Here, the occupation of the gadget varies around $n_{\rm gad}\approx0.5$ [see Fig.~\ref{fig:Crossing_FSS}(c)], and the Rydberg excitation is equally superpositioned among the four states depicted in the inset of Fig.~\ref{fig:Heff_appendix}(a). In the $\left|\Delta_g\right|\gg\Omega$ limit, the system can be studied perturbatively, generalizing the single-wire Hamiltonian in Eq.~\eqref{eq:eff_crossing_deltag0} to two vertex wires. In particular, the simplified three-level model is now adapted to a five-level one --- one empty gadget state plus the four states in Fig.~\ref{fig:Heff_appendix}(a) --- which predicts ground-state energy equal to $E_{\Delta_g>0}=-\left(\Delta_g+\sqrt{\Delta_g^2+16\Omega^2}\right)/2$.

In Fig~\ref{fig:Crossing_FSS}(b), we show that the ground-state energy obtained with the effective model in Eq.~\eqref{eq:eff_crossing_deltag0} (orange crosses) aligns with the one obtained using DMRG (blue dots). Furthermore,  it matches the two analytical predictions $E_{\Delta_g<0}=-8t$ and $E_{\Delta_g>0}=-\left(\Delta_g+\sqrt{\Delta_g^2+16\Omega^2}\right)/2$ in their respective limits. 
\subsection{Ising transition}
In Sec.~\ref{sec:Crossing}, we showed that the minimum energy gap, at the transition between the disordered phase and the active regime ($\mathbb{Z}_2 \times \mathbb{Z}_2$), scales as $ L^{-z}$ with $z\approx1$. To gain a better understanding of this phase transition, we perform a finite-size scaling analysis~\cite{Fisher_1972,Hamer_1980}. We use the following ansatz for the minimum gap:
\begin{equation}
    \delta E_m = L^{-z} f[L^{1/\nu}(\lambda-\lambda_c)],
    \label{eq:crossing_ising_fss}
\end{equation}
where $\lambda=\Delta/\Omega$, $\lambda_c$ is the critical point, $z$ and $\nu$ are the dynamical and the correlation length critical exponents, and $f$ is a universal scaling function. In Fig.~\ref{fig:Crossing_FSS}(a), we show the best-fitting data collapse of the energy gap for $z\approx0.91$, $\nu\approx1.07$, and $\lambda_c\approx1.32$. Due to the limited bond dimension ($D=16$) and system size, this estimation is not precise. These values are close to that of the Ising universality class with $z=\nu=1$~\cite{Suzuki_2013}, as in one-dimensional Rydberg chains with nearest-neighbor blockade~\cite{Sachdev_2002,Fendley_2004}. 

\section{Further analyses on the CWE setup}\label{sec:Appendix_CWE}
Here, we discuss how to study the CWE setup within the continuum model in Eq.~\eqref{eq:Helmoltz_2d} with homogeneous Dirichlet boundary conditions. This model reproduces the solution to the eigenvalue problem of the discrete effective Hamiltonian in Fig.~\ref{fig:effective_th}(c) for $\epsilon_{i,w}=\epsilon_{f,w}=0$ in the long-leads limit, i.e., ${L \gg \text{max}(\zeta_x, \zeta_y)}$. First, we present a matrix-transfer method that allows us to predict the occurrence of the bound state in an infinite $L$-shaped domain via the so-called \textit{single-mode} approximation. Second, we numerically predict the scaling of the minimum gap $\delta E_m$ as a function of $L$ and when varying the ratio between the width of the leads. Third, we present a semi-analytic treatment of the problem based on conformal mapping. 

\subsection{Matrix-transfer method and bound state}
\begin{figure}[t]
    \centering
    \includegraphics[width=1\columnwidth]{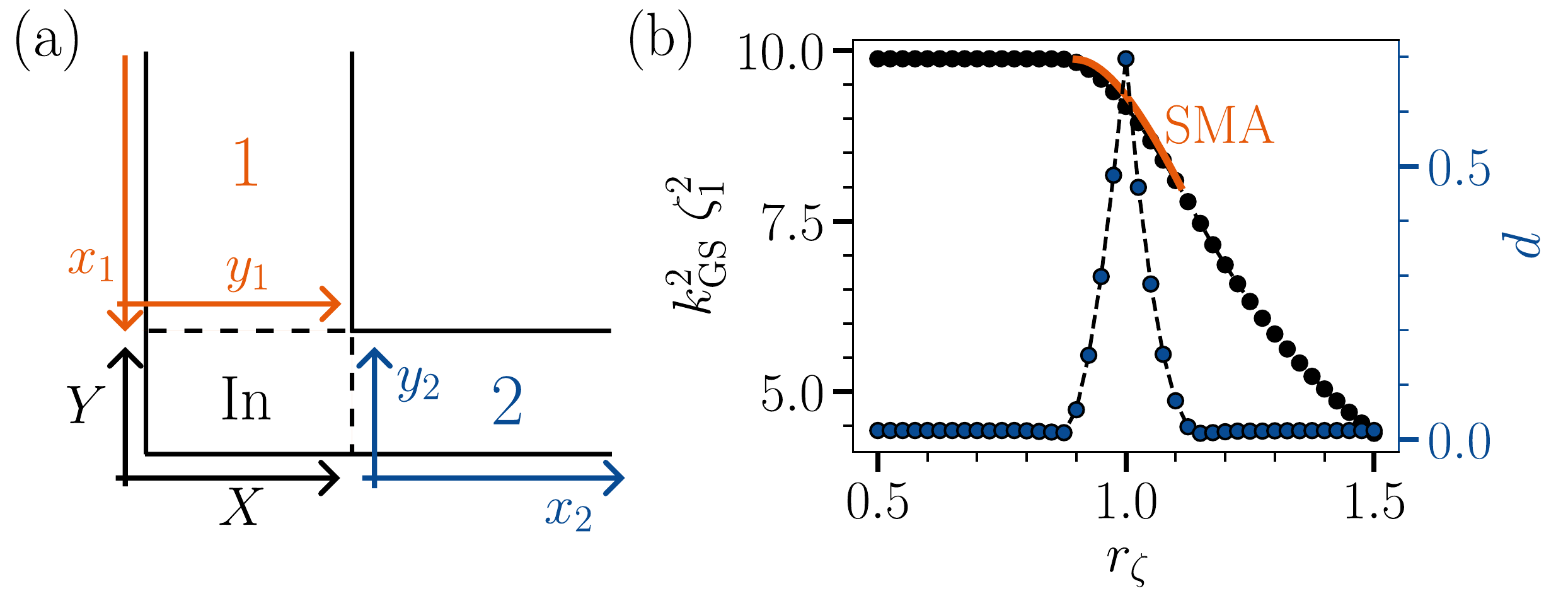}
    \caption{Analysis of the bound state in a infinite $L$-shaped domain. (a) For $L\to \infty$, the domain of the Helmholtz equation in Eq.~\eqref{eq:Helmoltz_2d} acquires an infinite $L$ shape. The two leads are denoted by $1$ and $2$, with widths $\zeta_1$ and $\zeta_2$, and the inner region with ``In". Three different reference frames are used for the three regions. (b) Ground-state energy, $k^2_{\rm GS}$, computed via (black) finite-element method (FEM) and (orange) single-mode approximation (SMA)  as a function of $r_{\zeta}=\zeta_2/\zeta_1$. The gap $d$ in Eq.~\eqref{eq:gap_dirichlet_eff} computed using FEM is shown in blue. Here, we set $L = 40 \zeta_1$. }
    \label{fig:cwe_gap_DBC_eff}
\end{figure}
Here, we generalize the matrix-transfer method presented in Ref.~\cite{Londergan_1999} to solve the Helmholtz equation with homogeneous DBC in a symmetric infinite $L$-shaped domain, to the asymmetric case (namely, leads of different widths). Following the notations of Fig.~\ref{fig:cwe_gap_DBC_eff}(a), we divide the Hilbert space into three distinct regions: two leads, denoted by $\Gamma \in \{ 1,2\}$, of different widths ($\zeta_1\equiv\zeta_x$ and $\zeta_2\equiv\zeta_y$); and an inner region, denoted by ``In." We employ three different reference frames: $\{x_1,y_1\}$ in the lead extending along $y$, $\{x_2,y_2\}$ in the lead extending along $x$, and $\{X,Y\}$ in the inner region.

Using the separation of variables, the solution of the Helmholtz equation in the lead $\Gamma \in \{ 1,2\}$ takes the form
\begin{equation}
\label{app:equation_leads}
    \psi_{\Gamma}(x,y) = \sum_{n=1}^{\infty} (C_{\Gamma n} e^{i \alpha_{\Gamma n} x}+\Tilde{C}_{\Gamma n} e^{-i \alpha_{\Gamma n} x}) \phi_{\Gamma n}(y),
\end{equation}
with 
\begin{equation}
    \phi_{\Gamma n}(y) = \sqrt{\frac{2}{\zeta_{\Gamma}}} \sin{\left(\frac{n \pi y}{\zeta_{\Gamma}}\right)},  \:\:\ 
    \alpha_{\Gamma n} = \sqrt{k^2-\left(\frac{n \pi}{\zeta_{\Gamma}}\right)^2}.
\end{equation}
In the inner region, it can instead be written as
\begin{equation}
    \psi_{in}(X,Y) = \sum_n D_n \chi_{n0}(X,Y)+E_n \chi_{0n}(X,Y),
\end{equation}
where $\chi_{n0}$ and $\chi_{0n}$ solve the Helmholtz equation in this region, satisfying the following boundary conditions:
\begin{equation}
\begin{aligned} 
&\chi_{n0}(X,\zeta_2)
=\phi_{1n}(X),\quad\quad \chi_{0n}(X,\zeta_2)
=0, \\
&\chi_{0n}(\zeta_1, Y) = \phi_{2n}(Y),\quad\quad\chi_{n0}(\zeta_1, Y) = 0.
\end{aligned}
\end{equation}

By imposing the continuity of the wavefunction and of its normal derivative at the interfaces between the inner region and the two leads, we derive the following relation among the expansion coefficients in the different regions: 
\begin{equation}
    \bm{S}_1=  \bm{M} \bm{S}_2 .
    \label{eq:scatteringmatrix}
\end{equation}
Here, $\bm{S}_1 = (\bm{C}_1, \Tilde{\bm{C}}_1)^{T}$ and $\bm{S}_2 = (\bm{C}_2, \Tilde{\bm{C}}_2)^{ T}$, where ${
\left(\bm{C}_\Gamma\right)_{mn} = C_{\Gamma n} \delta_{mn}}$ and ${(\Tilde{\bm{C}}_\Gamma)_{mn} = \Tilde{C}_{\Gamma n} \delta_{mn}}$ ($\delta_{mn}$ being the Kronecker delta); $T$ indicates transposed matrix; and
\begin{align}
\bm{M} = \begin{pmatrix}
\bm{B}+i\bm{\alpha}_1 & \bm{B}-i\bm{\alpha}_1\\
\bm{A}' & \bm{A}' 
\end{pmatrix}^{-1}
\begin{pmatrix}
\bm{A} & \bm{A}\\
\bm{B}'-i\bm{\alpha}_2 & \bm{B}'+i\bm{\alpha}_2
\end{pmatrix}
\end{align}
is the \textit{transfer} matrix.
Here, the different matrices are defined in the following way: $
{ \left(\bm{A}\right)_{mn} = A_{mn}}$, $
{(\bm{A}^{'})_{mn} = A_{mn}^{'}}$, ${ \left(\bm{B}\right)_{mn} = B_{m}\delta_{mn}}$, ${ (\bm{B}^{'})_{mn} = B_{m}^{'}\delta_{mn}}$, $\left(\bm{\alpha}_\Gamma\right)_{mn}=\alpha_{\Gamma n}\delta_{mn}$, where
\begin{widetext}
\begin{multline} \label{eq:HJB3qgates}
    A_{mn} = -\int_0^{\zeta_{1}} dx \phi_{1m}(x) \partial_{Y} \chi_{0n}(x,\zeta_2)= \frac{2nm \pi^2}{(\zeta_1 \zeta_2)^{3/2}} \frac{(-1)^{n+m}}{\left(\frac{m\pi}{\zeta_1}\right)^2 - \alpha^2_{2n}}, \:\:\:\
    B_{m} = \int_0^{\zeta_{1}} dx \phi_{1m}(x) \partial_{Y} \chi_{m0}(x,\zeta_2) = \alpha_{2m}  \cot{(\alpha_{2m}\zeta_1)}  \\
    A^{\prime}_{mn} = -\int_0^{\zeta_{2}} dy \phi_{2m}(y) \partial_{X} \chi_{n0}(\zeta_1,y) = \frac{2nm \pi^2}{(\zeta_1 \zeta_2)^{3/2}} \frac{(-1)^{n+m}}{\left(\frac{m\pi}{\zeta_2}\right)^2 - \alpha^2_{1n}}, \:\:\:\
    B^{\prime}_{m} = \int_0^{\zeta_{2}} dy \phi_{2m}(x) \partial_{X} \chi_{0m}(\zeta_1,y)= \alpha_{2m}  \cot{(\alpha_{1m}\zeta_2)}.
\end{multline}
\end{widetext}

The bound state is a stationary state, exponentially damped at infinity along the two leads. Thus, it requires all $\alpha_{\Gamma n}$ to be imaginary and can exist only for values of $k$ smaller than the threshold $k_{\rm th}=\min(\pi/\zeta_1,\pi/\zeta_2)$.
The bound state and its energy can be predicted efficiently using the single-mode approximation (SMA), which neglects all the modes with $n>1$. Within this approximation, the matrices $\bm{A}$, $\bm{B}$, $\bm{A}^{'}$ and $\bm{B}^{'}$ are scalars and the transfer matrix $\bm{M}$ is a $2 \times 2$ matrix. The bound state can thus be found from Eq.~\eqref{eq:scatteringmatrix} by setting the coefficients $C_{11}$ and $\Tilde{C}_{21}$ to zero:
\begin{equation}
    \begin{pmatrix}
        0 \\
        \Tilde{C}_{11}
    \end{pmatrix} = \begin{pmatrix}
        M_{11} & M_{12} \\
        M_{21} & M_{22}
    \end{pmatrix} \begin{pmatrix}
        C_{21} \\
        0 
    \end{pmatrix}.
\end{equation}

This equation implies ${M_{11}=0}$, from which we determine the value of $k$ for the bound state, $k_{\rm BS}$.
We rescale the system by setting (without loss of generality) $\zeta_1 =1$ and rescaling the energies accordingly. We find that $k_{\rm BS}$ satisfy the following equation:  
\begin{widetext}
\begin{eqnarray}\label{eq:SMA}
    &&\frac{4 \pi^4}{\zeta_2^3 \left(\pi^2\left(1+\zeta_2^{-2} \right) -k_{\rm BS}^2\right)^2\sqrt{\pi^2-k_{\rm BS}^2}\sqrt{\left(\pi/\zeta_2\right)^2-k_{\rm BS}^2}}  \\
   &=& 1+\coth\left(\zeta_2\sqrt{\pi^2-k_{\rm BS}^2} \right)\coth\left(\sqrt{\left(\pi/\zeta_2\right)^2-k_{\rm BS}^2}\right)+  \coth\left(\zeta_2\sqrt{\pi^2-k_{\rm BS}^2} \right)+ \coth\left(\sqrt{\left(\pi/\zeta_2\right)^2-k_{\rm BS}^2}\right). \nonumber
\end{eqnarray}
\end{widetext}

In Fig.~\ref{fig:cwe_gap_DBC_eff}(b), we compare the SMA prediction (orange line) with the results obtained by solving the Helmholtz equation numerically (black dots) for a sufficiently large domain ($L=40\zeta_1$), which approximate an infinite one. We observe that, in the range where the ground state is a bound state, the single-mode approximation correctly reproduces the value of the ground-state energy $k_{\rm GS}^2$. In particular, for the symmetric case ($\zeta_1=\zeta_2$), the SMA predicts ${k^2_{\mathrm{BS}} \zeta_1^2 = \left(1-0.054\right)\pi^2}$, which is known to overestimate the exact value of $\left(1-0.071\right)\pi^2$~\cite{Londergan_1999}.

\subsection{Energy gap scaling for arbitrary geometries}
Here, we employ a finite-element method (FEM) to solve the Helmholtz equation with homogeneous DBC, in the finite $L$-shaped domain. For different values of $r_{\zeta}$, we compute how the energy gap ${\delta E=t\left(k^2_{\rm th}-k^2_{\rm GS}\right)}$ scales with $L$. We choose, without loss of generality, $L=1$ and numerically solve the equation rescaling the frame of coordinates as $\left\{x'=x/L,y'=y/L\right\}$ and the energies via ${k'=k/L}$.  
We find the following scaling of the energy gap:
\begin{equation}
    \delta E/t \overset{L \to \infty}{\sim}    \frac{D\left(r_\zeta\right)}{L^2} + \frac{d\left(r_\zeta\right)}{\zeta_1^2} ,
    \label{eq:gap_dirichlet_eff}
\end{equation}
where $r_\zeta=\zeta_2/\zeta_1$, $d(r_{\zeta})$ is shown in Fig.~\ref{fig:cwe_gap_DBC_eff}(b), and $D(r_{\zeta})$ is a bounded function. In the infinite-lead limit, the bound state exists when  $d(r_{\zeta})\neq 0$, which is verified for geometry close to the symmetric one ($r_{\zeta}\sim1$).  

Eventually, we observe that in the $\Omega\ll\Delta$ limit, these results can be compared with the ones obtained using DMRG [see  Fig.~\ref{fig:gap_dirichlet_DMRG}]. For instance, given $\Omega/\Delta=0.5$, the DMRG method predicts $\delta E\sim 0.17 \Delta$, which agrees with the effective-model value $\delta E(r_\zeta=1) \sim 0.7 t$ for $t=0.5^2\Delta$.
\subsection{Conformal mapping and bound state}
\begin{figure}
	\centering
	\includegraphics[width=\columnwidth]{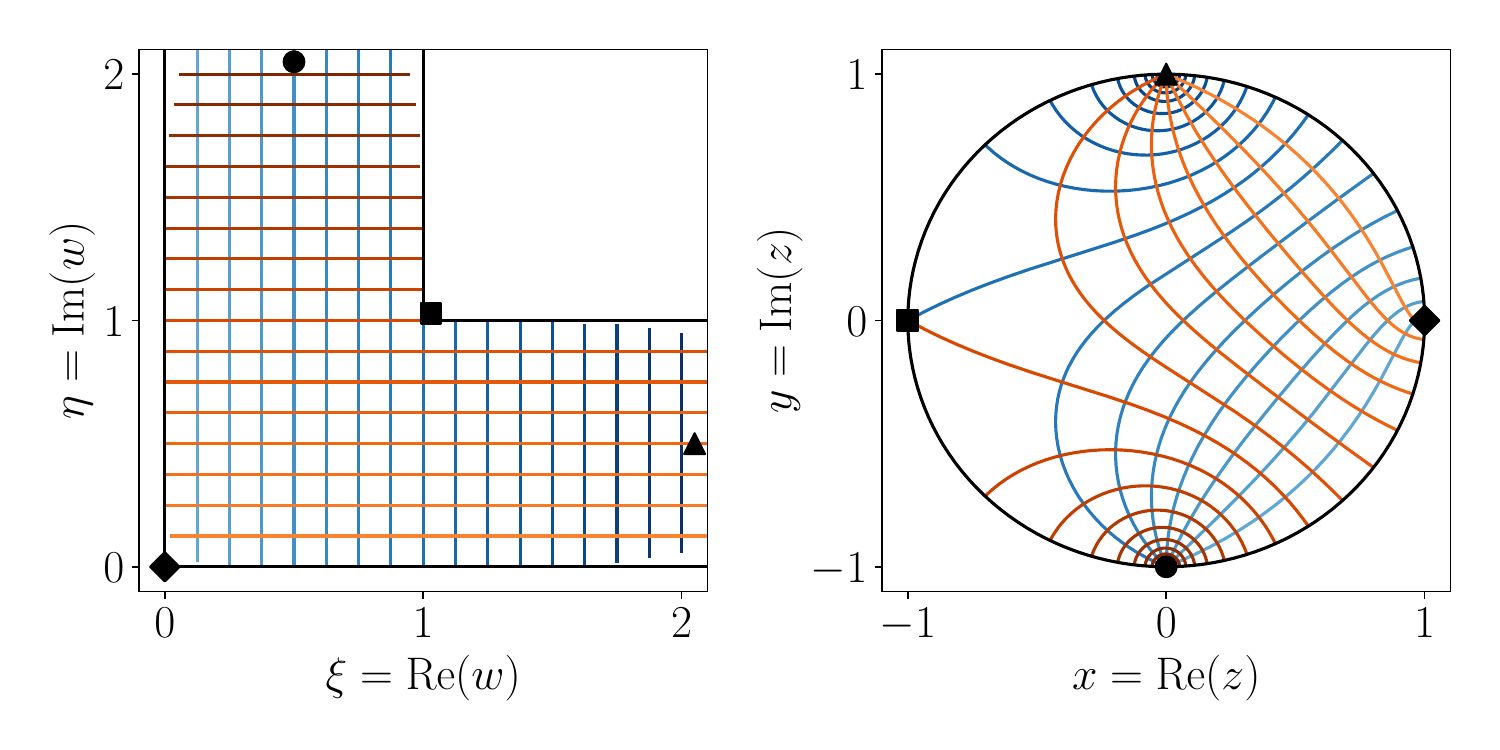}
 \caption{The conformal mapping $w\mapsto z$ from the $L$-shaped region $\mathcal{L}=\mathcal{L}_{p=\frac{1}{2}}$ to the unit disk $\mathcal{D}$ for $\zeta_x=\zeta_y=1$, whose inverse mapping is given by Eq.~\eqref{eq:conformal_map}. Conformal mapping of the Cartesian lines from $\mathcal{L}$ to  $\mathcal{D}$. The four black symbols show how the characteristic points are mapped between the two regions.}
 \label{fig:ConformalMap}
\end{figure}
The bound-state solution of the Helmholtz equation can be further investigated via a conformal mapping of the $L$-shaped domain into a unit disk. Here, the transformed equation admits a semianalytical treatment, which allows us to classify the eigenstates into bound states and propagating states at infinity. Without loss of generality, in the following discussion, we normalize the width of the two leads such that $\zeta_1\zeta_2=1$. 

We parameterize the $L$-shaped domain $\mathcal{L}$ with a complex coordinate $w=\xi + i \eta$, and the unit disk ${\mathcal{D}=\{(x,y)|\sqrt{x^2+y^2}<1\}}$ with  $z=x+iy$. We derive a conformal mapping between $w$ and $z$ in two steps, where we resort to an intermediate coordinate $u$ spanning the upper half of the complex plane. First, the mapping from $u$ to $w$ is achieved via a Schwarz-Christoffel transformation that sends the prevertices $u=-\zeta_1^{2},0,\zeta_1^{-2},\infty$ on the real axis $u$ to the vertices of the $L$-shaped polygon $w=+i\infty,0,+\infty,\zeta_1+i\zeta_1^{-1}$, respectively:
\begin{subequations}\label{eq:conformal_map}
\begin{eqnarray}
    w &=& -\frac{\zeta_1^2+\zeta_1^{-2}}{\pi}\int_0^u \frac{d \tilde{u}}{(\zeta_1^{-2}\tilde{u}-1)\tilde{u}^{1/2} (\zeta_1^2\tilde{u}+1)} \\
    &=& \frac{2}{\pi} \left[\zeta_1\arctan(\zeta_1\sqrt{u}) + \zeta_1^{-1} \artanh\left(\zeta_1^{-1}\sqrt{u}\right)\right]. \nonumber
\end{eqnarray}
Second, we map the unit disk, parameterized by $z$, to the upper half-plane, parameterized by $u$, via the following M\"{o}bius transformation: 
\begin{eqnarray}
    u = -i\frac{z-1}{z+1}.
\end{eqnarray}
\end{subequations}
In this way, we obtain a closed-form expression for the inverse conformal mapping from $z$ to $w$. In Fig.~\ref{fig:ConformalMap}, we show how the coordinate lines transform from $w$ to $z$ and highlight that the points $z=1$, $-1$, $i$, $-i$ on $\mathcal{D}$ corresponds to $w=0$, $\zeta_1+i\zeta_1^{-1}$,$+\infty+i/2$,$1/2+i\infty$ on $\mathcal{L}$, respectively.

To facilitate a later analysis, we generalize this procedure to ``bent" $L$-shaped domains. We introduce the following family of conformal maps, parameterized by ${p\in[0,1)}$:
\begin{equation}
\label{Eq:Generalized_conformal_mappings}
    w = -\frac{\zeta_1^2+\zeta_1^{-2}}{\pi}\int_0^u \frac{d \tilde{u}}{(\zeta_1^{-2}\tilde{u}-1)\tilde{u}^{p} (\zeta_1^2\tilde{u}+1)}.
\end{equation}
These transformations conformally map ``bent" $L$-shaped domains with internal angle of $(1-p)\pi$, namely, 
$\mathcal{L}_p=\{{(\xi,\eta)}|{\eta\ge0},\,{\xi\ge-\eta\cot p\pi}\}\setminus\{{(\xi,\eta)}|{\eta>\zeta_1^{2p}},\,{\xi>-\eta\cot p\pi+\zeta_1^{-2p}\csc p\pi}\}$, into the unit disk $\mathcal{D}$. 
In particular, the $p=1/2$ case corresponds to the $L$-shaped domain $\mathcal{L}$, while the $p=0$ case, where $z=\left(i+\cos(\pi w)\right)/\left(1+i\cos(\pi w)\right)$, corresponds to an infinitely long strip of width $1$: ${\mathcal{L}_0=\{(\xi,\eta)|0\le\eta\le1\}}$. 

Upon this mapping ($w \to z$), the Laplacian operator is transformed as $\nabla^2_w \mapsto \chi_p^{-1}\nabla^2_z,$
where $\chi_p=|\partial w/\partial z|^2$ is the Jacobian of the transformation. The Helmholtz equation on $\mathcal{L}$ is thus mapped to the following eigenvalue problem on $\mathcal{D}$:
\begin{eqnarray}\label{eq:mapped_eigenproblem}
	-\nabla^2_{x,y} \phi(x,y) = k^2 \chi_p(x,y)\phi(x,y),
\end{eqnarray}
with homogeneous Dirichlet boundary condition ${\phi(\partial \mathcal{D})=0}$. 
The Jacobians of the generalized mappings in Eq.~\eqref{Eq:Generalized_conformal_mappings} are given by
\begin{widetext}
\begin{equation}
    \chi_{p}(x,y) = \frac{4}{\pi^2}\left(\frac{(x+1)^2+y^2}{(x-1)^2+y^2}\right)^p \frac{1}{[(x-\cos\varphi)^2+(y+\sin\varphi)^2]} 
      \frac{1}{[(x+\cos\varphi)^2+(y-\sin\varphi)^2]},
\end{equation}
\end{widetext}
with ${\varphi=2\arccot\zeta_1^2}$. 
\subsubsection{Qualitative analysis of the  eigenstates}
\begin{figure}
	\centering
	\includegraphics[width=\columnwidth]{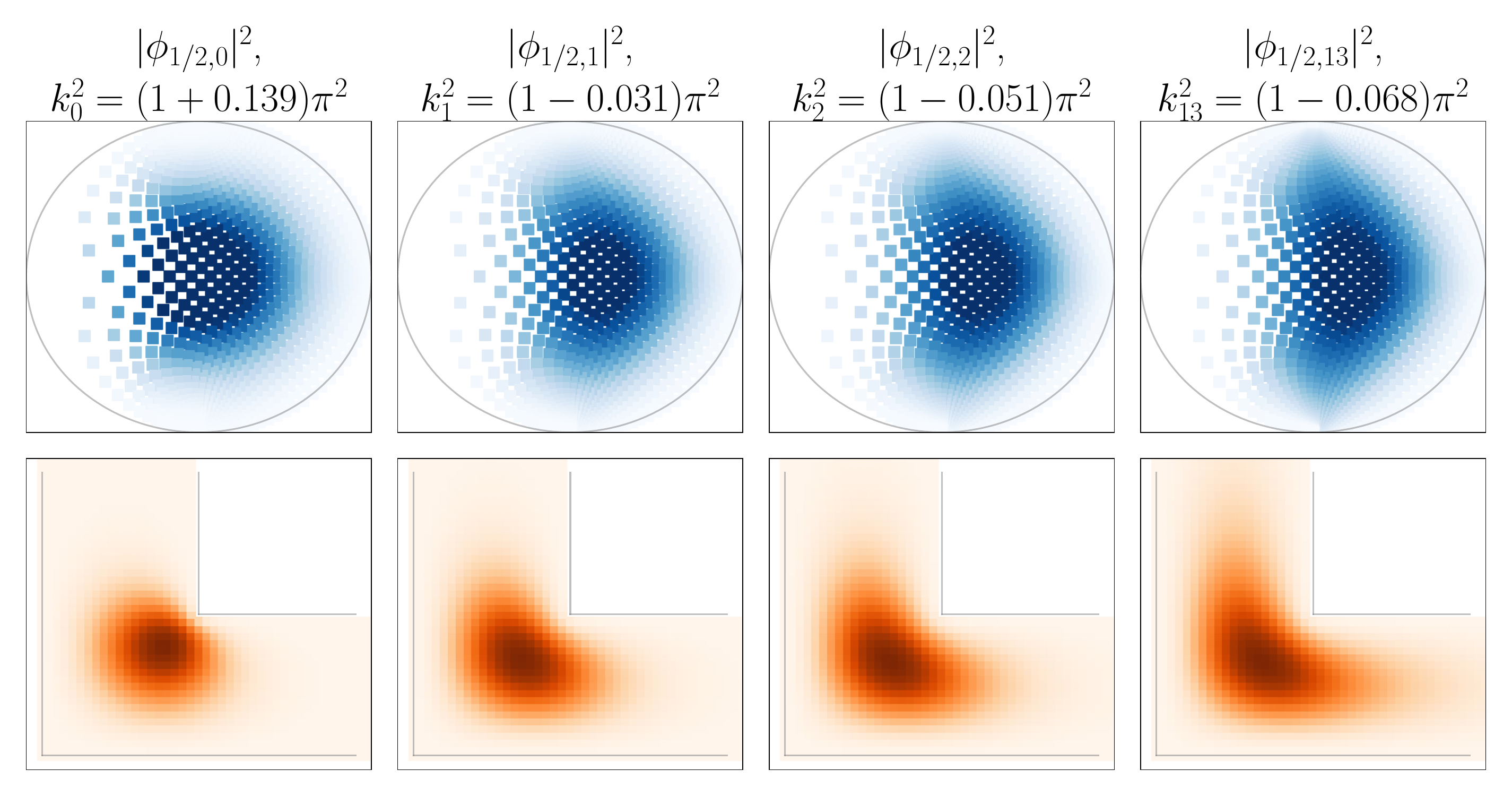}
    \caption{Homotopy perturbation method. $\left|\phi_{p=1/2,n}\right|^2$ is obtained iterating Eq.~\eqref{eq:conformal_iteration} and $k_{p=1/2,n}^2$ is extracted from Eqs.~(\ref{eq:mapped_eigenvalue},\ref{eq:HPM_Poisson}). Results for the first few iterations ($n=0,1,2$) and $n=13$ are shown (top) in the mapped unit disk $\mathcal{D}$ and (bottom) in the original $L$-shaped region $\mathcal{L}$. The initial guess $\phi_0$ is the ground state of the Laplacian in $\mathcal{D}$ (with homogeneous Dirichlet boundary conditions). The numerical mesh is chosen to distribute evenly in $\mathcal{L}$ for optimal numerical performance.
    As $n$ grows, $k_{p=1/2,n}^2$ approaches the exact value of $k^2/\pi^2=1-0.071$~\cite{Londergan_1999}.}
	\label{fig:Conformal_WF}
\end{figure}
The asymptotic behaviors of the eigenstates at infinity of $\mathcal{L}_p$ can be inferred from the analytical solution of Eq.~\eqref{eq:mapped_eigenproblem} in the vicinity of the transformed points. The points at infinity in the two leads $\Gamma\in\{1,2\}$, namely, $w_1=-e^{-ip\pi}\infty$ and $w_2=+\infty$, are, respectively, mapped to $z_1=-e^{-i\varphi}$ and $z_2=e^{-i\varphi}$ on the unit disk. In the vicinity of these poles, 
Eq.~\eqref{eq:mapped_eigenproblem} can be solved by separating the variables 
after switching to polar coordinates, i.e., $z-z_{\Gamma}=\rho e^{i\theta}$. To the lowest order in $\rho$, it results that $\chi_p\simeq\pi^2\zeta_{\Gamma}^{4p}/\rho^2$ has no angular dependence. By setting $\phi_p(\rho,\theta)=R(\rho)T(\theta)$,  Eq.~\eqref{eq:mapped_eigenproblem} returns ${T''+m^2T=0}$ and
\begin{eqnarray}\label{eq:mapped_eigenvalue_R}
    \rho^2 R'' + \rho R' + (k^2\zeta_{\Gamma}^{4p}/\pi^2-m^2) R = 0,
\end{eqnarray}
with boundary conditions $R(0)<\infty$ and $T(0)=T(\pi)=0$. 

Due to the angular boundary condition, $m$ is a positive integer, and, depending on its relation with $k$, two families of solutions can be identified. When $k^2<m^2\pi^2/\zeta_{\Gamma}^{4p}$, the two linearly independent solutions of Eq.~\eqref{eq:mapped_eigenvalue_R} are  $R_{\pm}=\rho^{\pm\sqrt{m^2-k^2\zeta_{\Gamma}^{4p}/\pi^2}}$ (with only $R_+$ satisfying the boundary condition). When $k^2>m^2\pi^2/\zeta_{\Gamma}^2$, the two linearly independent solutions are $R=\cos[\sqrt{k^2\zeta_{\Gamma}^{4p}/\pi^2-m^2}\log \rho],\quad \sin[\sqrt{k^2\zeta_{\Gamma}^{4p}/\pi^2-m^2}\log \rho]$. 
The first family corresponds to states exponentially decaying along the lead $\Gamma$, while the second corresponds to propagating states. This can be seen by observing that, for $z\simeq z_1$ ($z\simeq z_2$), the conformal mapping in Eq.~\eqref{eq:conformal_map} reduces to $\rho\approx e^{-\pi\eta}$ ($\rho\approx e^{-\pi\xi}$).

We proceed to characterize the ground-state energy $k_p^2$ for a general choice of $p$, in reference to $k_{p=0}^2$. To this end, we introduce the expectation value of the kinetic energy on an arbitrary state $\phi$, evaluated on $\mathcal{D}$, in the following way: 
\begin{eqnarray}\label{eq:mapped_eigenvalue}
	E_p[\phi] = \frac{-\int_{\mathcal{D}} dxdy \, \phi \nabla^2 \phi}{\int_{\mathcal{D}} dxdy \, \chi_p \phi^2},
\end{eqnarray}
where both the numerator and denominator are positive. We observe that the following inequalities hold for $p>0$:
\begin{eqnarray}\label{eq:mapping_ineq}
    k_{p=0}^2 = E_{p=0}[\phi_{p=0}] > E_p[\phi_{p=0}] \ge E_p[\phi_p] = k_p^2.
\end{eqnarray}

Here, the first inequality is derived from the spatial symmetries of $\phi_{p=0}$ and $\chi_p$. Indeed, the invariance of the operator $-\chi_0^{-1}\nabla^2$ upon the transformation $(x,y)\to(-x,-y)$ leads to  $\phi_{p=0}^2(x,y)=\phi_{p=0}^2(-x,-y)$, and the Jacobians verify ${\chi_p(x,y)+\chi_p(-x,-y)>2\chi_{p=0}(x,y)}$ for any $p>0$ (using $\mu+1/\mu>2$ for $\mu>0,\mu\neq1$). The second inequality stems from the variational principle for estimating the ground-state energy (i.e., $E_p[\phi]\ge E_p[\phi_p]$ for every $\phi$). Solving the Helmholtz equation directly in $\mathcal{L}_0$ results in $\phi_{p=0}=\frac{1-x^2-y^2}{\sqrt{[x^2+(y-1)^2][x^2+(y+1)^2]}}$ and $k_{p=0}^2=\pi^2$, and thus the ground-state energy in $\mathcal{L}_{p>0}$ is characterized by a strict upper bound: $k_{p>0}^2 < \pi^2$.

In the symmetric case (namely, $\zeta_1=1$), $\pi^2$ corresponds to the energy threshold at which propagating modes appear. Thus, the ground state is a bound state exponentially decaying in the two leads. This result is consistent with previous studies on similar ``bent'' $L$-shaped domains via variational methods~\cite{Avishai_1991}.

\subsubsection{Semianalytical solution}
We now present a possible semianalytical treatment of the ground state for $\mathcal{L}_p$ domains. In particular, the {\it homotopy perturbation method} (HPM)~\cite{Reck_11} can be applied to Eq.~\eqref{eq:mapped_eigenproblem}, leading to a recursive approximation of the bound state. 
Starting from an initial guess $\phi_{p,n=0}$,
this method generates a sequence of un-normalized wavefunctions $\phi_{p,n}$ and corresponding energies $k^2_{p,n}$, via the following Poisson equation:
\begin{eqnarray}\label{eq:HPM_Poisson}
	-\nabla^2 \phi_{p,n} = k^2_{p,n-1}\chi_p \phi_{p,n-1}, \:\:\:\:k_{p,n}^2 = E_{p}[\phi_{p,n}].
\end{eqnarray}
In the $n\to\infty$ limit, $\phi_{p,n}$ converges to $\phi_p$ and $k_{p,n}^2$ to $k_p^2$.

Specifically, as an initial guess, we choose the ground state of the Helmholtz equation in the disk (i.e., $\chi_p=1$):
\begin{eqnarray}
	\phi_{p,0}(x,y) = \frac{J_0(\beta \sqrt{x^2+y^2})}{\sqrt{\pi}J_1(\beta)},
\end{eqnarray}
where $J_n$ denotes the Bessel functions of the first kind and order $n$, and $\beta$ is the first zero of $J_0$. 
The Poisson equation in Eq.~\eqref{eq:HPM_Poisson} can then be solved using the Green function in the disk for homogeneous DBC:

\begin{widetext}
\begin{eqnarray}
	 \phi_{p,n}(x,y) &=& \int_{\mathcal{D}} d x_0 d y_0 \chi_p(x_0,y_0)\phi_{p,n-1}(x_0,y_0)G(x,y;x_0,y_0), 
    \label{eq:conformal_iteration}
    \\ 
     G(x,y;x_0,y_0) &=& \ln\left(r_0\sqrt{\frac{(x-x_0/r_0^2)^2+(y-y_0/r_0^2)^2}{(x-x_0)^2+(y-y_0)^2}}\right),
\end{eqnarray}
\end{widetext}
with $r_0^2=x_0^2+y_0^2$. In Fig.~\ref{fig:Conformal_WF}, we demonstrate the method in the symmetric $L$-shaped domain case (i.e., $p=1/2$ and $\zeta_1=1$) and show how the HPM approximates the exact ground-state energy as $n$ grows~\cite{Londergan_1999}.

\section{Analyses beyond the blockade limit}\label{sec:Appendix_InteractionTails}
Here, we discuss how the analyses conducted in the blockade limit also hold when considering the Van der Waals interaction $U_{v, u}$ in Eq.~\eqref{eq:hamiltonain_rydberg} beyond the blockade radius. Specifically, we focus on the domain-wall localization phenomenon occurring in the CWE setup (see Secs.~\ref{sec: effective th}C and \ref{sec:experiment}), but we mention that analogous reasoning applies to the vertex wire and the crossing gadget. 

Due to the tail interaction among nonblockaded atoms, the effective Hamiltonian in Eq.~\eqref{eq:Heff_2wires_general} acquires additional site-dependent diagonal terms. Specifically, the diagonal term $\epsilon(\bm{j})$ is given by the sum of the detuning $\Delta_v$ on the excited atoms of the domain-wall state $\ket{\bm{j}}$, plus the sum of the interactions $U_{v,u}$ between its pairs of excited atoms (at a distance $d_{v,u}$). In the case of the CWE setup of Fig.~\ref{fig:setups_gadgets}(d), due to its geometrical structure, these diagonal terms tend to localize the domain walls around the gadget. Indeed, to enforce the blockade constraints, the gadget atoms need to be placed at a shorter distance than those in the legs. This introduces an energetic penalty for states with more excited atoms near the gadget.

In the following, we focus on two possible strategies to mitigate this effect and recover the blockade-limit ground state with high fidelity. The first strategy relies on adjusting the position of the atoms along the vertex wires, while the second one relies on adjusting the local detuning on each atom. In both strategies, we aim to minimize the spreading of the diagonal terms $\delta \epsilon \equiv \max_{\bm{j}}{\left[\epsilon(\bm{j})\right]} - \min_{\bm{j}}{\left[\epsilon(\bm{j})\right]}$, since the blockade-limit is recovered in the $\delta\epsilon\ll t$ limit.

In the numerical simulations, we consider the parameters of Sec.~\ref{sec:experiment}A: $\Omega=\Delta/3 = 7.5$~rad/\textmu s, $C_6 = 5.42 \times 10^6$~\textmu m$^6$~$\times$~rad/\textmu s, $R_b=\left(C_6/\sqrt{(2\Omega)^2 + \Delta^2}\right)^{1/6}=7.65$~\textmu m, and the constraint of $d_{v,u}>4$~\textmu m; include the tail interaction between nonblockade atoms; and employ the exact diagonalization method.
In the implementation of Sec.~\ref{sec:experiment}A, we adopted the position-based mitigation strategy. 

\subsection{Position optimization to mitigate the tail interaction}
\begin{figure}
    \centering   \includegraphics[width=\columnwidth]{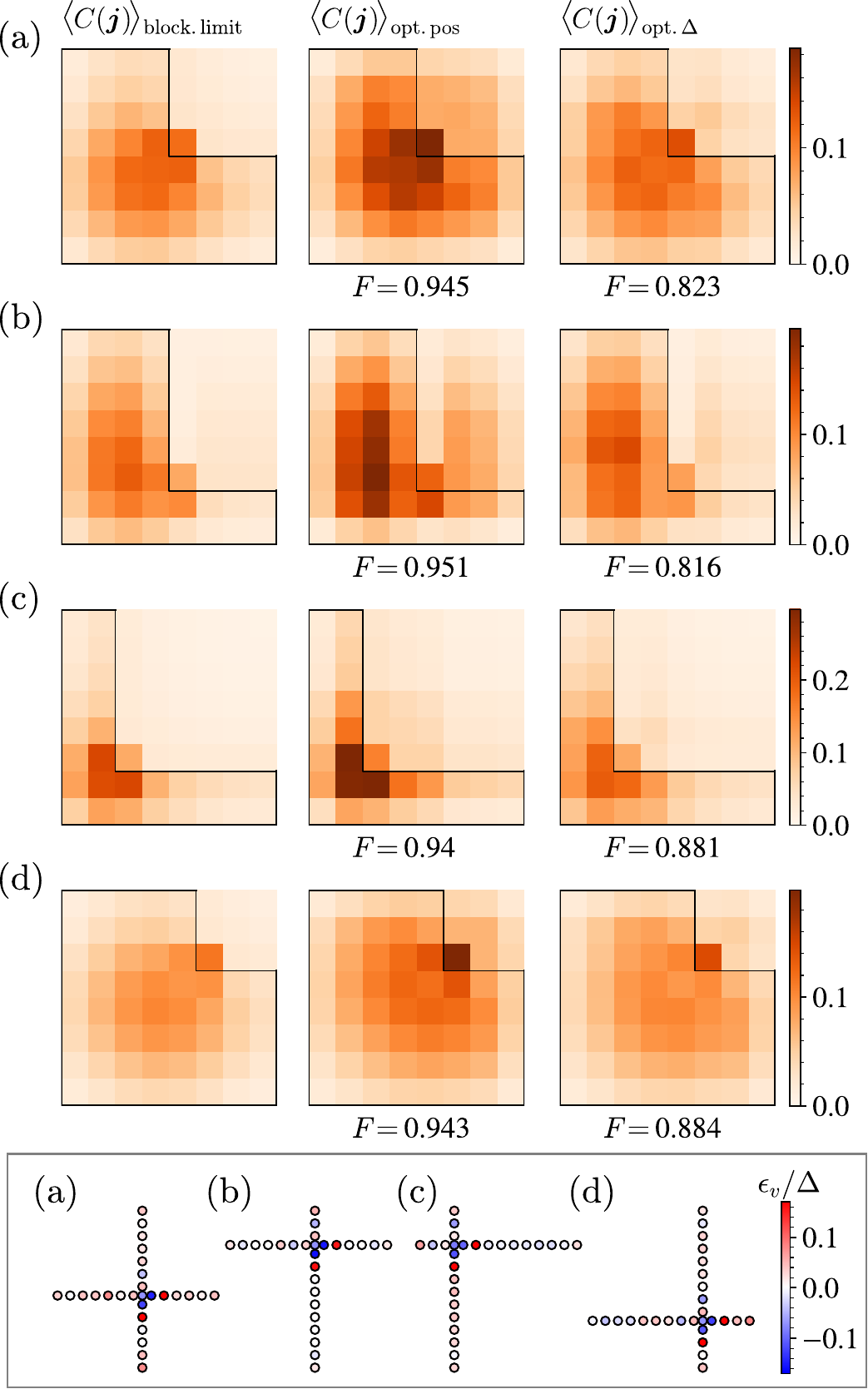}
    \caption{Mitigation of the tail interaction in the four configurations of the crossing-with-edge setup in Figs.~\ref{fig:protocol1_scaling_different_geometries} and \ref{fig:protocol2_scalings}, for $\Omega=\Delta/3=7.5$ rad/\textmu s. Expectation value of the domain-wall correlation operator in Eq.~\eqref{eq:dw_correlation} for the ground state: (left column) in the blockade limit, denoted as $\langle C(\bm{j})\rangle_{\rm block.\,limit}$; (central column)  when considering the optimized-position patterns in the left column of Fig.~\ref{fig:expermient}, denoted as $\langle C(\bm{j})\rangle_{\rm opt. \,dist}$; and (right column) when considering the optimal local-detuning patterns shown in the bottom inset, denoted as $\langle C(\bm{j})\rangle_{{\rm opt.}\,\Delta}$. $F$ indicates the fidelity between the optimized ground states and those in the blockade limit. The fidelity between two states $\ket{\psi_1}$ and $\ket{\psi_2}$ is defined as $F=\left|\langle \psi_1 | \psi_2\rangle \right|^2$. (Bottom inset) Optimal local-detuning patterns $\Delta_v = \Delta-\epsilon_v$ to mitigate the tail interaction.}
\label{fig:tails_detuning_distances}
\end{figure}
Here, we optimize the position of the vertex wire atoms to minimize $\delta \epsilon$, while fixing the position of the five atoms at the intersection of the two vertex wires. In particular, we set the distance between $G_c$ and its neighbors [see inset in Fig.~\ref{fig:setups_gadgets}(d)] to $0.9R_b/\sqrt{2}$ and the other distances (e.g., between $G_x$ and $G_y$) to $0.9R_b$. Then, we search for an optimal position pattern along the vertex wires, constraining the distance between neighboring atoms to be smaller than $0.9 R_b=6.88$~\textmu m (and $>4$~\textmu m) and the one between the second nearest neighbors to be greater than $1.2 R_b$. The optimal solutions are shown in the left column of Fig.~\ref{fig:expermient}. 

For $\Omega=\Delta/3$, states with more domain walls cannot be neglected, and the two-domain-wall wavefunction is generalized by the two-domain-wall correlation operator $C(\bm{j})$ in Eq.~\eqref{eq:dw_correlation}. 
In Fig.~\ref{fig:tails_detuning_distances}, we compare the expectation value of $C(\bm{j})$ for the ground state (left column) in the blockade-limit case and (central column) when including the tail interaction for the optimized configurations. In all cases, we achieve a fidelity $F\gtrsim0.8$, proving that the local rearrangement of the position of the atoms can efficiently be used to mitigate the tail interaction. 

\subsubsection{Measurement errors}
\begin{figure}
    \centering   \includegraphics[width=\columnwidth]{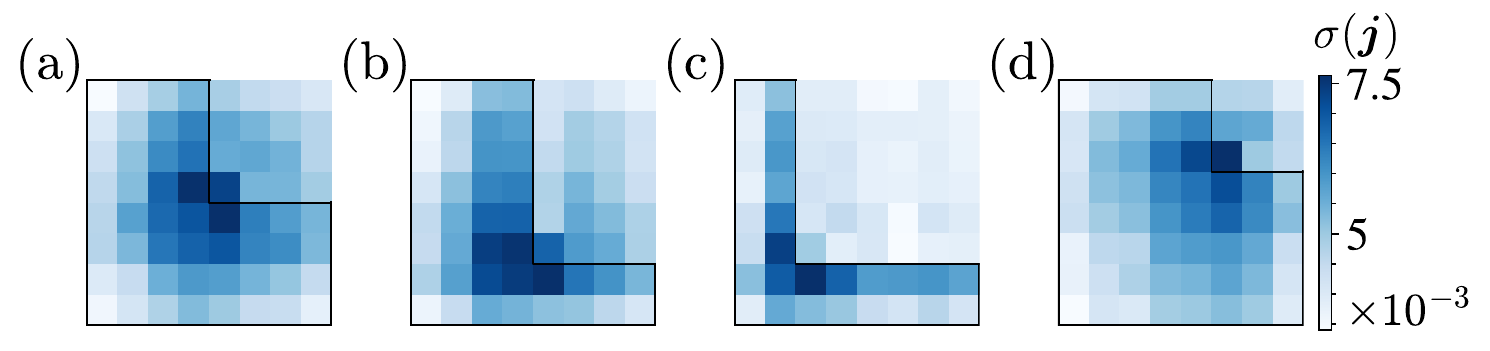}
    \caption{Standard deviation $\sigma(\bm{j})$ of the expectation value of the domain-wall correlation operator $C(\bm{j})$ (right column in Fig.~\ref{fig:expermient}). For each configuration, we perform 4500 measurements in the $\left\{\ket{0_v},\ket{1_v}\right\}$ basis.}
    \label{fig:exp_errors}
\end{figure}

In Fig.~\ref{fig:exp_errors}, we show the standard deviation  $\sigma(\bm{j})$ associated with the measurement of $\langle C(\bm{j})\rangle$ presented in the right column of Fig.~\ref{fig:expermient}.  For each configuration, we perform $N=4500$ measurements in the $\left\{\ket{0_v},\ket{1_v}\right\}$ basis. Each measurement returns a binary string of $0$ and $1$. Indicating with $C(\bm{j})_s$ the expectation value of $C(\bm{j})$ computed on the string $s$, we have $\sigma(\bm{j})=\sqrt{\frac{1}{N(N-1)}\sum_{s=1}^N \left(C(\bm{j})_s-\bar{C}(\bm{j})\right)^2}$, where ${\bar{C}(\bm{j})=\frac{1}{N}\sum_{s=1}^NC(\bm{j})_s}$.
\subsection{Local-detuning optimization to mitigate the tail interaction}
The spreading $\delta \epsilon$ can also be minimized by adjusting the local detuning on the atoms while keeping their position fixed. By parameterizing the detuning as $\Delta_v = \Delta - \epsilon_v$, with $\epsilon_v \ll \Delta$, the $\epsilon(\bm{j})$ term in the 
effective Hamiltonian in Eq.~\eqref{eq:Heff_2wires_general} acquires an additional contribution given by the sum of the detuning $\epsilon_v$ associated to the excited atoms of the corresponding state $\ket{\bm{j}}$. 

For all geometrical configurations, we search for a pattern of local detuning $\epsilon_v$ to minimize $\delta\epsilon$. The optimal patterns are shown in the bottom inset of Fig.~\ref{fig:tails_detuning_distances}, and the expectation value of the domain-wall correlation operator $C(\bm{j})$ computed on the associated ground states is shown in the right column. In all cases, we achieve a fidelity \(F \gtrsim 0.95\) between these ground states and the ones in the blockade limit, proving that the local adjustment of the detuning is an efficient strategy to mitigate the tail interaction. Furthermore, using local detuning instead of position optimization to mitigate tail interactions would allow us to accommodate longer wires on the current platform area (${75\:\mbox{\textmu m} \,\times\, 75\: \mbox{\textmu m}}$) by `snaking' the wires on the available 2D grid.

\section{Data analysis for the success probability}\label{App:exp_MIS}
\begin{figure}
    \centering
    \includegraphics[width=1\columnwidth]{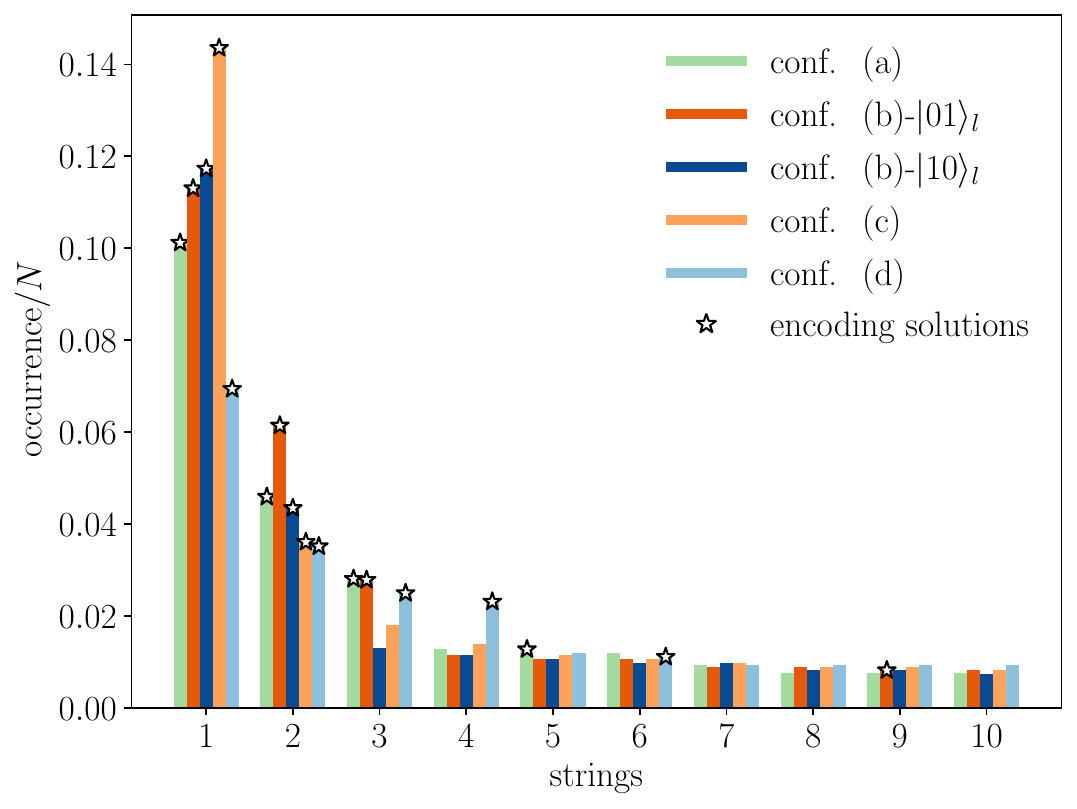}
    \caption{String occurrence of the raw data for the four configurations of the CWE setup (see Figs.~\ref{fig:expermient} and~\ref{fig:MWIS_experiment}). We report the occurrence of the most probable ten strings divided by the number of measurements $N$. We mark the strings encoding the solution to the encoded MWIS problem with a star.} 
    \label{fig:exp_MIS_occurrences}
\end{figure}
\begin{table}
\centering
\begin{tabular}{|c|c|c|c|}
    \hline
     Config.& N & $P_{\rm raw}$ & $P_B$  \\ \hline
    \:\:(a)\:\: & \:\:1176\:\: & \:\:$0.19\pm0.01$\:\: & \:\:$0.24\pm0.01$\:\:\\ \hline
    (b)-$\ket{01}_l$ & 1221 & $0.21\pm 0.01$ & $0.25\pm0.01$\\ \hline
    (b)-$\ket{10}_l$ & 1219 & $0.16\pm0.01$ & $0.20\pm 0.01$\\ \hline
    (c) & 1219 & $0.18\pm0.01$ & $0.22\pm0.01$ \\ \hline
    (d) & 1081 & $0.16\pm0.01$ & $0.22\pm0.01$\\
    \hline
\end{tabular}
\caption{Postprocessing for the success probabilities in Fig.~\ref{fig:MWIS_experiment}. For the four configurations of the CWE setup, we report the number of measurements $N$ where the atoms are correctly loaded. We report the probability of measuring the strings encoding the solution to the MWIS problem when: considering all the measurements, denoted as $P_{\rm raw}$; discarding the strings that violate the blockade, denoted as $P_{\rm B}$.}
\label{tab:post_processing}
\end{table}
Here, we discuss the data analysis for the success probabilities shown in Fig.~\ref{fig:MWIS_experiment}. 

For each configuration and target state, we load the atoms into optical tweezers 1500 times,  retaining only the instances where all the atoms are correctly loaded (whose number $N$ is given in Table~\ref{tab:post_processing}). For each of these instances, we perform the standard protocol shown in Fig.~\ref{fig:MWIS_experiment} (with $T=4$~\textmu s) and measure the final state in the $\{\ket{0_v},\ket{1_v}\}$ basis. In Fig.~\ref{fig:exp_MIS_occurrences}, we show the occurrence of the most probable strings divided by $N$, marking with a star the strings encoding the solution to the optimization problem. The success probability is then obtained by summing it over these strings encoding the solution. We denote this probability as $P_{\rm raw}$. Specifically, since we targeted the final state without local detuning control, but rather removing an atom from the system, the logical states are encoded in more than one classical string. In particular, while the logical state $\ket{1}_l$ on the odd wire is encoded by the string $1010\dots01$, the logical state $\ket{0}_l$ on the even wire is encoded by all the strings corresponding to a domain-wall before the gadget. 

In the raw data, we observe both strings that violate the blockade constraint and strings with multiple domain walls per wire. We attribute these effects to a combination of factors, including nonadiabatic evolution, detection errors, and decoherence. To minimize these effects, we perform the following postprocessing scheme. First, we discard the strings violating the blockade condition, obtaining a refined success probability, denoted as $P_B$ in Table~\ref{tab:post_processing}. Then, we perform a greedy vertex addition strategy following Ref.~\cite{Ebadi_2022}, which yields the final refined success probability $P$, shown in the top left of Fig.~\ref{fig:MWIS_experiment}.

\section{Ancillary atoms for the CWE gadget}\label{sec:Appendix_CWE_ancillae} 
Here, we discuss the \textit{extended} two-domain-wall space obtained when modifying the CWE-gadget geometry via the introduction of ancillary atoms. The \textit{original} two-domain-wall space [see Fig.~\ref{fig:effective_th}(c)] is effectively extended at the nonlogical boundary through the so-called \textit{copy} states, as shown in Fig.~\ref{fig:CWE_ancilla_combined_text}. In this extended two-domain-wall space, we can control the diagonal terms on all the boundary states. This allows us to exactly satisfy the hopping-detuning balance condition in Eq.~\eqref{eq:condition_flat_GS} on the entire two-domain-wall space, leading to the equal-superposition state as the ground state.

Particularly, we introduce the ancillary atoms $A_w$, $A_{Nw}$, and $B_w$ (with $w={x,y}$) to the CWE-gadget geometry, as shown in Fig.~\ref{fig:CWE_ancilla_combined_text}.
We then parameterize the detuning and Rabi frequency on $A_w$, respectively, as $\Delta+\epsilon_{A_w}$ and $\Omega_{A_w}$; on $A_{Nw}$ as $\Delta$ and $\Omega_{A_{Nw}}$; and on $B_{w}$ as $\epsilon_{B_w}$ and $\Omega_{B_w}$. Then, we work in the perturbative limit of $\epsilon_{i,w},\epsilon_{f,w},\epsilon_{A_w},\epsilon_{B_w}\ll\Delta$ and $\Omega, \Omega_{A_{w}}, \Omega_{A_{Nw}}, \Omega_{B_{w}}\ll\Delta$.

\subsection{Positioning of the ancillae \texorpdfstring{$A$}{A}}
First, we discuss the effective theory when introducing only the ancillary atoms $A_{w}$ and $A_{Nw}$ to the CWE gadget geometry. In the perturbative limit, we consider $H_0 = -\Delta  \sum_{v} n_{v}$ as unperturbed Hamiltonian, which has homogeneous detuning $\Delta$ on all atoms. 

The lowest-energy subspace of $H_0$ is an \textit{extended} two-domain-wall space.
The states are now labeled by the coordinates of the domain walls and the occupation numbers $n_{A_{x}}$ and $n_{A_{y}}$ of $A_{x}$ and $A_{y}$ as $\ket{j_x,j_y; n_{A_x}, n_{A_y}}$. For $n_{A_{x}}=n_{A_{y}}=0$, we retrieve the original two-domain-wall space; while, for $n_{A_{x}}=1$ and $n_{A_{y}}=0$ ($n_{A_{x}}=0$ and $n_{A_{y}}=1$), we retrieve the copy states of the nonlogical boundary line $\{j_x>\zeta_x+1,j_y=\zeta_y\}$ ($\{j_x=\zeta_x,j_y>\zeta_y+1$\}). The states located at the nonlogical boundary line $\{j_x>\zeta_x+1,j_y=\zeta_y\}$ ($\{j_x=\zeta_x,j_y>\zeta_y+1\}$) have an additional copy state, characterized by the same values of $j_x$ and $j_y$ but with $n_{A_{x}}=1-n_{A_{y}}=1$ ($n_{A_{y}}=1-n_{A_{x}}=1$).
At this point, the ancillae $A_{Nx}$ and $A_{Ny}$ do not further extend the lowest-energy subspace, but they are needed in combination with the ancillae $B_w$, as explained in the next subsection. Indeed, the occupation numbers $n_{A_{Nx}}$ and $n_{A_{Ny}}$ are fixed to be the negation of $n_{A_{x}}$ and $n_{A_{y}}$, respectively: $A_w$ and $A_{Nw}$ are blockaded with each other, and the states with $n_{A_{w}}=n_{A_{Nw}}=0$ are energetically unfavorable. 

In the lowest-energy subspace of $H_0$, the dynamics are determined by the effective Hamiltonian in Eq.~\eqref{eq:Heff} with perturbation term given by ${H_{1,A}=H_{\epsilon}+H_{\epsilon, {\rm A}}+H_\Omega+H_{\Omega,\rm A}}$, where
\begin{align}
\label{eq:ancilla_perturb_A}
    & H_{\epsilon} =   \sum_{w=x,y}  \left(\epsilon_{i,w} n_{i,w} + \epsilon_{f,w} n_{f,w}\right) , \nonumber \\
    & H_{\epsilon, \rm A} = -  \sum_{w=x,y} \epsilon_{A_w} n_{A_{w}}, \:\:\
    H_\Omega = \Omega \sum_{v \notin \{A_w, A_{Nw}, B_w\} } \sigma^x_{v} , \nonumber \\
    & H_{\Omega, \rm A} = \sum_{w=x,y}\left(\Omega_{A_{w}} \sigma^x_{A_{w}} +\Omega_{A_{Nw}} \sigma^x_{A_{Nw}}\right).
\end{align} 
Here, $H_\Omega$ and $H_\epsilon$ give rise to the \textit{original} effective Hamiltonian in  Eq.~\eqref{eq:Heff_2wires_general}; $H_{\epsilon,{\rm A}}$ contributes to the first-order corrections, introducing additional diagonal terms $-\epsilon_{A_w}$ on the copy states where $n_{A_{w}}=1$; and $H_{\Omega, \rm A}$ contributes to second-order corrections, leading to the following off-diagonal couplings: 
\begin{widetext}
\begin{equation}
\begin{split}
    \Omega_{A_x} \sigma^x_{A_x} \Omega_{A_{Nx}} \sigma^x_{A_{Nx}} \ket{x,y; 0, 0;  0,0} &= \Omega_{A_x} \Omega_{A_{Nx}}  \ket{ x,y;1,0;0,0}, \\
    \Omega_{A_y} \sigma^x_{A_y}\Omega_{A_{Ny}}\sigma^x_{A_{Ny}}\ket{x,y; 0, 0; 0,0 } &= \Omega_{A_y}\Omega_{A_{Ny}} \ket{ x,y; 0, 1;0,0}.
\end{split}
\end{equation}
\end{widetext}

We can thus control the diagonal terms also along the nonlogical boundary via the detuning on the ancillae. By setting ${\Omega_{A_w}=\Omega_{A_{Nw}}=\Omega}$, we homogeneously extend the hopping to the copy states and thus obtain the Hamiltonian represented in Fig.~\ref{fig:CWE_ancilla_combined_text}, with the addition of the orange sites only (and no blue sites).

Via this gadget modification, by setting $\epsilon_{A_x}=\epsilon_{A_y}=t$, we impose the hopping-detuning balance condition in Eq.~\eqref{eq:condition_flat_GS} on the whole nonlogical component of the two-domain-wall space boundary, apart from the internal corner states $\ket{\zeta_x+2, \zeta_y; 1,0}$ and $\ket{\zeta_x, \zeta_y+2;0,1}$. This results in an incomplete delocalization of the ground-state wavefunction, motivating us to add further ancillary atoms as discussed below.

\subsection{Positioning of the ancillae \texorpdfstring{$B$}{B}}
Second, we discuss the effective theory when further introducing the ancillary atoms $B_{w}$ (with $w={x,y}$), as shown in Fig.~\ref{fig:CWE_ancilla_combined_text}. These ancillae allow us to create an extra copy of the corner states $\ket{\zeta_x+2, \zeta_y; 1,0}$ and $\ket{\zeta_x, \zeta_y+2; 0,1}$, which enables us to impose the hopping-detuning balance condition in Eq.~\eqref{eq:condition_flat_GS} on the whole two-domain-wall space. 

In the perturbative limit, the extended two-domain-wall subspace (the lowest-energy subspace of $H_0$) is now also labeled by the occupation number of $B_w$: $\ket{j_x,j_y; n_{A_x}, n_{A_y}; n_{B_x}, n_{B_y}}$. 
We strategically position the $B_{x}$ ancilla in a way that it can only be excited for $j_x=\zeta_x+2$, $j_y=\zeta_y$, $n_{A_x}= 1$ and $n_{A_y}=0$. This effectively creates a copy, labeled by $\ket{{\zeta_x+2}, \zeta_y; 1,0; 1,0}$, of the corner state $\ket{\zeta_x, {\zeta_y+2};1,0; 0,0}$. Symmetrically, $B_y$ generates a copy, labeled by $\ket{\zeta_x, {\zeta_y+2};0,1; 0,1}$, of the other corner state $\ket{\zeta_x, {\zeta_y+2};0,1; 0,0}$.

In the lowest energy subspace of $H_0$, the dynamics are determined by the effective Hamiltonian in Eq.~\eqref{eq:Heff} with perturbation term given by  ${H_{1, AB} = H_{1,A}+H_{\epsilon, {\rm B}}+H_{\Omega,\rm B}}$. Here, $H_{1,A}$ is given by Eq.~\eqref{eq:ancilla_perturb_A}, and
\begin{equation}
    H_{\epsilon, \rm B} =  -  \sum_{w=x,y} \epsilon_{B_w} n_{B_{w}}, \
    H_{\Omega, \rm B} = \sum_{w=x,y} \Omega_{B_{w}} \sigma^x_{B_{w}}.
\end{equation}
These two terms contribute to first-order corrections.
$H_{\epsilon,{\rm B}}$ introduces additional diagonal terms $-\epsilon_{B_w}$ on the new copy states (where ${n_{B_{w}}=1}$), and $H_{\Omega, \rm B}$ extends the hopping to these copies via the following off-diagonal couplings: $\Omega_{B_x} \sigma^x_{B_x}\ket{ x,y;1,0;0,0}= \Omega_{B_x}\ket{ x,y;1,0;1,0}$, and $\Omega_{B_y} \sigma^x_{B_y}\ket{ x,y;0,1;0,0}= \Omega_{B_y}\ket{ x,y;0,1;0,1}$.

Finally, by setting ${\Omega_{A_w}=\Omega_{A_{Nw}}=\Omega}$ and ${\Omega_{B_w}=t}$, with $w=x,y$, we homogeneously extend the hopping to the copy states and obtain the Hamiltonian represented in Fig.~\ref{fig:CWE_ancilla_combined_text}(b).
The hopping-detuning balance condition in Eq.~\eqref{eq:condition_flat_GS} can now be satisfied for every site in the Hilbert space by selecting $\epsilon_{i,w} = \epsilon_{f,w} = t$, $\epsilon_{A_w}=t$ on $A_w$, and $\epsilon_{B_w} = 2 t$ on $B_w$, for $w=x,y$.

\section{Tackling exponential gaps via boundary Rabi frequency control}\label{sec:CWE_Rabi}

 \begin{figure}
     \centering
     \includegraphics[width=1.0\columnwidth]{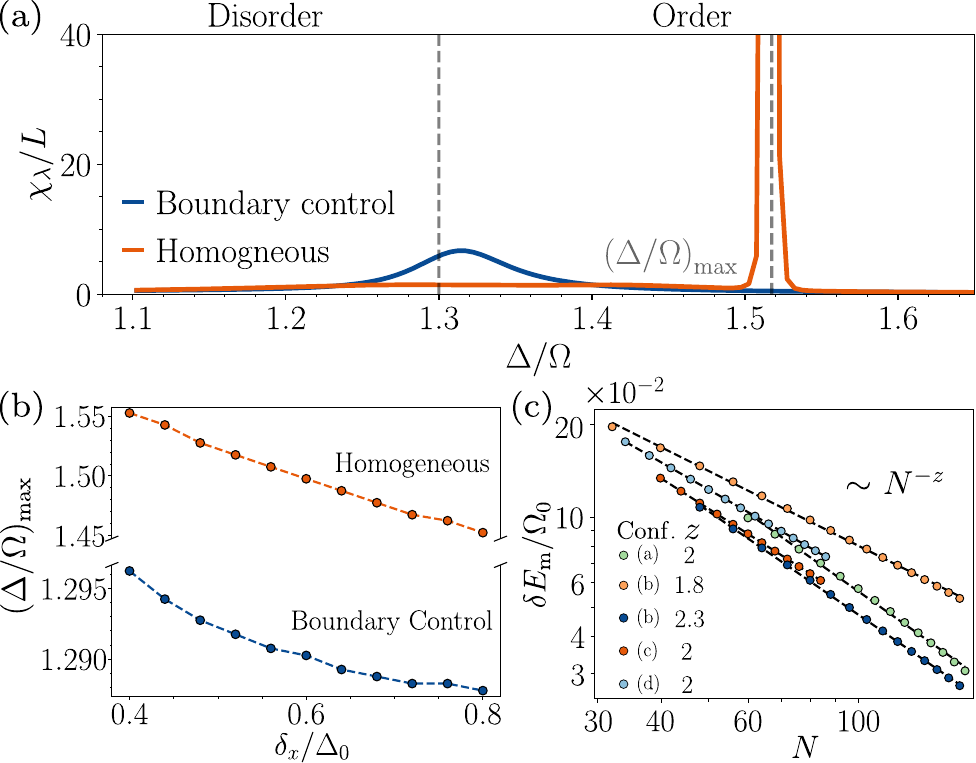}
     \caption{Boundary Rabi frequency control method. (a) Fidelity susceptibility [Eq.~\eqref{eq:crossing_chi} with $\delta\lambda=0.005$] along the standard protocol for the configuration in Fig.~\ref{fig:protocol1_scaling_different_geometries}(b) with $N=100$ atoms per vertex wire, setting $\delta_x/\Delta_0 = 0.6$ and $\delta_y/\Delta_0 = 0$. We compare the two cases of (orange) homogeneous Rabi frequency and of (blue) applying the boundary Rabi frequency control method. The dashed gray lines indicate the transition point known for one-dimensional Rydberg chains with nearest-neighbor blockade, $\Omega/\Delta \sim 1/1.3$~\cite{Bernien_2017}, and the point where the fidelity is maximum given homogeneous Rabi frequency, $(\Delta/\Omega)_{\rm max}\sim1.5$. (b) Position of the maximum of $\chi$, $\left(\Delta/\Omega\right)_{\rm max}$, along the standard protocol as a function of the wire weight $\delta_x/\Delta_0$ (for $\delta_y=0$), with $N=60$. 
     We compare the two cases of homogeneous Rabi frequency ($\delta\lambda=0.005$) and of applying the boundary Rabi frequency control method ($\delta\lambda=0.0005$).
     (c) Minimum gap scaling as a function of $N$ for the four configurations of Fig.~\ref{fig:protocol1_scaling_different_geometries} after applying the boundary Rabi frequency control method along the standard protocol. For the configuration in Fig.~\ref{fig:protocol1_scaling_different_geometries}(b), the data corresponding to the two different final ground states $\ket{01}_l$ and $\ket{10}_l$ are shown in orange and blue, respectively. 
    }
     \label{fig:CWE_improvement_rabi_1}
\end{figure}
Here, we present a strategy to mitigate exponentially closing gaps, making use of local control of the Rabi frequency at the boundary of the vertex wires. This approach is driven by numerical observations in the standard protocol.

We illustrate the method for the (b) configuration of Fig.~\ref{fig:protocol1_scaling_different_geometries} when targeting the $\ket{10}_l$ state. We first discuss the ground-state properties along the standard protocol. As $\Delta/\Omega$ increases above 1.3, the ground-state continuously transitions from the disordered to the ordered phase. Within the ordered phase, we detect a dominant peak in the fidelity susceptibility at around $(\Delta/\Omega)_{\rm max}\sim1.5$ (where $\chi\simeq1500$), corresponding to the exponentially closing gap dominating the adiabatic dynamics [see Fig.~\ref{fig:CWE_improvement_rabi_1}(a)]. 

We can circumvent this unfavorable gap by exploiting the boundary control of the Rabi frequency. In particular, we set the ratio between boundary and bulk Rabi frequencies to be the same as the one between boundary and bulk detuning, i.e., $\Omega_{1,w}/\Omega = \Delta_{1,w}/\Delta$ and $\Omega_{N,w}/\Omega = \Delta_{N,w}/\Delta$ (with $w=x,y$). In Fig.~\ref{fig:CWE_improvement_rabi_1}, we consider different values of the wire weight $\delta_x$ (with $\delta_y=0$) and show the position of the maximum of the fidelity susceptibility $(\Delta/\Omega)_{\rm max}$ in the two cases of homogeneous Rabi frequency and of boundary control. 
Notably, for the above choice of the boundary Rabi frequency control, the previously dominant peak disappears, and the fidelity presents only one peak around $(\Delta/\Omega)\sim1.3$ (where $\chi\simeq7$) [see Fig.~\ref{fig:CWE_improvement_rabi_1}(a)]. The dominant peak is therefore located at the continuous transition from the disordered to the ordered phase and thus results in a polynomial closing of the minimum gap. 
In Fig.~\ref{fig:CWE_improvement_rabi_1}(c), we show numerically how this strategy results in a polynomial scaling of the minimum gap along the standard protocol for all the configurations and target states of Fig.~\ref{fig:protocol1_scaling_different_geometries}. 

\bibliography{bibliography}
\end{document}